\documentclass[11pt]{article}
\usepackage[table,xcdraw]{xcolor}
\usepackage[a4paper]{geometry}
\usepackage[T1]{fontenc}
\usepackage[utf8x]{inputenc} 
\usepackage[affil-it]{authblk}
\usepackage{lmodern}
\usepackage[english]{babel}
\usepackage{amsfonts}
\usepackage{amssymb}
\usepackage{stmaryrd}
\usepackage{textcomp}
\usepackage{bbold}
\usepackage{amsmath}
\usepackage{authblk}
\usepackage{subfig}
\usepackage{caption}
\usepackage[toc,page]{appendix}
\usepackage{bbm} 
\usepackage{hyperref}
\usepackage{mathrsfs}
\usepackage[all]{xy} 
\usepackage{amsmath}
\usepackage{relsize}
\usepackage{tikz} 
\usepackage{graphicx}
\usepackage{wrapfig}
\usepackage{lscape}
\usepackage{rotating}
\usepackage{epstopdf}
\usepackage{babel}
\usepackage{amsthm}
\usepackage{amsmath}
\usepackage{verbatim}
\usepackage{float}
\usepackage{svg}
\usepackage{setspace}
\usepackage{listings}
\PrerenderUnicode{é}
\usepackage{amsmath}

\date{May 11th 2020}
\title{Influence Of Climate Change On The Corn Yield In Ontario And Its Impact On Corn Farms Income At The 2068 Horizon.}
\author[1]{Antoine Kornprobst  \thanks{Corresponding author: \texttt{antoinekor9042@gmail.com}}}
\author[2]{Matt Davison}
\affil[1]{University of Western Ontario, London (ON, Canada)}
\affil[2]{University of Western Ontario, London (ON, Canada)}

\begin{document}
\maketitle
\begin{abstract}
Our study aims at quantifying the impact of climate change on corn farming in Ontario under several warming scenarios at the 2068 horizon. It is articulated around a discrete-time dynamic model of corn farm income with an annual time-step, corresponding to one agricultural cycle from planting to harvest.  At each period, we compute the income given the corn yield, which is highly dependent on weather variables. We also provide a reproducible forecast of the yearly distribution of corn yield for the region around ten cities in Ontario. The price of corn futures at harvest time is taken into account and we fit our model by using 49 years of historical data. We then conduct out-of-sample Monte-Carlo simulations to obtain the farm income forecasts under a given climate change scenario.
\end{abstract}


\section{Introduction}
Climate change is now an accepted scientific fact and its denial is increasingly becoming an intellectually untenable position in the scientific community, as described in Björnberg, Karlsson, Gilek and Hansson (2017). All sectors of the economy are affected by climate change, but agriculture is naturally among the most exposed. The literature focusing on the economic and financial aspect of climate change is extensive, with numerous papers like Tol (2009) focusing on how the global economy should adapt to climate change and how climate change will impact the stability of the global financial system. For instance, Kolk and Pinkse (2004) explore how companies in many different sectors of activity adapt their financial and corporate strategy with respect to climate change, both from a purely operational point of view, since climate change is expected to directly or indirectly influence their business activities, and from the perspective of government policies and the regulatory response to the climate threat, which will also affect businesses. Dafermos, Nikolaidi and Galanis (2018) studied from a macro-economic point of view how climate change will impact global financial stability and monetary policy. How to hedge climate risk in a long term investment strategy is also a much discussed topic, as detailed in Andersson, Bolton and Samama (2016). The influence of climate change on farming from the point of view of agronomy and agricultural yields is well studied, for instance in Bootsma, Gameda and McKenney (2005), in Deryng, Sacks, Barford and Ramankutty (2011) or in Lobell and Field (2007). The impact of climate change on food production, though its influence on crop yields, has also been discussed in many research papers for decades, as in Katz (1977) or Almaraz, Mabood, Zhou, Gregorich and Smith (2008). On the other hand, the question of how climate change will impact the financial situation of farmers is still a relatively unexplored topic. Kaiser, Riha, Wilks, Rossiter and Sampath (1993) developed a farm-level analysis of a gradual climate warming on the economic situation of grain farmers in southern Minnesota under various climate scenarios and we took inspiration from their discrete-time dynamic model. Wang, Mendelsohn, Dinar and Huang (2010) created a multinomial logit model to study how farmers in China choose the optimal crop under several warming scenarios and use that model to make previsions at the 2100 horizon. Our own novel approach is focused on the financial health of corn farms in Ontario from a credit risk point of view. We study the income of farms, which directly impacts the owners' ability to repay their loans. In our whole study, we limit ourselves to grain corn, excluding fodder varieties. We study how several climate change  scenarios, from no warming at all (+0°C) to +4°C over the next 49 years at the horizon 2068, might impact the probability of default on loans granted to a corn farmers in Ontario. Our model is fitted using available historical data between 1970 and 2019 for the temperature in order to generate the climate change scenarios and for rainfall that will enable us to determine the start and the end of the corn growing season for each year. We took our inspiration from the work of McDermid, Fera and Hogg (2015) for the climate change scenarios. The price of corn futures is assumed to be constant and equal to the average price between 2009 and 2019 of a generic corn price future. This approximation is made in order to focus exclusively on the influence of climate change in our model. We then conduct Monte-Carlo simulations at the 2068 horizon in order to estimate the average income forecasts of the corn farms in the region around ten cities in Ontario, where most of the corn farming is concentrated in the province. This approach, which mixes together both climate variables and financial aspects has, to the best of our knowledge, never been conducted before. Our results are expected to be of great interest to both the financial institutions providing the loans and to the farmers receiving them, as well as to government planners at the local, national and international levels who are tasked with mitigating the harmful effects of climate change on the agricultural sector. While our numerical study is focused on corn farming in Ontario, our farm income model and Monte-Carlo techniques could be applied to any region and any crop, provided that financial, agricultural and climate data is available in order to fit the model and then conduct the simulations.

\section{Simulated Climate Change Paths}
We articulate our corn farm income simulations study around ten cities in Ontario: Brockville, Cornwall, Fergus, Kapuskasing, Kingsville, North Bay, Ottawa, Toronto, Trenton and Woodstock. Those cities are representative of the corn farming regions in Ontario according to the Ontario Ministry of Agriculture, Food and Rural Affairs (OMAFRA) census of land use conducted in 2011.\footnote{\url{http://www.omafra.gov.on.ca/english/landuse/gis/maps/Census2011/corn_cd.png}} The first step is to create, for each city, simulated daily temperature and rainfall paths under a given climate change scenario between 2019 and 2068. We need to simulate the daily maximum temperature, the daily minimum temperature and the daily rainfall. The temperature values will enable us to compute the corn heat units, which in turn will give us the simulated corn yield. The rainfall value will enable us to decide, through a set of rules explained later in Section 3, the dates for the start and the end of the corn growing season on a given year of the simulation between 2019 and 2068. All our historical weather data is obtained from the Global Historical Climatology Network Daily (GHCND) database of the National Oceanic and Atmospheric Administration (NOAA). The global identification number and precise location of the weather stations which have created the data that was used in our study is provided in Appendix A. The ten cities were chosen because they were representative of corn growing regions in Ontario and also because of practical considerations. Indeed, we were able to find for them uninterrupted and good quality weather data with sufficient depth in the NOAA database. In the rare instances when a temperature or rainfall entry was missing for a particular date, we carried over the last previous valid entry. For a given climate change scenario, we create 1500 paths. We will see later that this number is sufficient to obtain a stable and reproducible distribution of the corn yield for a given city and a given year of the simulation in the future. To create an individual climate path, we adopt the step by step block bootstrap method detailed below:

\begin{enumerate}
\item The 49 years of historical temperature and rainfall data are sliced by blocks of one year, from January 1st to December 31st. For each of the ten cities (Brockville: j=1; Cornwall: j=2; Fergus: j=3; Kapuskasing: j=4; Kingsville: j=5; North Bay: j=6; Ottawa: j=7; Toronto: j=8; Trenton: j=9 and Woodstock: j=10), the blocks are called $TMAX^{j}(i)$, $TMIN^{j}(i)$, $RAIN^{j}(i)$, for $i \in \llbracket1,49 \rrbracket$. The year 1970 corresponds to $i=1$ and the year 2019 corresponds to $i=49$.
\item For each city $j$ and for each year $i$ of the historical data, the average maximum daily temperature, minimum daily temperature and daily rainfall is computed. We call them $\overline{TMAX^{j}(i)}$, $\overline{TMIN^{j}(i)}$, $\overline{RAIN^{j}(i)}$. We then perform, for each city, a linear regression by the least squares method on the 49 values of $\overline{TMAX^{j}(i)}$, $\overline{TMIN^{j}(i)}$, $\overline{RAIN^{j}(i)}$. We therefore obtain yearly trends $\mathscr{T}_{tmax}^{j}$, $\mathscr{T}_{tmin}^{j}$ and $\mathscr{T}_{rain}^{j}$ for the minimum daily temperature, maximum daily temperature and daily rainfall respectively.  Those trends from 1970 to 2019 represent the historical climate change. We assume that they continue unchanged for rainfall and they are replaced by our climate change scenarios, from 0°C to +4°C, for the maximum and the minimum temperature in the future between 2019 and 2068. The values we obtained for the historical climate trends and the variance $\mathscr{V}_{tmax}^{j}$, $\mathscr{V}_{tmin}^{j}$ and $\mathscr{V}_{rain}^{j}$ of the series of $\overline{TMAX^{j}(i)}$, $\overline{TMIN^{j}(i)}$, $\overline{RAIN^{j}(i)}$ are displayed in Table 1 and Table 2.

\begin{table}[H]

\begin{tabular}{|l|l|l|l|}
\hline
  & $\mathscr{V}_{tmax}^{j}$ (tenth of °C)  & $\mathscr{V}_{tmin}^{j}$ (tenth of °C)  & $\mathscr{V}_{rain}^{j}$ (tenth of mm)   \\ \hline
Brockville & 0.456 & 0.457 & -0.116 \\ \hline
Cornwall & 0.547 & 0.433 & 0.052  \\ \hline
Fergus & 0.246 & 0.784 & 0.072  \\ \hline
Kapuskasing & 0.291 & 0.348 & 0.078  \\ \hline
Kingsville & 0.219 & 0.364 & 0.020  \\ \hline
North Bay & 0.472 & 0.255 & -0.139 \\ \hline
Ottawa & 0.402 & 0.415 & 0.114  \\ \hline
Toronto & 0.450 & 0.617 & 0.033  \\ \hline
Trenton & 0.264 & 0.326 & 0.132  \\ \hline
Woodstock & 0.323 & 0.245 & 0.143  \\ \hline
\end{tabular}
\caption{Historical climate trends per year in Ontario (1970-2019)}

\bigskip

\begin{tabular}{|l|l|l|l|}
\hline
  & $\mathscr{T}_{tmax}^{j}$ (tenth of °C)  & $\mathscr{T}_{tmin}^{j}$  (tenth of °C) & $\mathscr{T}_{rain}^{j}$ (tenth of mm) \\ \hline
Brockville & 91.6  & 106.4 & 13.2 \\ \hline
Cornwall & 97.0  & 91.3  & 12.3 \\ \hline
Fergus & 100.2 & 103.8 & 13.1 \\ \hline
Kapuskasing & 111.5 & 137.5 & 7.4  \\ \hline
Kingsville & 75.5  & 169.5 & 18.3 \\ \hline
North Bay & 106.9 & 124.5 & 19.9 \\ \hline
Ottawa & 92.0  & 83.7  & 10.2 \\ \hline
Toronto & 114.3 & 158.7 & 9.9  \\ \hline
Trenton & 81.0  & 81.7  & 11.4 \\ \hline
Woodstock & 98.6  & 86.3  & 21.1 \\ \hline
\end{tabular}
\caption{Historical climate variance in Ontario (1970-2019)}

\end{table}

Those values for our ten cities in Ontario, most of which are located in the south of the province where most of the corn production is concentrated, are consistent with the findings of an April 2019 report by the Canadian Government \footnote{Canada’s Changing Climate Report. \url{https://changingclimate.ca/CCCR2019/}}. They underline the scale of climate change in Canada, with warming trends as high as tree times the global average. Kapuskasing and North Bay are located more to the north and their emerging corn farming industry could benefit from warmer conditions, while of course the increased variability of the climate and the possibility of extreme events, both in terms of temperature and rainfall, could negatively offset any positive aspects of a warmer climate from the financial point of view of corn farming.

\item For each city $j$, the 49 years of a simulated climate path, under a given climate change scenario that assumes a warming of $+W$°C ($W \in \llbracket 0,4 \rrbracket$) and no extra rainfall besides the historical trend over the next 49 years, are sliced by blocks of one year from January 1st to December 31st. The new blocks are called $TMAX\_S^{j}(i)$, $TMIN\_S^{j}(i)$, $RAIN\_S^{j}(i)$, $i \in \llbracket1,49 \rrbracket$. The year 2020 corresponds to $i=1$ and the year 2068 corresponds to $i=49$.We perform a random permutation $\mathscr{P}$ of the integers between 1 and 49 and choose $TMAX\_S^{j}(i) =TMAX^{j}(\mathscr{P}(i))$; $TMIN\_S^{j}(i) =TMIN^{j}(\mathscr{P}(i))$ and $RAIN\_S^{j}(i) =RAIN^{j}(\mathscr{P}(i))$.

\item  We remove the historical trend, to be replaced by our scenarios in the next step, for the temperatures from each block, according to its former place in the historical data: $TMAX\_S^{j}(i)=TMAX^{j}(\mathscr{P}(i)) - \mathscr{T}_{tmax}^{j} \times \mathscr{P}(i)$ ;  $TMIN\_S^{j}(i) =TMIN^{j}(\mathscr{P}(i)) - \mathscr{T}_{tmin}^{j} \times \mathscr{P}(i)$. For the rain, we add to each block the historical trend according to its place in the simulation: $$RAIN\_S^{j}(i) =RAIN^{j}(\mathscr{P}(i)) +\mathscr{T}_{rain}^{j}\times (49-\mathscr{P}(i)+i) \ (1)$$

\item For the maximum and minimum temperature, we add to each block a random Gaussian perturbation term $\mathscr{N}(m,v)$, with mean $m$ and variance $v$, according to our chosen climate scenario and the block's position in the simulation. We lastly add a corrective term to account for the realized warming trends in the historical data. This is done in order to avoid a discontinuity in our climate paths at the interface between the historical and simulated parts. 
$$TMAX\_S^{j}(i)=TMAX^{j}(\mathscr{P}(i)) - \mathscr{T}_{tmax}^{j} \times \mathscr{P}(i) + \mathscr{N}(\dfrac{W\times i}{49},\sqrt{\mathscr{V}_{tmax}^{j}}) + \mathscr{T}_{tmax}^{j} \times 49 \ (2)$$ 
$$TMIN\_S^{j}(i) =TMIN^{j}(\mathscr{P}(i)) - \mathscr{T}_{tmin}^{j} \times \mathscr{P}(i) + \mathscr{N}(\dfrac{W \times i}{49},\sqrt{\mathscr{V}_{tmin}^{j}}) +  \mathscr{T}_{tmin}^{j} \times 49 \ (3)$$

\end{enumerate}
Our climate scenarios assume the value of the variable $W$ to be an integer between $0$ and $4$ degrees Celsius. According to the values in Table 1, the historical realized maximum temperature warming for the 49 years between 1070 and 2019 is between $1.2$ °C for Fergus and almost $2.7$ °C for Cornwall with an average of $1.8$ °C for the whole province. The historical realized minimum temperature warming for the 49 years is generally higher, from $1.2$ °C for Woodstock to more than $3.8$°C for Fergus with an average of $2$°C for the province. Roughly speaking, we can say that our historical climate data shows that, on average, the corn growing regions of Ontario have experience a $2$°C warming over the past five decades. Since we have removed the historical trend between 1970 and 2019 at the fourth step of the climate path creation method, a climate scenario at the 2068 horizon defined by $W=0$°C in our framework corresponds to a break of the historical trend and no warming at all over the length of our simulations. It is obviously not meant to be a realistic depiction of a possible future for the climate in Ontario but it will provide us with a useful limit case. Similarly, the climate scenario defined by $W=1$°C corresponds to a slowing down of the climate warming trend, possibly through climate change mitigation programs at the local, national and international levels. The climate scenario defined by $W=2$°C represents essentially a continuation of the warming trend that has been going on since 1970 and the climate scenarios corresponding to $W=3$°C  and $W=4$°C describe at the 2068 horizon an accelerating warming of the climate to the level of up to twice the historical trend. The rainfall aspect of a climate scenario is modeled differently since we always assume of a continuation of the historical trend, which is very small for all cities considered. As an illustration, we have represented, in Figure 1 to Figure 9, the three components of a climate path (maximum temperature, minimum temperature and rainfall) for the city of Brockville. We assume first that $W=0$°C (reversal of the historical warming trend, no more warming), then that $W=2$°C (continuation of the historical warming trend, status quo) and finally that $W=4$°C (doubling of the historical warming trend, acceleration of climate change). All the other results for each of the ten cities and each of the five values of $W$ are available as supplementary online material. 

\begin{figure}[]
	\begin{minipage}{0.30\textwidth}
		\centering
		\includegraphics[width=52mm]{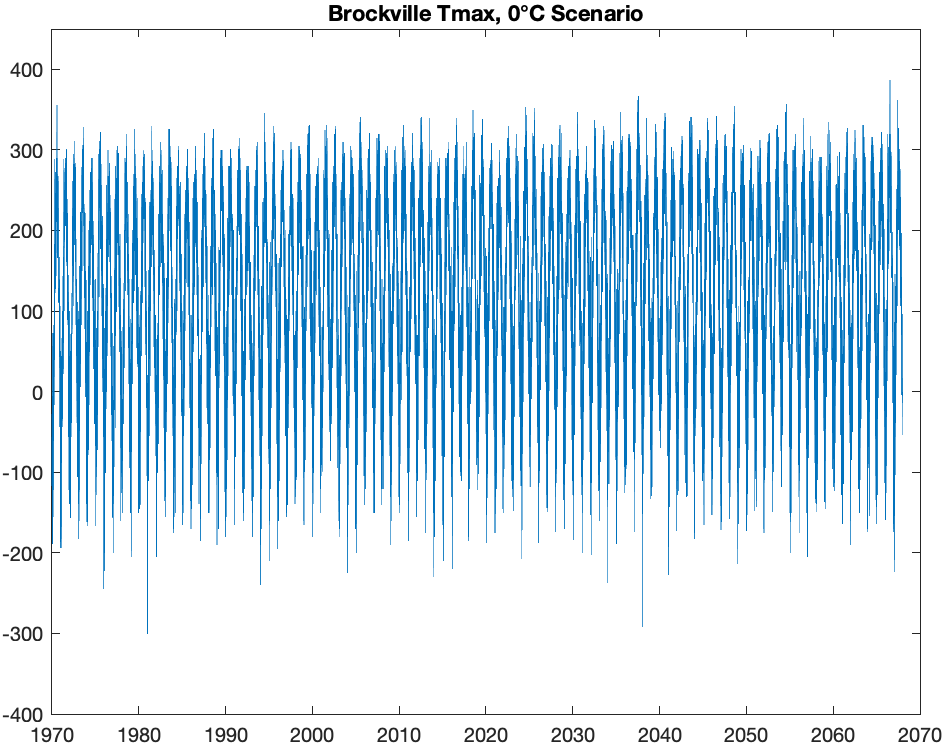}
		\caption{Maximum daily temperature ($W=0$°C)}
	\end{minipage}\hfill
	\begin{minipage}{0.30\textwidth}
		\centering
		\includegraphics[width=52mm]{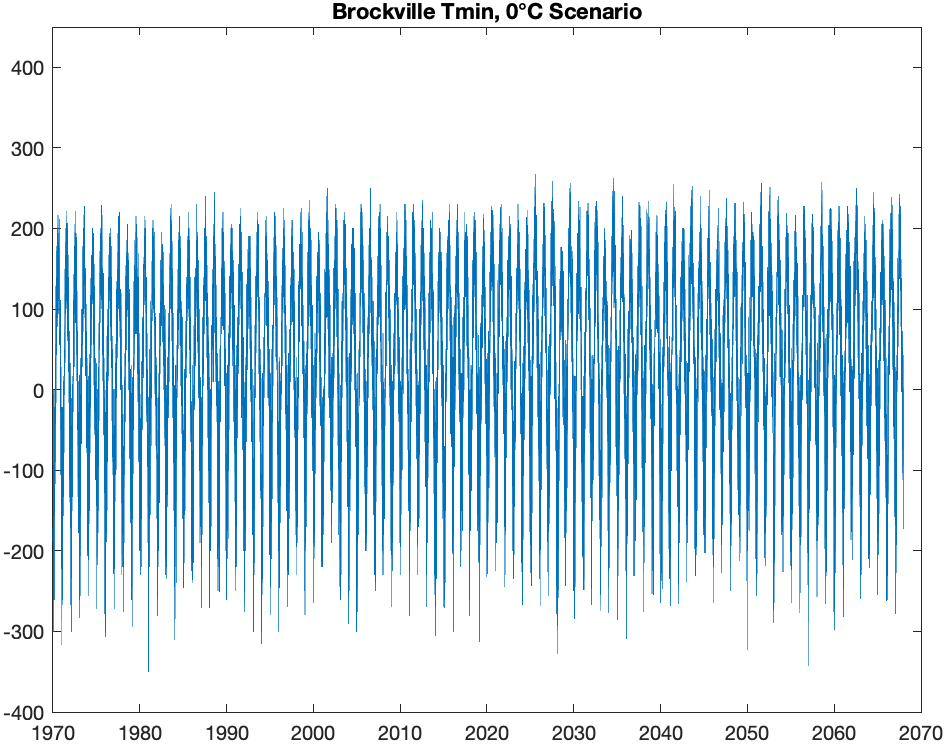}
		\caption{Minimum daily temperature ($W=0$°C)}
	\end{minipage}\hfill
	\begin{minipage}{0.30\textwidth}
		\centering
		\includegraphics[width=52mm]{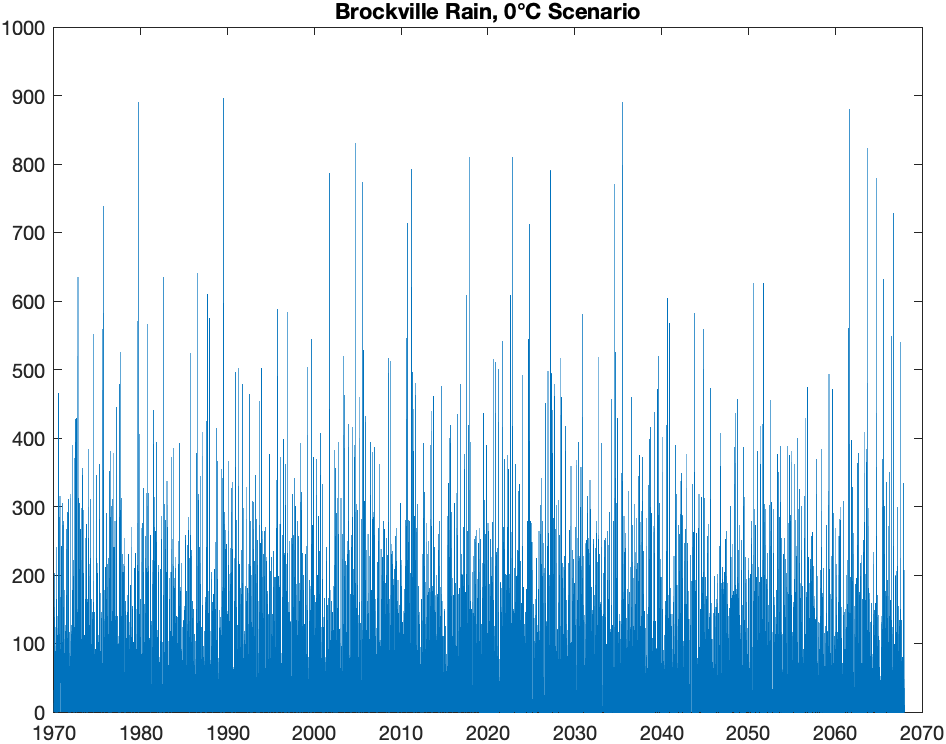}
		\caption{Daily rainfall ($W=0$°C)}
	\end{minipage}\hfill
	\begin{minipage}{0.30\textwidth}
		\centering
		\includegraphics[width=52mm]{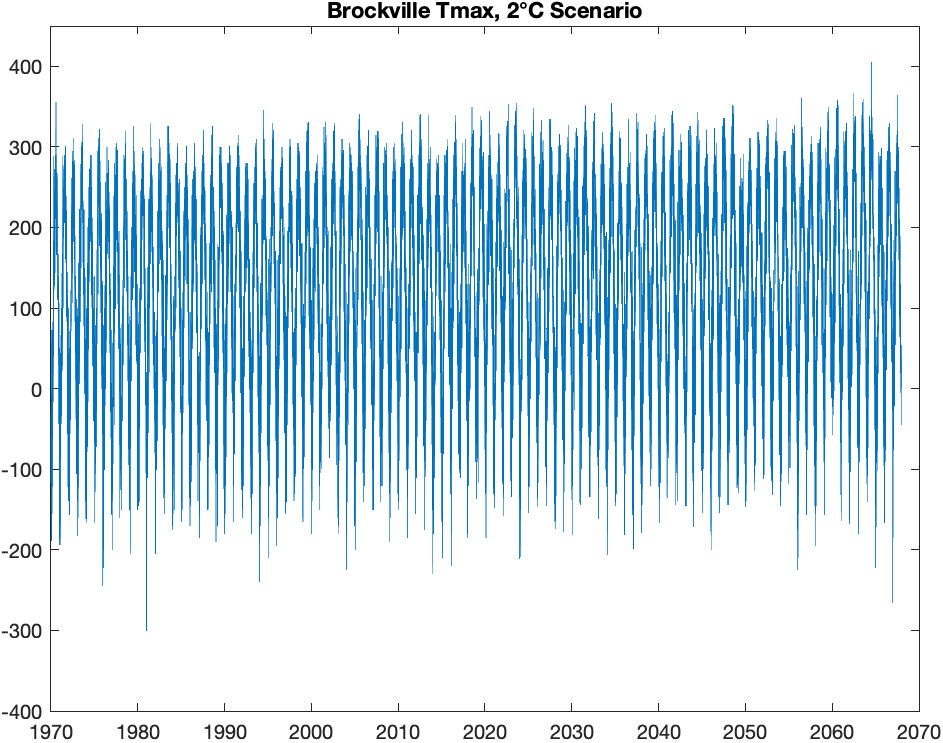}
		\caption{Maximum daily temperature ($W=2$°C)}
	\end{minipage}\hfill
	\begin{minipage}{0.30\textwidth}
		\centering
		\includegraphics[width=52mm]{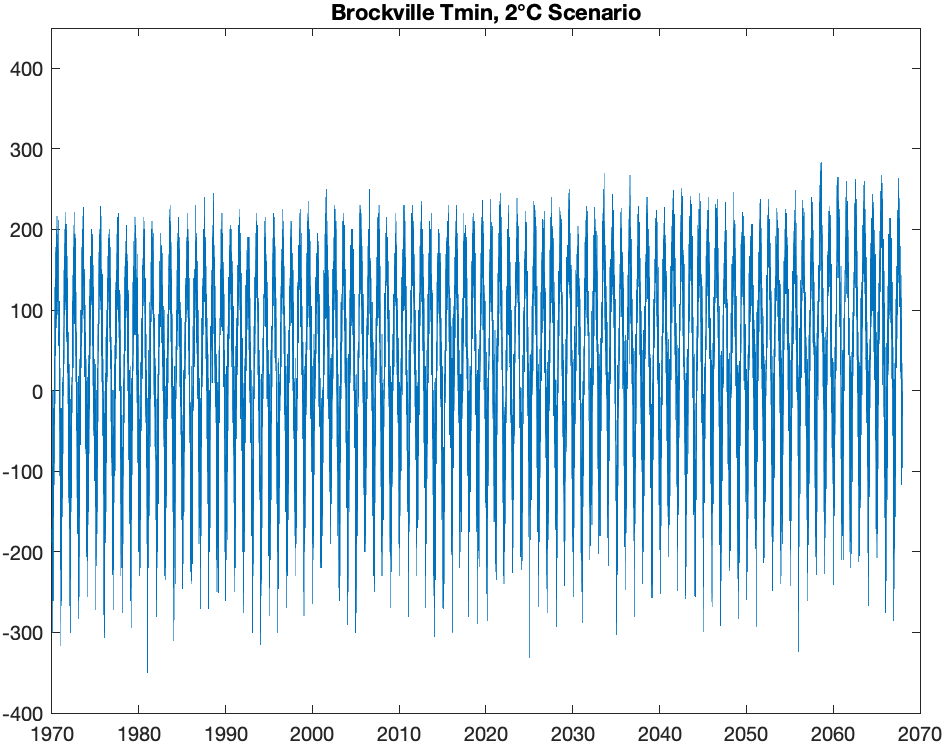}
		\caption{Minimum daily temperature ($W=2$°C)}
	\end{minipage}\hfill
	\begin{minipage}{0.30\textwidth}
		\centering
		\includegraphics[width=52mm]{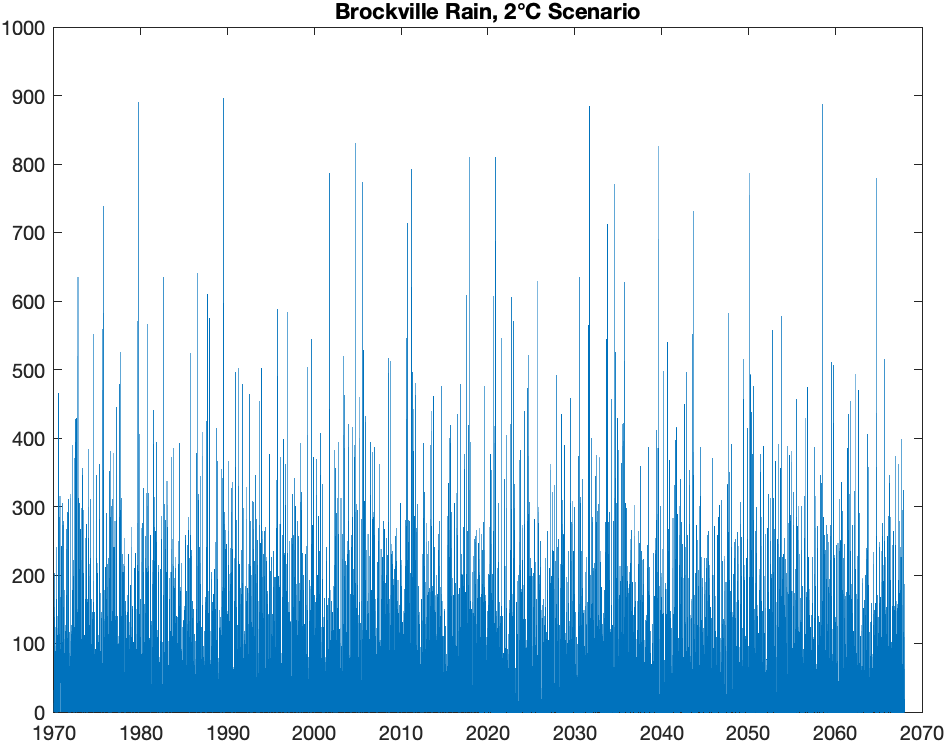}
		\caption{Daily rainfall ($W=2$°C)}
	\end{minipage}\hfill
	\begin{minipage}{0.30\textwidth}
		\centering
		\includegraphics[width=52mm]{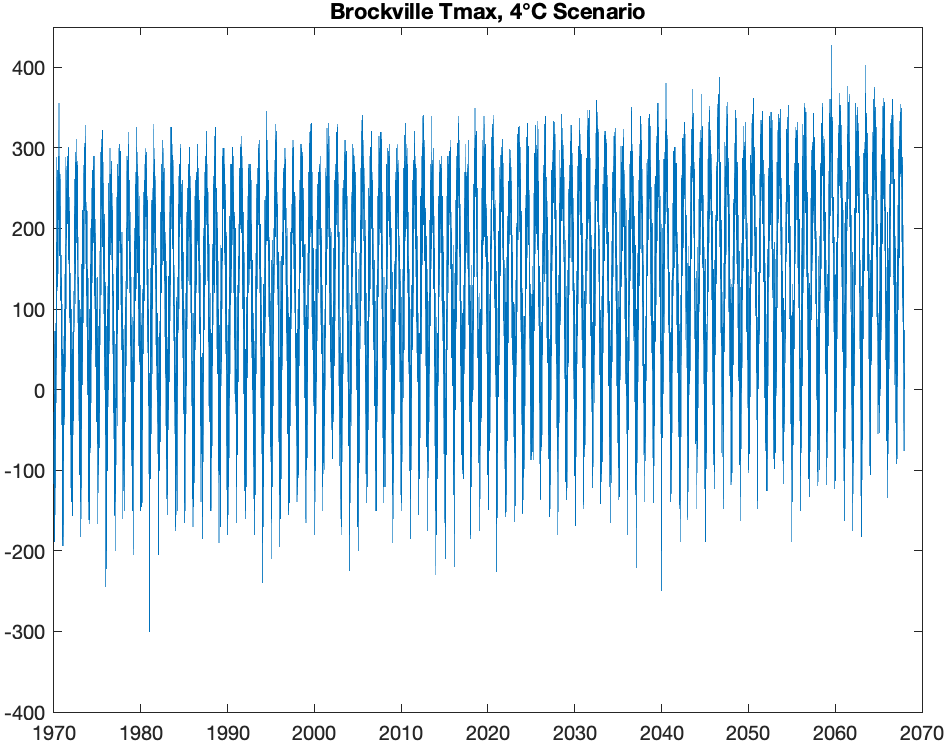}
		\caption{Maximum daily temperature ($W=4$°C)}
	\end{minipage}\hfill
	\begin{minipage}{0.30\textwidth}
		\centering
		\includegraphics[width=52mm]{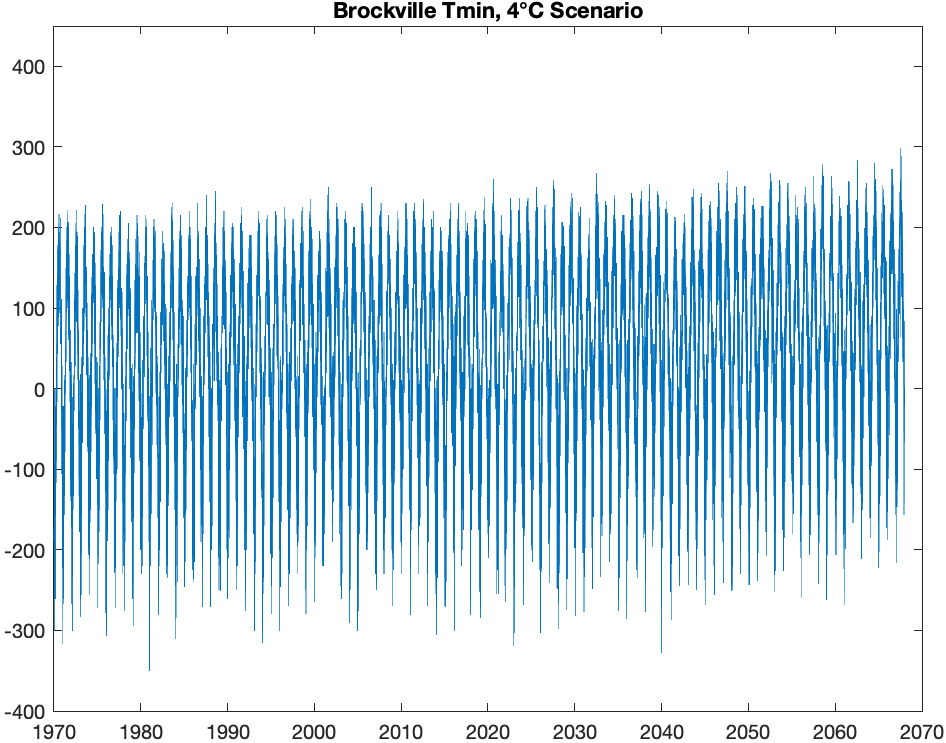}
		\caption{Minimum daily temperature ($W=4$°C)}
	\end{minipage}\hfill
	\begin{minipage}{0.30\textwidth}
		\centering
		\includegraphics[width=52mm]{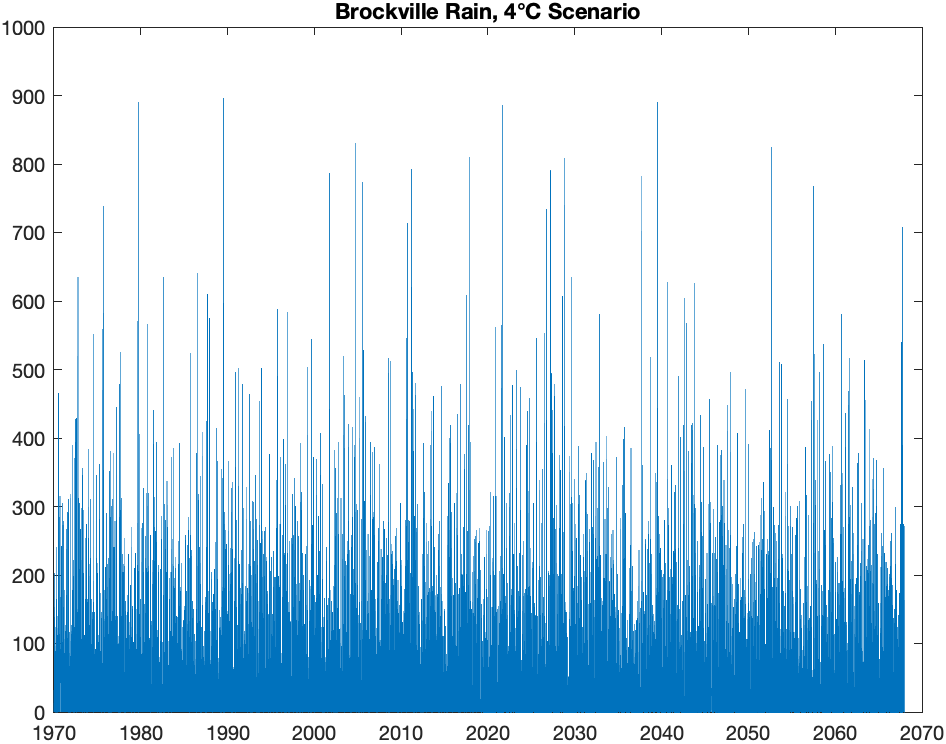}
		\caption{Daily rainfall ($W=4$°C)}
	\end{minipage}\hfill
\end{figure}

\section{Simulated Corn Yield Paths}
Now that we have simulated paths for the climate variables, we switch our attention to creating corn yield paths for our farm income model. The first step toward the creation of simulated corn yield paths is  to compute for each year in the future the sum over the growing season of the daily corn heat units (CHU) received for each city by the corn crop. Let us consider one climate path, constituted of the daily maximum temperature, the daily minimum temperature and the daily rainfall,  obtained by the method describe above. For each year $i \in \llbracket1,49 \rrbracket$ of the simulation and for each city $j \in \llbracket1,10 \rrbracket$, we can compute the daily CHU on each day. The corn heat units depend only on the temperature maximum and minimum. We call $H^{j}_{i}$ is the sum of the daily corn heat units over the corn growing season for the city $j$ at time step $i$. The computation of $H_{i}^{j}$ is achieved by using a well established method, given in Formula (4). It is used both in purely academic papers like Kwabiah, MacPherson and McKenzie (2003) as well as in industry reports and handbooks like Brinkman, McKinnon and Pitblado (2008). The numerical coefficients in the formula are computed for corn farming in Ontario, which is the focus of our numerical study, but as explained in Kwabiah, MacPherson and McKenzie (2003), we believe that the formula would still be valid for corn farming in similar cool climate ecosystems like the U.S Midwest, Northern Europe and Eastern China, which together represent the overwhelming majority of global commercial corn farming. In Formula (4), the summation is over each day $k$ of the growing season of length $N_{i}^{j}$ for the year corresponding to our time step $i$ and for the city $j$. The length of the growing season has been studied as an important indicator of climate change for agriculture, as explained in Brinkmann (1979). According to Cabas, Weersink and Olale (2010), the length of the growing season, which depends only on rainfall considerations in our framework, has a very strong impact on several crop yields, especially corn, in southwestern Ontario. The effects of climate change on crop yields in Ontario are also studied in details in Smit, Brklacich, Stewart, McBride, Brown and Bond (1989). Again, the length of the growing season is one of the determining factors.
$$H_{i}^{j}=\mathlarger{\mathlarger{\sum}}_{k=1}^{N_{i}^{j}}\frac{1}{2} [1.8(Tmin_{k}^{j}-4.4)+3.3 (Tmax_{k}^{j}-10))-0.084  (Tmax_{k}^{j}-10)^{2}] \  (4)$$
Since precise county level historical data was not available to us for the growing seasons in Ontario, which are defined as the the time between the planting date and the harvest date, we adopted an approach that is based on published agronomic studies that we modified to include the influence of our climate change scenarios through a set of rules based on rainfall. We do not claim that this model is very realistic, but it serves our purposes for this study and it relies on the common sense consideration that corn farmers need a relatively wet soil to plant their seeds at the end of Spring  and a firm ground to harvest their crop, especially with motorized equipment,  at the beginning of Autumn. We grounded our approach in average historical planting and harvesting dates for corn discussed in Sacks, Deryng, Foley and Ramankutty (2010). In this wide ranging paper about planting and harvesting patterns for a variety of crops, the authors state that corn planting in the northern hemisphere generally occurs in April and May, while harvesting takes place in mid to late October. They also found that soil moisture often determines the length of the growing season, much more than temperature related considerations. This led us to the establishment our rules based on rainfall in order to trigger planting and harvest. The work of Kucharik (2006) about corn planting trends in the United States was also inspirational to us. To determine the length of the growing on a given year of the simulated path for one of our ten cities in Ontario, we started from the time averaged historical corn planting and harvesting dates provided by Sacks, Deryng, Foley and Ramankutty (2010) and we used the online database associated with the paper as well \footnote{\url{https://nelson.wisc.edu/sage/data-and-models/crop-calendar-dataset/index.php}}. That is June 1st ($D_{1}$) for planting and October 25th ($D_{2}$) for harvest. This is a simplification of the author's work for the purpose of our study. Indeed, in Sacks, Deryng, Foley and Ramankutty (2010), they differentiate between the date when planting (resp, harvesting) start and the date when planting (resp, harvesting) stops, making the boundaries of the  growing season a little blurry, as it is of course in real life. We chose  our $D_{1}$ and $D_{2}$ as the average of the start and end  dates provided in Sacks, Deryng, Foley and Ramankutty (2010).  Starting from those dates that we use to anchor our simulated growing seasons in the future, we add the following rules based on the simulated rainfall data inside the climate path being considered:
\begin{itemize}
	\item The growing season starts (planting) $\pm 15$ days around $D_{1}$, after the first occurrence of three consecutive days with a strictly positive rainfall, or at $D_{1}+15$.
	\item The growing season ends (harvest) $\pm 15$ days around $D_{2}$, after the first occurrence of three consecutive days with zero rainfall, or at $D_{2}+15$.
\end{itemize}
Experimentally while doing our computations, we found that the length of the growing season, which drives the size of the $H_{i}^{j}$ and therefore  the corn yield, as we will see below, upon which the farm income depends, has a large influence. It was not possible to create a meaningful impact on farm income using the temperature variability of a given climate change scenario alone. Rainfall, through its influence on the length of the growing season, was essential in order to properly model the impact of climate change on the income of corn farms in Ontario. That is a very interesting result because, given that the CHU increases with heat and the corn yield in turn increases with CHU, one could be tempted to draw the simplistic conclusion that a warming climate is purely beneficial for the corn farming industry in Ontario. It is essentially true that a warmer climate helps corn crops, especially for the northern cities, Kapuskasing and North Bay, but the true influence of climate change, for example through shorter growing seasons due to extreme rainfall, or absence thereof, events could overall be much more unpredictable, especially at time frames beyond the 49 years horizon of our study. Increased uncertainty is always a bad thing from a financial point of view because uncertainty means risk and risk costs farmers money through higher option prices, lower future prices and insurance premiums.\\

\noindent Now that we know how to compute the CHU, we move to the computation of the corn yield itself. The corn yield $Y$  is defined in Formula (5) and Formula (6).

\begin{itemize}
\item For any year $h  \in \llbracket1,49 \rrbracket$ of the historical data, the yield for the city $j$ is : $$Y^{j}_{h}=C_{0}^{j}+C_{1}^{j}\times h + C_{2}^{j} \times H^{j}_{h}\  (5) $$ The computation of the CHU does not change when considering historical climate data ($h  \in \llbracket1,49 \rrbracket$) from 1970 to 2019, instead of one of our simulated climate paths from 2019 to 2068 ($i  \in \llbracket1,49 \rrbracket$). It is still achieved by using Formula (4). The coefficients $C_{0}^{j}$, $C_{1}^{j}$ and $C_{2}^{j}$ are obtained by linear regression using the least square method on historical yield data that we possess for Ontario down the county level. They are independent of the climate path being  considered since they are obtained only from the historical climate data. They depend only on the city $j$. The constant $C_{1}^{j}$ represents the technology trend that is responsible for most of the increase in yearly corn yield over the last five decades through the progressive availability over time of better fertilizers, motorized farming equipment, better seeds for corn varieties adapted to the local environment as well as genetic engineering. The influence of the warming climate on the corn yield since 1970, as we have seen with the temperature trends contained in Table 1, is realized through the CHU. The database of historical corn yields at county level in Ontario is presented in Appendix B. The data was provided by the Ministry of Agriculture, Food and Rural Affairs (OMAFRA) \footnote{\url{www.omafra.gov.on.ca/english}}. The data from 2004 onward is published online and the older entries back to 1970 were graciously obtained through a direct inquiry to the Ministry's Economic Department. The data provided in appendix is expressed in bushel per acre that we converted in our computations to metric tonnes per hectare. The coefficients that we obtained for each city as well as the goodness of fit of the linear regression are contained in Table 3.

\begin{table}[H]
	\begin{center}
	\begin{tabular}{|l|l|l|l|l|}
		\hline
		& $C_{0}^{j}$      & $C_{1}^{j}$      & $C_{2}^{j}$         & gof     \\ \hline
		Brockville  & 1.19 & 1.21E-01 & 8.68E-04  & 82.86\% \\ \hline
		Cornwall    & 6.37E-01 & 1.44E-01 & 1.18E-03  & 88.22\% \\ \hline
		Fergus      & 5.67E-01 & 1.19E-01 & 1.44E-03  & 82.07\% \\ \hline
		Kapuskasing & 3.00 & 8.60E-02 & 4.68E-06  & 42.03\% \\ \hline
		Kingsville  & 4.36& 1.48E-01 & -4.00E-05 & 62.32\% \\ \hline
		North Bay   & 1.59 & 7.67E-02 & 8.13E-04  & 43.44\% \\ \hline
		Ottawa      & 4.26 & 1.45E-01 & -1.50E-04 & 82.26\% \\ \hline
		Toronto     & 3.90 & 1.14E-01 & 7.33E-05  & 76.75\% \\ \hline
		Trenton     & 2.81 & 1.08E-01 & 3.30E-04  & 67.85\% \\ \hline
		Woodstock   & 3.49 & 1.36E-01 & 5.91E-04  & 86.45\% \\ \hline
	\end{tabular}
	\end{center}
\caption{Regressed coefficients for the historical yields and goodness of fit.}
\end{table}

The goodness of fit is excellent for all cities, which validates our approach, except for the two northern ones. It was to be expected given the gaps in the historical data, which produced  plateaus once we carried over the last valid entry, degrading the quality of the linear regression. As expected the technological trend $C_{1}^{j}$ dominates the influence of climate change expressed through the CHU. The coefficient $C_{2}^{j}$ is always small relative to $C_{1}^{j}$, which reflects the fact that most of the yield gain over the past five decades were due to technological improvements in the corn farms. More surprising, $C_{2}$ is extremely small but negative for Kingsville and Ottawa. It is also extremely small for Toronto, rendering the effect of climate change almost imperceptible for this city in our framework. This shows that in our study, the yield does not necessarily always increase with the CHU, which may sound strange at first but it does reflects the fact that we have included both the temperature and the rainfall in our computation of the CHU under a given climate change scenario. As we have already underlined, more heat (within reason) tends to help corn crops, but increased variability of rainfall, accompanied by the possibility of more frequent extreme events, may shorten the growing season and interfere with the planting or harvest dates. These competing effects of temperature and rainfall  on corn farming in Ontario renders the real influence of climate change difficult to predict for the province as a whole and there seems to be a lot of variability depending on the city considered. Besides those considerations, the choice of Formula (5) as a bilinear function of the CHU and time was not the only one available to us.  Liang, MacKemie, Kirby  and  Remillard  (1991) propose a more elaborate model of corn yield computation that explicitly includes rainfall, while our approach keeps the influence of rainfall limited to the computation of the CHU itself, through the length of the growing season. Their model for the corn yield  does not feature a technology trend however. Fitting it to our historical data over the last five decades  would therefore have implied that the large increase of corn yield in Ontario over the past decades was due only to the evolution of climate variables. This was clearly unreasonable, since the influence of climate change on corn yield is subtle and most of the increase that we witness over the past decades in the historical data is due to technological advances and more efficient methods of farming.

\item For any year $i  \in \llbracket1,49 \rrbracket$ in the future, given a climate path under a chosen climate scenario of $+W$°C ($W \in \llbracket 0,4 \rrbracket$), the yield for the city $j$ is defined is given in Formula 6. $$Y^{j}_{i}=C_{0}^{j} + C_{2}^{j} \times H^{j}_{i}\  (6) $$ While Formula 6 may seem simplistic, modeling the corn yield as a linear function of CHU is often used in agronomic studies, particularly in the context of climate change. This is for example the case in the reports from Agriculture and Agri-Food Canada (AAFC) about climate change scenarios for agriculture \footnote{Climate Change Scenarios for Agriculture \url{www.mcgill.ca/brace/files/brace/Gameda.pdf}}. Our purpose in this study is to measure the influence of climate change only. We therefore assume that the technology will not improve after 2019 and thus we removed the term that contained the technology trend $C_{1}^{j}$ in Formula (6). This is of course a simplification. Indeed, while the corn yield will necessarily tend to plateau in the future because the big technological changes in agriculture, like the advent of pesticides, fertilizers and machines, are in the past, it is very conceivable that technological advances will still drive a large  increase of farms efficiency for many years. The coefficients $C_{0}^{j}$ and $C_{2}^{j}$ are those that were  computed for a given city $j$ by fitting Formula (5) to the historical county level corn yield data provided in Annex B. Since $C_{0}^{j}$ and $C_{2}^{j}$ are now known, we can now compute the corn yield for all the dates $i$ in the future. After computing a climate path under a given climate change scenario, the CHU $H^{j}_{i}$ is obtained by using Formula (4) and we compute $Y^{j}_{i}$ by applying Formula (6).
\end{itemize}

\section{Corn Yield Distributions and Farm Income}

We now have successfully created corn yield paths from our temperature and rainfall paths under a given climate change scenario. Given one of our ten cities in Ontario and a warming factor $W$, we now wonder how many climate paths are needed in order to obtain stable and reproducible results. More precisely, we need a stable and reproducible distribution of the simulated corn yield for each year between 2020 and 2069. In our framework, we have limited ourselves to 1500 climate paths for one realization of the model. This is mostly due to hardware limitations in our computations but this number of paths is enough for our purposes and we now demonstrate that fact by studying four independent realizations of the model for a given city $j \in \llbracket 1, 10 \rrbracket$ and a given scenario $W \in \llbracket 0, 4 \rrbracket$. We decided to work with generalized extreme value distributions (GEV). We initially considered fitting our simulated data to a  Gaussian distribution for simplicity, however even though the log-likelihood of a Gaussian fit was of the same order of magnitude as the one obtained for a GEV fit, the versatility of this latter type of densities and its ability to fit data with heavy shifting skew and fat tails, as this is often needed to describe the distribution of the corn yield, made us decide to abandon a normal approach. The probability density function $Psi$ of a GEV is provided in Equation (7). Parameter $\mu$ is the mean, $\sigma$ is the scale and $k$ is the shape. We assume $k \neq 0$ and $(1+k\dfrac{x-\mu}{\sigma})>0$. For each of our ten cities and for each of the 49 years of the simulation at the 2068 horizon, we look at the evolution of those three coefficients and the reproducibility of the results over four distinct realizations of our model, consisting of 1500 climate paths each.

$$\Psi(x)=(1+k\dfrac{x-\mu}{\sigma})^{(-1-\dfrac{1}{k})} \dfrac{1}{\sigma} e^{-(1+k\dfrac{x-\mu}{\sigma})^{-\dfrac{1}{k}}}\  (7) $$

\noindent For each year of the simulation from 2020 to 2068, for each city and for each of the four realizations of 1500 corn yield paths, we fit a GEV distribution to our simulated data constituted of 1500 points. We obtain four sets of three coefficients ($k$, $\sigma$, $\mu$) each year for each city. We compute the coefficient of variation, defined as the quotient of the  standard deviation by the mean and expressed in percentage, of the four values at hand for each of the three coefficients. We finally take the average of the 49 coefficients of variation over the whole simulation and obtain a measure of the stability of the GEV fit for the corn yield over the four independent realization of the model. The results are presented in Table 4 and they are excellent for each of the ten cities in Ontario. The average variability of the mean of the distribution of corn yield fitted to a GEV density  is very small, between one hundredth and one thousandth of a percent. The mean of the yield is the most important parameter from the point of view of farm income.  The average variability of the shape and scale of the fitted GEV is always below 10\%, which is remarkable given the natural unpredictability of agricultural yields and weather patterns. This underlines the quality of the simulated weather paths and climate scenarios within our framework. Given that the four realizations lead to stable fits of a GEV density to the simulated yield paths, we are confident that limiting ourselves to 1500 paths per realization was  indeed valid approach. In the following of this study, we will therefore consider only one realization constituted of 1500 yield paths in order to study the impact of our five climate scenarios on the income of corn farms in Ontario. To further visualize the stability of the GEV fits using 1500 paths per realization, we plotted in Figure 10 to Figure 24 for each of the five climate scenarios $W$,  the evolution of the three GEV coefficients over four independent realizations of 1500 paths for the city of Brockville. The plots are  between the first year of the simulation (2020) and the 49th year (2068). Similar plots for the other cities are available as supplementary online material. The stability of the fits for each the 49 years is striking and further validates our approach.

\begin{table}[]
	\begin{center}
		\small
\begin{tabular}{|llll
		>{\columncolor[HTML]{C0C0C0}}l llll}
	\hline
	\multicolumn{1}{|l|}{$W=0$°C}                   & \multicolumn{1}{l|}{$k$}                         & \multicolumn{1}{l|}{$\sigma$}                    & \multicolumn{1}{l|}{$\mu$}                           & \multicolumn{1}{l|}{\cellcolor[HTML]{C0C0C0}{\color[HTML]{C0C0C0} }} & \multicolumn{1}{l|}{$W=1$°C}                    & \multicolumn{1}{l|}{$k$}                         & \multicolumn{1}{l|}{$\sigma$}                    & \multicolumn{1}{l|}{$\mu$}                                           \\ \cline{1-4} \cline{6-9} 
	\multicolumn{1}{|l|}{Brockville}                & \multicolumn{1}{l|}{{\color[HTML]{000000} 6.80}} & \multicolumn{1}{l|}{{\color[HTML]{000000} 1.97}} & \multicolumn{1}{l|}{{\color[HTML]{000000} 6.67E-02}} & \multicolumn{1}{l|}{\cellcolor[HTML]{C0C0C0}{\color[HTML]{C0C0C0} }} & \multicolumn{1}{l|}{Brockville}                 & \multicolumn{1}{l|}{{\color[HTML]{000000} 7.32}} & \multicolumn{1}{l|}{{\color[HTML]{000000} 1.91}} & \multicolumn{1}{l|}{{\color[HTML]{000000} 6.69E-02}}                 \\ \cline{1-4} \cline{6-9} 
	\multicolumn{1}{|l|}{Cornwall}                  & \multicolumn{1}{l|}{{\color[HTML]{000000} 7.96}} & \multicolumn{1}{l|}{{\color[HTML]{000000} 1.78}} & \multicolumn{1}{l|}{{\color[HTML]{000000} 6.57E-02}} & \multicolumn{1}{l|}{\cellcolor[HTML]{C0C0C0}{\color[HTML]{C0C0C0} }} & \multicolumn{1}{l|}{Cornwall}                   & \multicolumn{1}{l|}{{\color[HTML]{000000} 7.83}} & \multicolumn{1}{l|}{{\color[HTML]{000000} 1.98}} & \multicolumn{1}{l|}{{\color[HTML]{000000} 5.86E-02}}                 \\ \cline{1-4} \cline{6-9} 
	\multicolumn{1}{|l|}{Fergus}                    & \multicolumn{1}{l|}{{\color[HTML]{000000} 7.44}} & \multicolumn{1}{l|}{{\color[HTML]{000000} 2.12}} & \multicolumn{1}{l|}{{\color[HTML]{000000} 9.66E-02}} & \multicolumn{1}{l|}{\cellcolor[HTML]{C0C0C0}{\color[HTML]{C0C0C0} }} & \multicolumn{1}{l|}{Fergus}                     & \multicolumn{1}{l|}{{\color[HTML]{000000} 6.87}} & \multicolumn{1}{l|}{{\color[HTML]{000000} 2.18}} & \multicolumn{1}{l|}{{\color[HTML]{000000} 1.03E-01}}                 \\ \cline{1-4} \cline{6-9} 
	\multicolumn{1}{|l|}{Kapuskasing}               & \multicolumn{1}{l|}{{\color[HTML]{000000} 6.77}} & \multicolumn{1}{l|}{{\color[HTML]{000000} 2.03}} & \multicolumn{1}{l|}{{\color[HTML]{000000} 5.44E-04}} & \multicolumn{1}{l|}{\cellcolor[HTML]{C0C0C0}{\color[HTML]{C0C0C0} }} & \multicolumn{1}{l|}{Kapuskasing}                & \multicolumn{1}{l|}{{\color[HTML]{000000} 6.61}} & \multicolumn{1}{l|}{{\color[HTML]{000000} 1.98}} & \multicolumn{1}{l|}{{\color[HTML]{000000} 6.09E-04}}                 \\ \cline{1-4} \cline{6-9} 
	\multicolumn{1}{|l|}{Kingsville}                & \multicolumn{1}{l|}{{\color[HTML]{000000} 8.60}} & \multicolumn{1}{l|}{{\color[HTML]{000000} 1.97}} & \multicolumn{1}{l|}{{\color[HTML]{000000} 2.36E-03}} & \multicolumn{1}{l|}{\cellcolor[HTML]{C0C0C0}{\color[HTML]{C0C0C0} }} & \multicolumn{1}{l|}{Kingsville}                 & \multicolumn{1}{l|}{{\color[HTML]{000000} 7.49}} & \multicolumn{1}{l|}{{\color[HTML]{000000} 1.75}} & \multicolumn{1}{l|}{{\color[HTML]{000000} 2.52E-03}}                 \\ \cline{1-4} \cline{6-9} 
	\multicolumn{1}{|l|}{North Bay}                 & \multicolumn{1}{l|}{{\color[HTML]{000000} 6.74}} & \multicolumn{1}{l|}{{\color[HTML]{000000} 1.86}} & \multicolumn{1}{l|}{{\color[HTML]{000000} 8.93E-02}} & \multicolumn{1}{l|}{\cellcolor[HTML]{C0C0C0}{\color[HTML]{C0C0C0} }} & \multicolumn{1}{l|}{North Bay}                  & \multicolumn{1}{l|}{{\color[HTML]{000000} 7.45}} & \multicolumn{1}{l|}{{\color[HTML]{000000} 1.91}} & \multicolumn{1}{l|}{{\color[HTML]{000000} 9.55E-02}}                 \\ \cline{1-4} \cline{6-9} 
	\multicolumn{1}{|l|}{Ottawa}                    & \multicolumn{1}{l|}{{\color[HTML]{000000} 7.99}} & \multicolumn{1}{l|}{{\color[HTML]{000000} 2.05}} & \multicolumn{1}{l|}{{\color[HTML]{000000} 7.87E-03}} & \multicolumn{1}{l|}{\cellcolor[HTML]{C0C0C0}{\color[HTML]{C0C0C0} }} & \multicolumn{1}{l|}{Ottawa}                     & \multicolumn{1}{l|}{{\color[HTML]{000000} 7.47}} & \multicolumn{1}{l|}{{\color[HTML]{000000} 1.88}} & \multicolumn{1}{l|}{{\color[HTML]{000000} 7.62E-03}}                 \\ \cline{1-4} \cline{6-9} 
	\multicolumn{1}{|l|}{Toronto}                   & \multicolumn{1}{l|}{{\color[HTML]{000000} 8.17}} & \multicolumn{1}{l|}{{\color[HTML]{000000} 1.90}} & \multicolumn{1}{l|}{{\color[HTML]{000000} 6.13E-03}} & \multicolumn{1}{l|}{\cellcolor[HTML]{C0C0C0}{\color[HTML]{C0C0C0} }} & \multicolumn{1}{l|}{Toronto}                    & \multicolumn{1}{l|}{{\color[HTML]{000000} 6.82}} & \multicolumn{1}{l|}{{\color[HTML]{000000} 1.90}} & \multicolumn{1}{l|}{{\color[HTML]{000000} 5.16E-03}}                 \\ \cline{1-4} \cline{6-9} 
	\multicolumn{1}{|l|}{Trenton}                   & \multicolumn{1}{l|}{{\color[HTML]{000000} 7.47}} & \multicolumn{1}{l|}{{\color[HTML]{000000} 1.93}} & \multicolumn{1}{l|}{{\color[HTML]{000000} 2.09E-02}} & \multicolumn{1}{l|}{\cellcolor[HTML]{C0C0C0}{\color[HTML]{C0C0C0} }} & \multicolumn{1}{l|}{Trenton}                    & \multicolumn{1}{l|}{{\color[HTML]{000000} 6.36}} & \multicolumn{1}{l|}{{\color[HTML]{000000} 1.99}} & \multicolumn{1}{l|}{{\color[HTML]{000000} 2.03E-02}}                 \\ \cline{1-4} \cline{6-9} 
	\multicolumn{1}{|l|}{Woodstock}                 & \multicolumn{1}{l|}{{\color[HTML]{000000} 6.46}} & \multicolumn{1}{l|}{{\color[HTML]{000000} 1.89}} & \multicolumn{1}{l|}{{\color[HTML]{000000} 3.14E-02}} & \multicolumn{1}{l|}{\cellcolor[HTML]{C0C0C0}{\color[HTML]{C0C0C0} }} & \multicolumn{1}{l|}{Woodstock}                  & \multicolumn{1}{l|}{{\color[HTML]{000000} 6.14}} & \multicolumn{1}{l|}{{\color[HTML]{000000} 1.78}} & \multicolumn{1}{l|}{{\color[HTML]{000000} 2.83E-02}}                 \\ \cline{1-4} \cline{6-9} 
	\cellcolor[HTML]{C0C0C0}{\color[HTML]{C0C0C0} } & \cellcolor[HTML]{C0C0C0}{\color[HTML]{C0C0C0} }  & \cellcolor[HTML]{C0C0C0}{\color[HTML]{C0C0C0} }  & \cellcolor[HTML]{C0C0C0}{\color[HTML]{C0C0C0} }      & {\color[HTML]{C0C0C0} }                                              & \cellcolor[HTML]{C0C0C0}{\color[HTML]{C0C0C0} } & \cellcolor[HTML]{C0C0C0}{\color[HTML]{C0C0C0} }  & \cellcolor[HTML]{C0C0C0}{\color[HTML]{C0C0C0} }  & \multicolumn{1}{l|}{\cellcolor[HTML]{C0C0C0}{\color[HTML]{C0C0C0} }} \\ \cline{1-4} \cline{6-9} 
	\multicolumn{1}{|l|}{$W=2$°C}                   & \multicolumn{1}{l|}{$k$}                         & \multicolumn{1}{l|}{$\sigma$}                    & \multicolumn{1}{l|}{$\mu$}                           & \multicolumn{1}{l|}{\cellcolor[HTML]{C0C0C0}{\color[HTML]{C0C0C0} }} & \multicolumn{1}{l|}{$W=3$°C}                    & \multicolumn{1}{l|}{$k$}                         & \multicolumn{1}{l|}{$\sigma$}                    & \multicolumn{1}{l|}{$\mu$}                                           \\ \cline{1-4} \cline{6-9} 
	\multicolumn{1}{|l|}{Brockville}                & \multicolumn{1}{l|}{{\color[HTML]{000000} 7.31}} & \multicolumn{1}{l|}{{\color[HTML]{000000} 1.99}} & \multicolumn{1}{l|}{{\color[HTML]{000000} 6.16E-02}} & \multicolumn{1}{l|}{\cellcolor[HTML]{C0C0C0}{\color[HTML]{C0C0C0} }} & \multicolumn{1}{l|}{Brockville}                 & \multicolumn{1}{l|}{{\color[HTML]{000000} 7.11}} & \multicolumn{1}{l|}{{\color[HTML]{000000} 1.75}} & \multicolumn{1}{l|}{{\color[HTML]{000000} 6.67E-02}}                 \\ \cline{1-4} \cline{6-9} 
	\multicolumn{1}{|l|}{Cornwall}                  & \multicolumn{1}{l|}{{\color[HTML]{000000} 7.64}} & \multicolumn{1}{l|}{{\color[HTML]{000000} 2.13}} & \multicolumn{1}{l|}{{\color[HTML]{000000} 6.04E-02}} & \multicolumn{1}{l|}{\cellcolor[HTML]{C0C0C0}{\color[HTML]{C0C0C0} }} & \multicolumn{1}{l|}{Cornwall}                   & \multicolumn{1}{l|}{{\color[HTML]{000000} 7.34}} & \multicolumn{1}{l|}{{\color[HTML]{000000} 1.83}} & \multicolumn{1}{l|}{{\color[HTML]{000000} 5.70E-02}}                 \\ \cline{1-4} \cline{6-9} 
	\multicolumn{1}{|l|}{Fergus}                    & \multicolumn{1}{l|}{{\color[HTML]{000000} 6.61}} & \multicolumn{1}{l|}{{\color[HTML]{000000} 1.94}} & \multicolumn{1}{l|}{{\color[HTML]{000000} 8.50E-02}} & \multicolumn{1}{l|}{\cellcolor[HTML]{C0C0C0}{\color[HTML]{C0C0C0} }} & \multicolumn{1}{l|}{Fergus}                     & \multicolumn{1}{l|}{{\color[HTML]{000000} 7.03}} & \multicolumn{1}{l|}{{\color[HTML]{000000} 2.12}} & \multicolumn{1}{l|}{{\color[HTML]{000000} 8.99E-02}}                 \\ \cline{1-4} \cline{6-9} 
	\multicolumn{1}{|l|}{Kapuskasing}               & \multicolumn{1}{l|}{{\color[HTML]{000000} 6.84}} & \multicolumn{1}{l|}{{\color[HTML]{000000} 2.03}} & \multicolumn{1}{l|}{{\color[HTML]{000000} 5.98E-04}} & \multicolumn{1}{l|}{\cellcolor[HTML]{C0C0C0}{\color[HTML]{C0C0C0} }} & \multicolumn{1}{l|}{Kapuskasing}                & \multicolumn{1}{l|}{{\color[HTML]{000000} 7.38}} & \multicolumn{1}{l|}{{\color[HTML]{000000} 2.11}} & \multicolumn{1}{l|}{{\color[HTML]{000000} 5.71E-04}}                 \\ \cline{1-4} \cline{6-9} 
	\multicolumn{1}{|l|}{Kingsville}                & \multicolumn{1}{l|}{{\color[HTML]{000000} 7.73}} & \multicolumn{1}{l|}{{\color[HTML]{000000} 1.99}} & \multicolumn{1}{l|}{{\color[HTML]{000000} 2.21E-03}} & \multicolumn{1}{l|}{\cellcolor[HTML]{C0C0C0}{\color[HTML]{C0C0C0} }} & \multicolumn{1}{l|}{Kingsville}                 & \multicolumn{1}{l|}{{\color[HTML]{000000} 6.93}} & \multicolumn{1}{l|}{{\color[HTML]{000000} 1.83}} & \multicolumn{1}{l|}{{\color[HTML]{000000} 2.27E-03}}                 \\ \cline{1-4} \cline{6-9} 
	\multicolumn{1}{|l|}{North Bay}                 & \multicolumn{1}{l|}{{\color[HTML]{000000} 7.57}} & \multicolumn{1}{l|}{{\color[HTML]{000000} 1.91}} & \multicolumn{1}{l|}{{\color[HTML]{000000} 8.45E-02}} & \multicolumn{1}{l|}{\cellcolor[HTML]{C0C0C0}{\color[HTML]{C0C0C0} }} & \multicolumn{1}{l|}{North Bay}                  & \multicolumn{1}{l|}{{\color[HTML]{000000} 7.05}} & \multicolumn{1}{l|}{{\color[HTML]{000000} 1.99}} & \multicolumn{1}{l|}{{\color[HTML]{000000} 8.85E-02}}                 \\ \cline{1-4} \cline{6-9} 
	\multicolumn{1}{|l|}{Ottawa}                    & \multicolumn{1}{l|}{{\color[HTML]{000000} 7.72}} & \multicolumn{1}{l|}{{\color[HTML]{000000} 1.85}} & \multicolumn{1}{l|}{{\color[HTML]{000000} 7.49E-03}} & \multicolumn{1}{l|}{\cellcolor[HTML]{C0C0C0}{\color[HTML]{C0C0C0} }} & \multicolumn{1}{l|}{Ottawa}                     & \multicolumn{1}{l|}{{\color[HTML]{000000} 8.85}} & \multicolumn{1}{l|}{{\color[HTML]{000000} 1.92}} & \multicolumn{1}{l|}{{\color[HTML]{000000} 7.20E-03}}                 \\ \cline{1-4} \cline{6-9} 
	\multicolumn{1}{|l|}{Toronto}                   & \multicolumn{1}{l|}{{\color[HTML]{000000} 8.38}} & \multicolumn{1}{l|}{{\color[HTML]{000000} 1.94}} & \multicolumn{1}{l|}{{\color[HTML]{000000} 4.83E-03}} & \multicolumn{1}{l|}{\cellcolor[HTML]{C0C0C0}{\color[HTML]{C0C0C0} }} & \multicolumn{1}{l|}{Toronto}                    & \multicolumn{1}{l|}{{\color[HTML]{000000} 7.20}} & \multicolumn{1}{l|}{{\color[HTML]{000000} 2.11}} & \multicolumn{1}{l|}{{\color[HTML]{000000} 4.97E-03}}                 \\ \cline{1-4} \cline{6-9} 
	\multicolumn{1}{|l|}{Trenton}                   & \multicolumn{1}{l|}{{\color[HTML]{000000} 7.68}} & \multicolumn{1}{l|}{{\color[HTML]{000000} 1.96}} & \multicolumn{1}{l|}{{\color[HTML]{000000} 2.07E-02}} & \multicolumn{1}{l|}{\cellcolor[HTML]{C0C0C0}{\color[HTML]{C0C0C0} }} & \multicolumn{1}{l|}{Trenton}                    & \multicolumn{1}{l|}{{\color[HTML]{000000} 7.49}} & \multicolumn{1}{l|}{{\color[HTML]{000000} 2.16}} & \multicolumn{1}{l|}{{\color[HTML]{000000} 1.94E-02}}                 \\ \cline{1-4} \cline{6-9} 
	\multicolumn{1}{|l|}{Woodstock}                 & \multicolumn{1}{l|}{{\color[HTML]{000000} 6.98}} & \multicolumn{1}{l|}{{\color[HTML]{000000} 1.97}} & \multicolumn{1}{l|}{{\color[HTML]{000000} 3.04E-02}} & \multicolumn{1}{l|}{\cellcolor[HTML]{C0C0C0}{\color[HTML]{C0C0C0} }} & \multicolumn{1}{l|}{Woodstock}                  & \multicolumn{1}{l|}{{\color[HTML]{000000} 7.68}} & \multicolumn{1}{l|}{{\color[HTML]{000000} 1.89}} & \multicolumn{1}{l|}{{\color[HTML]{000000} 3.14E-02}}                 \\ \cline{1-4} \cline{6-9} 
	\cellcolor[HTML]{C0C0C0}{\color[HTML]{C0C0C0} } & \cellcolor[HTML]{C0C0C0}{\color[HTML]{C0C0C0} }  & \cellcolor[HTML]{C0C0C0}{\color[HTML]{C0C0C0} }  & \cellcolor[HTML]{C0C0C0}{\color[HTML]{C0C0C0} }      & {\color[HTML]{C0C0C0} }                                              & \cellcolor[HTML]{C0C0C0}{\color[HTML]{C0C0C0} } & \cellcolor[HTML]{C0C0C0}{\color[HTML]{C0C0C0} }  & \cellcolor[HTML]{C0C0C0}{\color[HTML]{C0C0C0} }  & \multicolumn{1}{l|}{\cellcolor[HTML]{C0C0C0}{\color[HTML]{C0C0C0} }} \\ \cline{1-4} \cline{6-9} 
	\multicolumn{1}{|l|}{$W=4$°C}                   & \multicolumn{1}{l|}{$k$}                         & \multicolumn{1}{l|}{$\sigma$}                    & \multicolumn{1}{l|}{$\mu$}                           & \multicolumn{1}{l|}{\cellcolor[HTML]{C0C0C0}{\color[HTML]{C0C0C0} }} &                                                 &                                                  &                                                  &                                                                      \\ \cline{1-4}
	\multicolumn{1}{|l|}{Brockville}                & \multicolumn{1}{l|}{{\color[HTML]{000000} 7.98}} & \multicolumn{1}{l|}{{\color[HTML]{000000} 2.05}} & \multicolumn{1}{l|}{{\color[HTML]{000000} 6.59E-02}} & \multicolumn{1}{l|}{\cellcolor[HTML]{C0C0C0}{\color[HTML]{C0C0C0} }} &                                                 &                                                  &                                                  &                                                                      \\ \cline{1-4}
	\multicolumn{1}{|l|}{Cornwall}                  & \multicolumn{1}{l|}{{\color[HTML]{000000} 7.31}} & \multicolumn{1}{l|}{{\color[HTML]{000000} 1.81}} & \multicolumn{1}{l|}{{\color[HTML]{000000} 4.83E-02}} & \multicolumn{1}{l|}{\cellcolor[HTML]{C0C0C0}{\color[HTML]{C0C0C0} }} &                                                 &                                                  &                                                  &                                                                      \\ \cline{1-4}
	\multicolumn{1}{|l|}{Fergus}                    & \multicolumn{1}{l|}{{\color[HTML]{000000} 6.48}} & \multicolumn{1}{l|}{{\color[HTML]{000000} 1.95}} & \multicolumn{1}{l|}{{\color[HTML]{000000} 8.88E-02}} & \multicolumn{1}{l|}{\cellcolor[HTML]{C0C0C0}{\color[HTML]{C0C0C0} }} &                                                 &                                                  &                                                  &                                                                      \\ \cline{1-4}
	\multicolumn{1}{|l|}{Kapuskasing}               & \multicolumn{1}{l|}{{\color[HTML]{000000} 6.71}} & \multicolumn{1}{l|}{{\color[HTML]{000000} 1.91}} & \multicolumn{1}{l|}{{\color[HTML]{000000} 5.21E-04}} & \multicolumn{1}{l|}{\cellcolor[HTML]{C0C0C0}{\color[HTML]{C0C0C0} }} &                                                 &                                                  &                                                  &                                                                      \\ \cline{1-4}
	\multicolumn{1}{|l|}{Kingsville}                & \multicolumn{1}{l|}{{\color[HTML]{000000} 7.71}} & \multicolumn{1}{l|}{{\color[HTML]{000000} 1.70}} & \multicolumn{1}{l|}{{\color[HTML]{000000} 2.31E-03}} & \multicolumn{1}{l|}{\cellcolor[HTML]{C0C0C0}{\color[HTML]{C0C0C0} }} &                                                 &                                                  &                                                  &                                                                      \\ \cline{1-4}
	\multicolumn{1}{|l|}{North Bay}                 & \multicolumn{1}{l|}{{\color[HTML]{000000} 8.17}} & \multicolumn{1}{l|}{{\color[HTML]{000000} 1.77}} & \multicolumn{1}{l|}{{\color[HTML]{000000} 8.50E-02}} & \multicolumn{1}{l|}{\cellcolor[HTML]{C0C0C0}{\color[HTML]{C0C0C0} }} &                                                 &                                                  &                                                  &                                                                      \\ \cline{1-4}
	\multicolumn{1}{|l|}{Ottawa}                    & \multicolumn{1}{l|}{{\color[HTML]{000000} 8.03}} & \multicolumn{1}{l|}{{\color[HTML]{000000} 1.79}} & \multicolumn{1}{l|}{{\color[HTML]{000000} 6.76E-03}} & \multicolumn{1}{l|}{\cellcolor[HTML]{C0C0C0}{\color[HTML]{C0C0C0} }} &                                                 &                                                  &                                                  &                                                                      \\ \cline{1-4}
	\multicolumn{1}{|l|}{Toronto}                   & \multicolumn{1}{l|}{{\color[HTML]{000000} 8.61}} & \multicolumn{1}{l|}{{\color[HTML]{000000} 1.89}} & \multicolumn{1}{l|}{{\color[HTML]{000000} 4.62E-03}} & \multicolumn{1}{l|}{\cellcolor[HTML]{C0C0C0}{\color[HTML]{C0C0C0} }} &                                                 &                                                  &                                                  &                                                                      \\ \cline{1-4}
	\multicolumn{1}{|l|}{Trenton}                   & \multicolumn{1}{l|}{{\color[HTML]{000000} 7.24}} & \multicolumn{1}{l|}{{\color[HTML]{000000} 2.08}} & \multicolumn{1}{l|}{{\color[HTML]{000000} 2.06E-02}} & \multicolumn{1}{l|}{\cellcolor[HTML]{C0C0C0}{\color[HTML]{C0C0C0} }} &                                                 &                                                  &                                                  &                                                                      \\ \cline{1-4}
	\multicolumn{1}{|l|}{Woodstock}                 & \multicolumn{1}{l|}{{\color[HTML]{000000} 7.09}} & \multicolumn{1}{l|}{{\color[HTML]{000000} 2.05}} & \multicolumn{1}{l|}{{\color[HTML]{000000} 2.64E-02}} & \multicolumn{1}{l|}{\cellcolor[HTML]{C0C0C0}{\color[HTML]{C0C0C0} }} &                                                 &                                                  &                                                  &                                                                      \\ \cline{1-5}
\end{tabular}
	\end{center}
	\caption{Average coefficients of variability (in \%) over 49 years.}
\end{table}

\begin{figure}[]
	\begin{minipage}{0.30\textwidth}
		\centering
		\includegraphics[width=51mm]{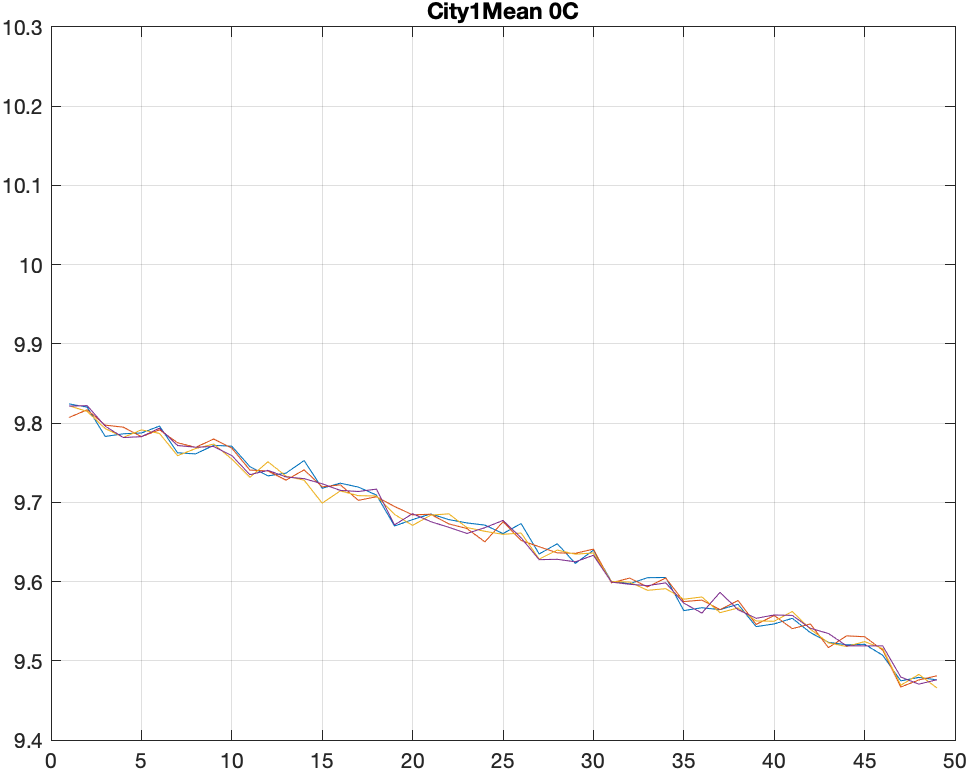}
		\caption{$\mu$ for $W=0$ °C}
	\end{minipage}\hfill
	\begin{minipage}{0.30\textwidth}
		\centering
		\includegraphics[width=51mm]{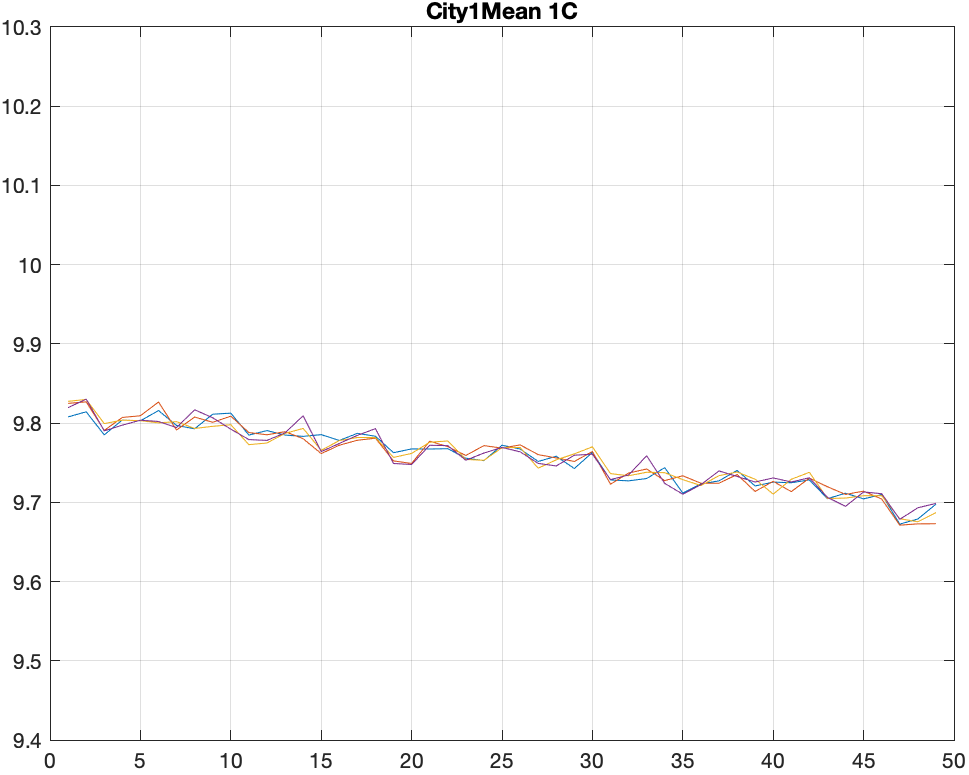}
		\caption{$\mu$ for $W=1$ °C}
	\end{minipage}\hfill
	\begin{minipage}{0.30\textwidth}
		\centering
		\includegraphics[width=51mm]{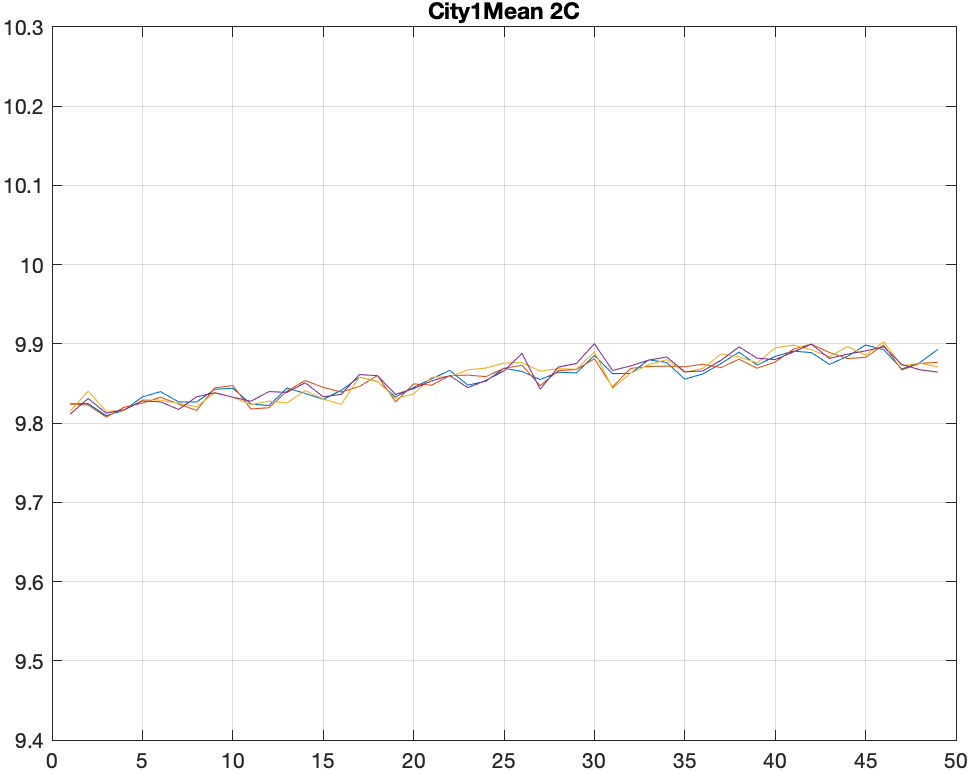}
		\caption{$\mu$ for $W=2$ °C}
	\end{minipage}\hfill
	\begin{minipage}{0.50\textwidth}
		\centering
		\includegraphics[width=51mm]{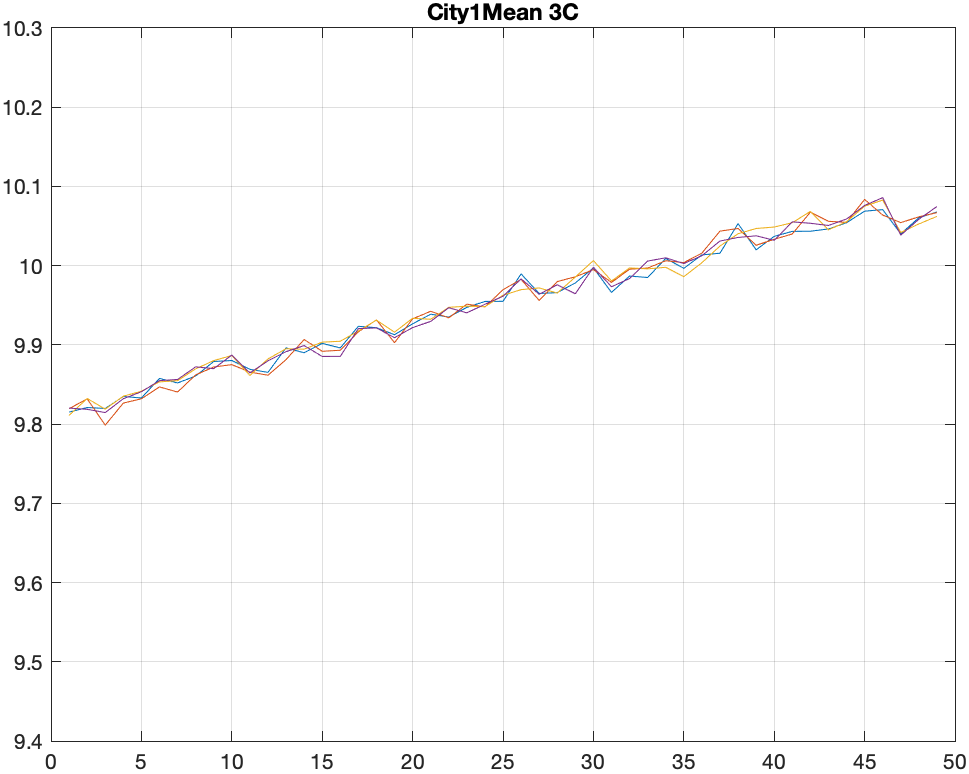}
		\caption{$\mu$ for $W=3$ °C}
	\end{minipage}\hfill
	\begin{minipage}{0.50\textwidth}
		\centering
		\includegraphics[width=51mm]{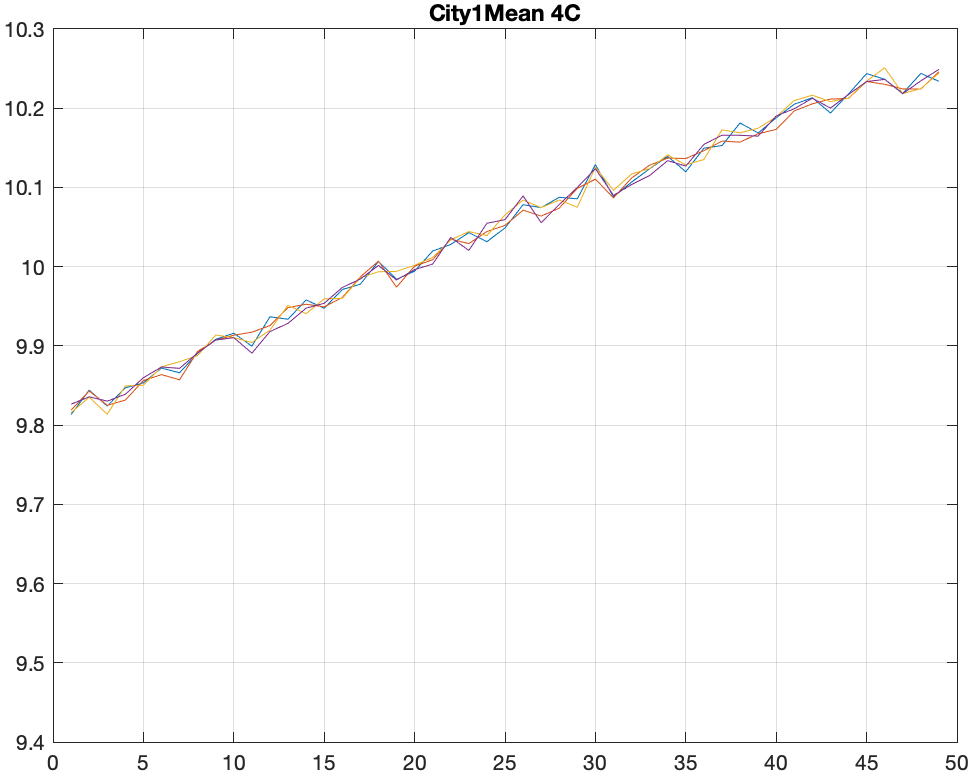}
		\caption{$\mu$ for $W=4$ °C}
	\end{minipage}\hfill
\end{figure}

\begin{figure}[]
	\begin{minipage}{0.30\textwidth}
		\centering
		\includegraphics[width=50mm]{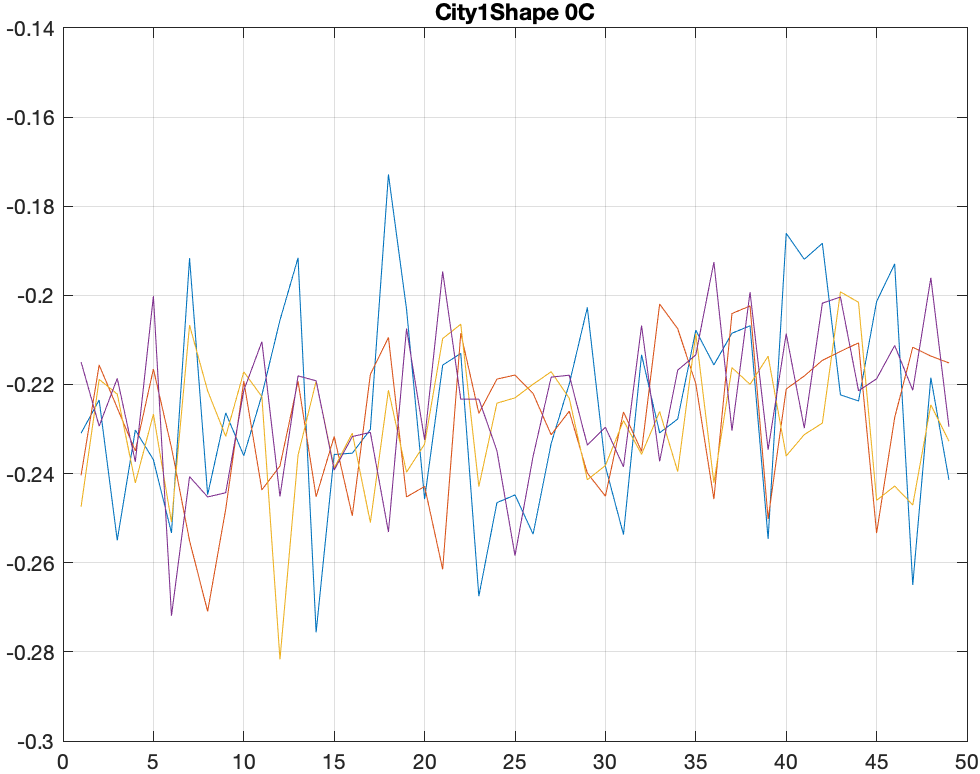}
		\caption{$k$ for $W=0$ °C}
	\end{minipage}\hfill
	\begin{minipage}{0.30\textwidth}
		\centering
		\includegraphics[width=50mm]{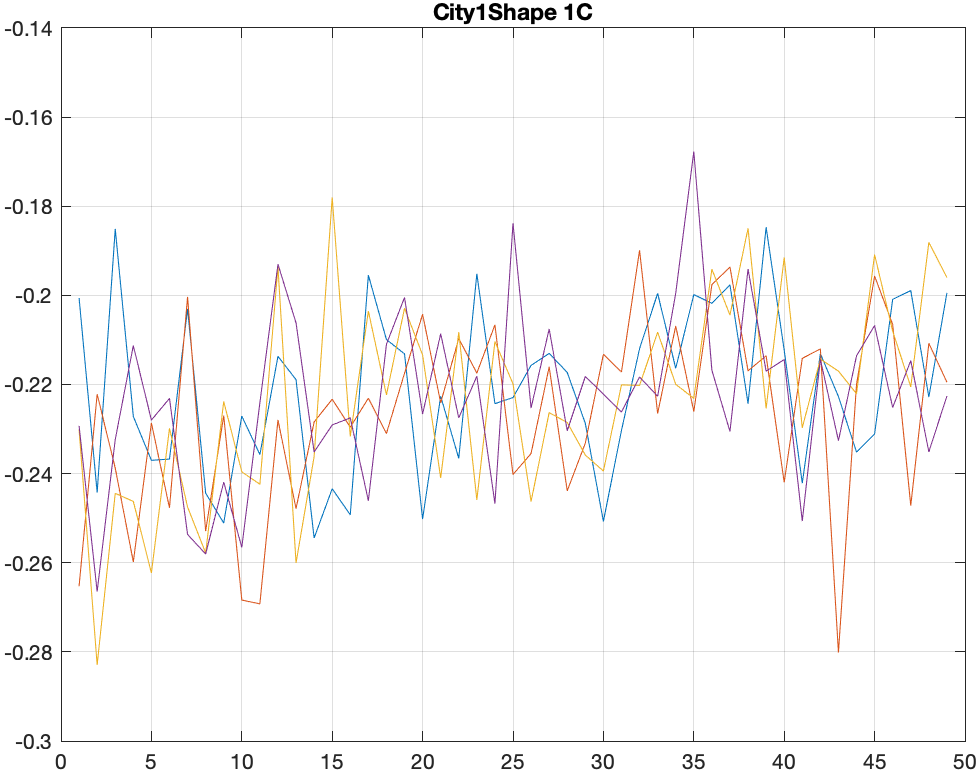}
		\caption{$k$ for $W=1$ °C}
	\end{minipage}\hfill
	\begin{minipage}{0.30\textwidth}
		\centering
		\includegraphics[width=50mm]{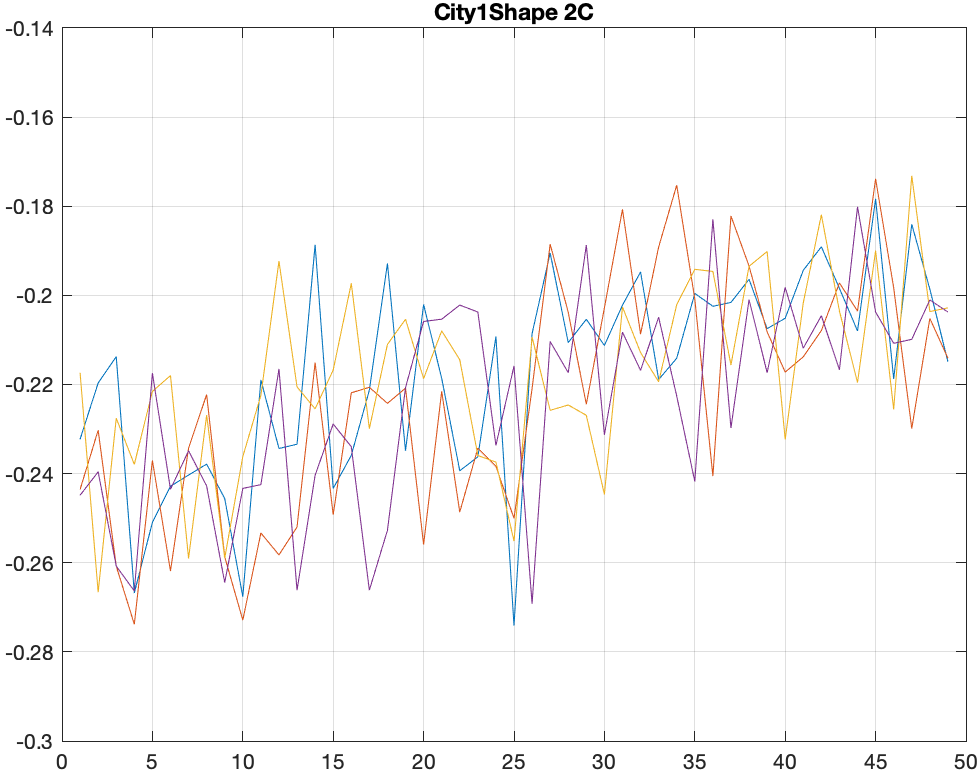}
		\caption{$k$ for $W=2$ °C}
	\end{minipage}\hfill
	\begin{minipage}{0.50\textwidth}
		\centering
		\includegraphics[width=50mm]{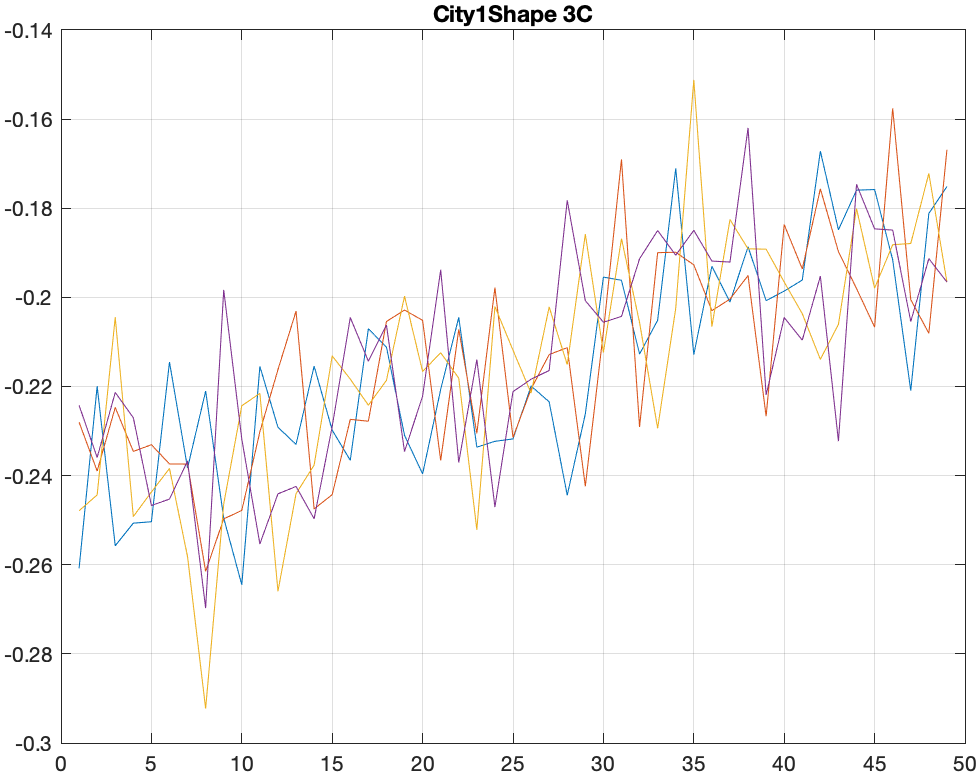}
		\caption{$k$ for $W=3$ °C}
	\end{minipage}\hfill
	\begin{minipage}{0.50\textwidth}
	\centering
	\includegraphics[width=50mm]{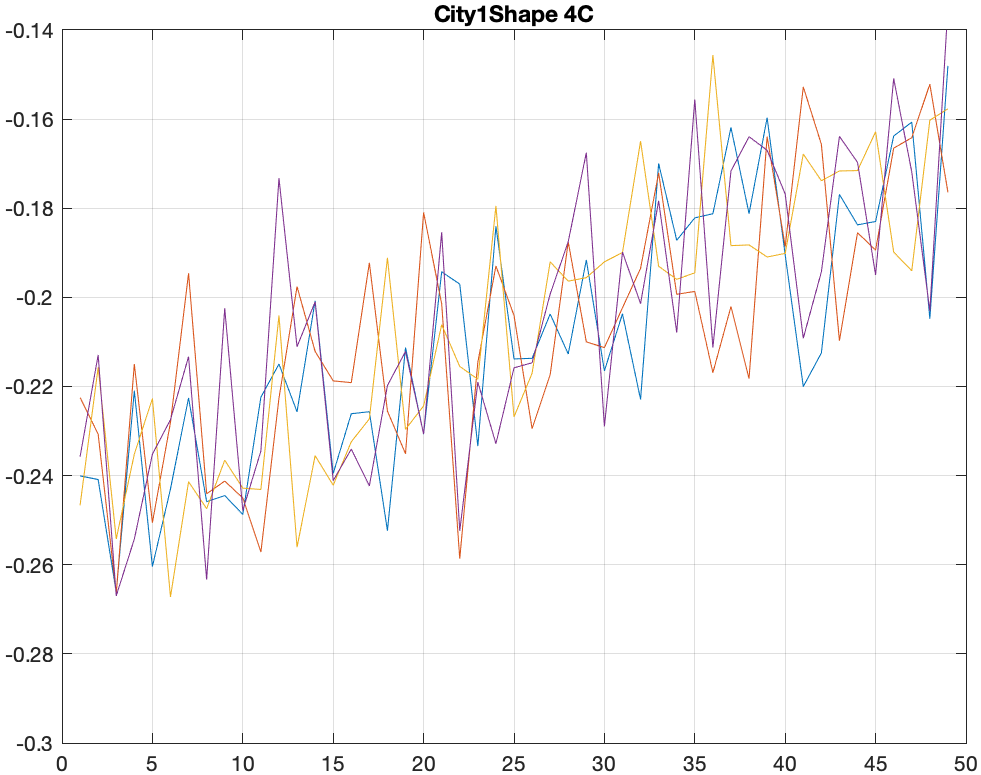}
	\caption{$k$ for $W=4$ °C}
\end{minipage}\hfill
\end{figure}

\begin{figure}[]
	\begin{minipage}{0.30\textwidth}
	\centering
	\includegraphics[width=50mm]{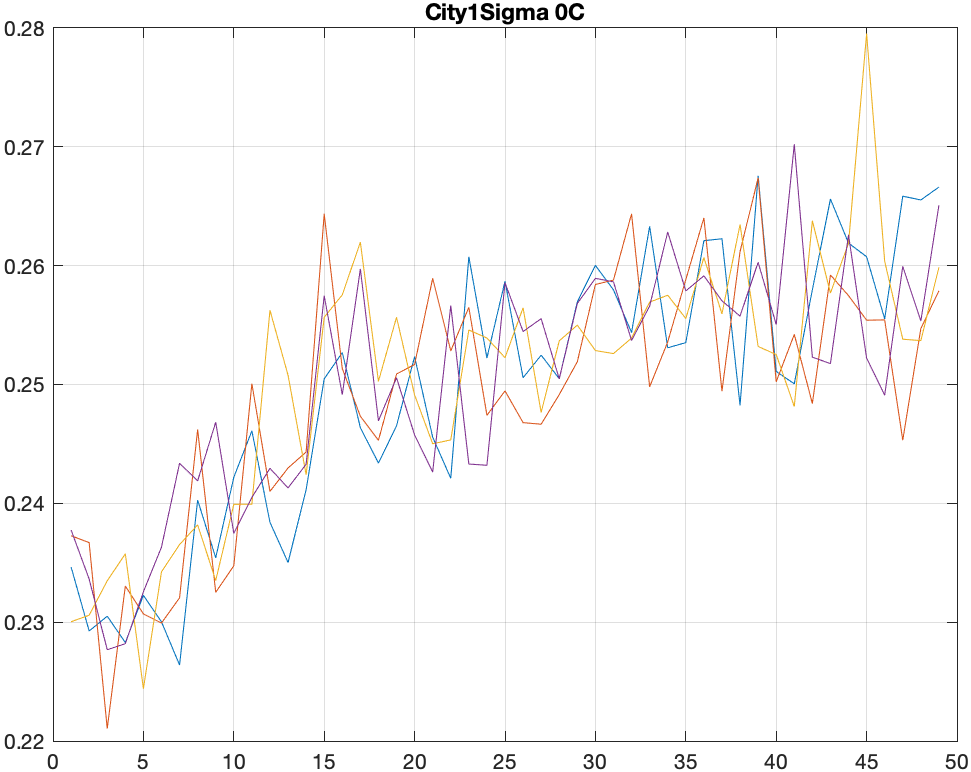}
	\caption{$\sigma$ for $W=0$ °C}
\end{minipage}\hfill
	\begin{minipage}{0.30\textwidth}
	\centering
	\includegraphics[width=50mm]{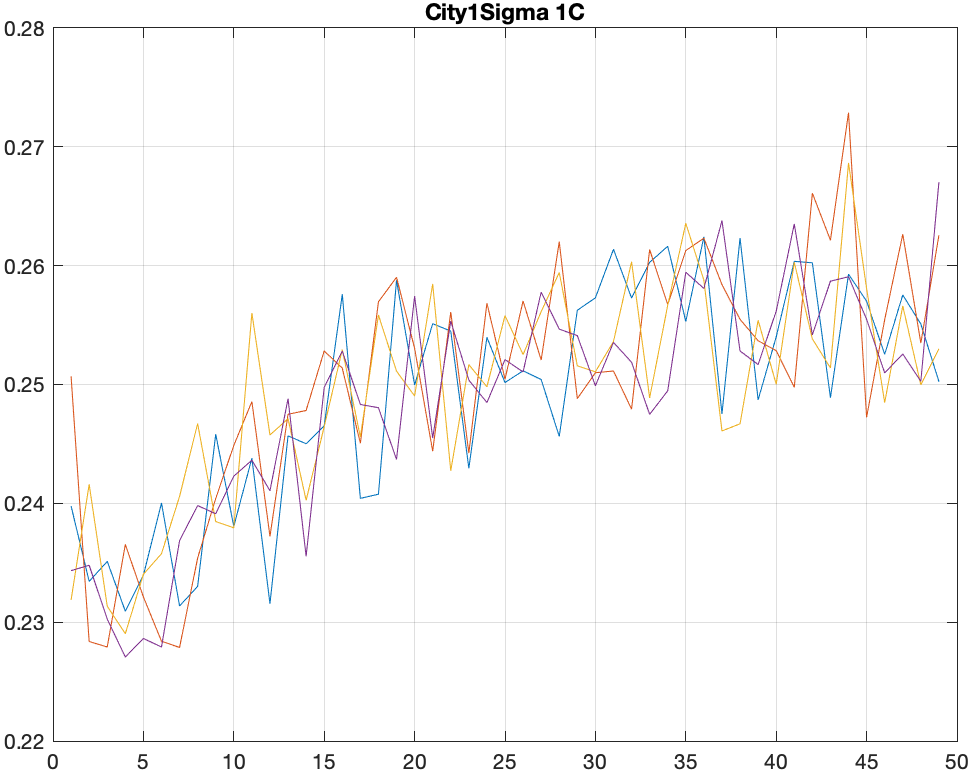}
	\caption{$\sigma$ for $W=1$ °C}
\end{minipage}\hfill
	\begin{minipage}{0.30\textwidth}
	\centering
	\includegraphics[width=50mm]{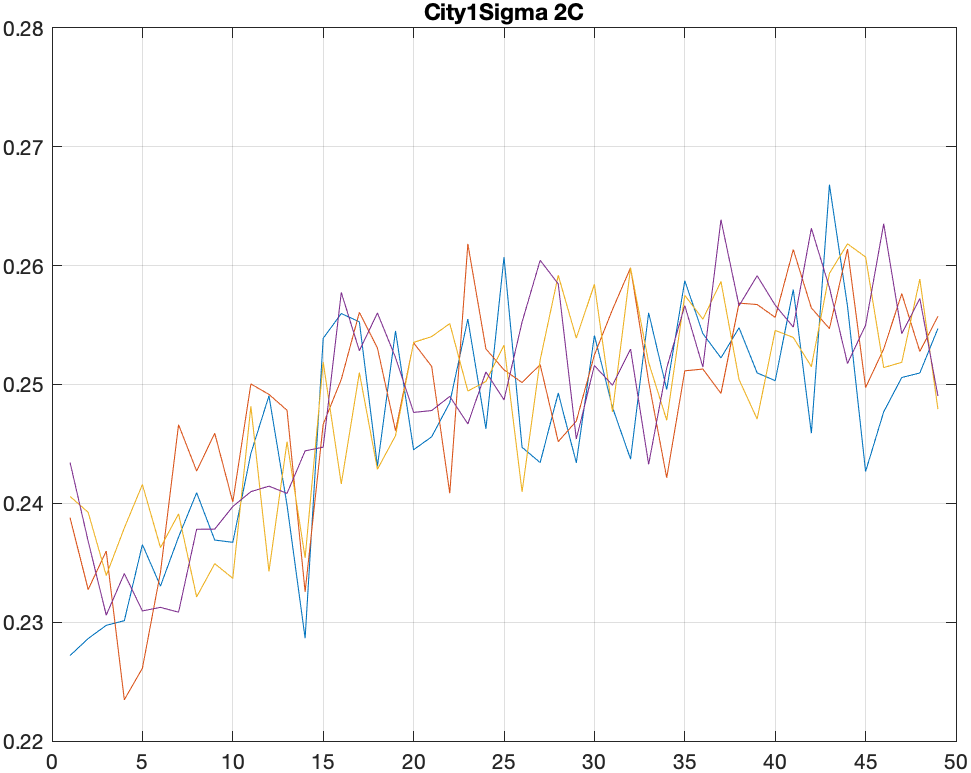}
	\caption{$\sigma$ for $W=2$ °C}
\end{minipage}\hfill
	\begin{minipage}{0.50\textwidth}
	\centering
	\includegraphics[width=50mm]{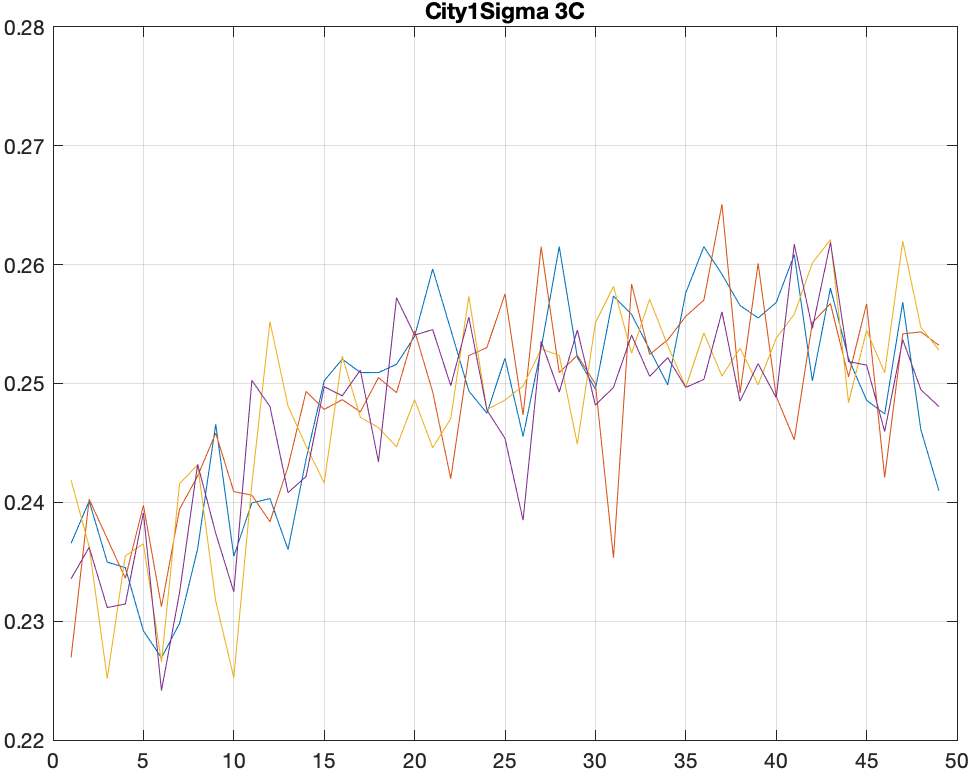}
	\caption{$\sigma$ for $W=3$ °C}
\end{minipage}\hfill
	\begin{minipage}{0.50\textwidth}
	\centering
	\includegraphics[width=50mm]{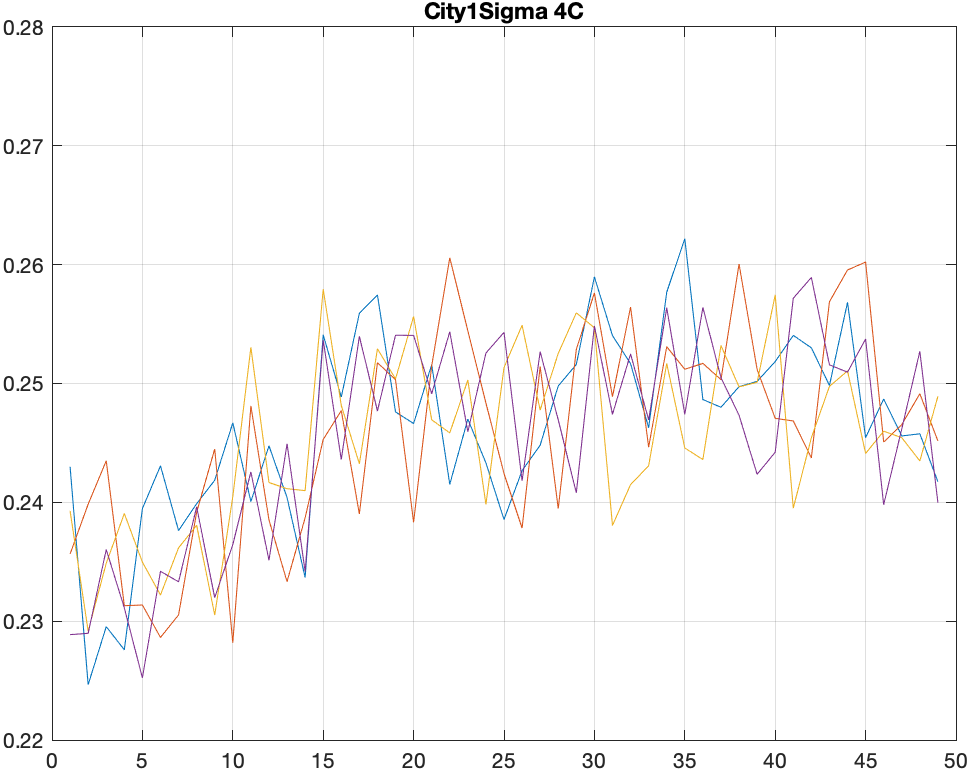}
	\caption{$\sigma$ for $W=4$ °C}
\end{minipage}\hfill
\end{figure}

\bigskip

\noindent We are now in a position to compute the income $\mathscr{I}_{i}^{j}$ of an average corn farm in the region around the city $j \in  \llbracket 1, 10 \rrbracket$  at each step $i \in \llbracket 1, 49 \rrbracket$ in the future. The computation of the farm income is given in Formula 8.

$$\mathscr{I}_{i}^{j} = A \times Y_{i}^{j} \times P  \  (8)$$

\begin{itemize}
	\item $A$ is a constant scale factor representing the size of the farm in hectares. We assume that it does not change over time. $A$ is chosen as the average farm size in Ontario. According to Statistics Canada in a  report entitled \textit{Farm and Farm Operator Data, 2016 Census of Agriculture} \footnote{\url{www.statcan.gc.ca/eng/ca2016}}, the average farm size in Ontario is presently 249 acres, which is approximately 100 hectares. Assuming that the typical size of a corn farm in Ontario matches the provincial average, which is a reasonable assumption given that corn is the dominant crop in the province, we choose $A=100$ for the duration of our study. While this is an approximation, the statistical study of Eastwood, Lipton and Newell (2010) shows that on the North American continent, the mean farm size, despite a slight trend toward larger values over the years, has not dramatically changed since 1970.
	
	\item $Y_{i}^{j}$ is the simulated corn yield for the city $j$ at the year $i$ of a given path among the 1500 that constitute of a realization of the model.
	
	\item The value of $P$, expressed in Canadian Dollar, is derived from the historical price of the Generic First Corn Future (C1 Comdty) corresponding to one metric tonne of grain corn. First we obtained from Bloomberg a time series of C1 Comdty in U.S Dollar between 2009 and 2019. We compensated for inflation using a time series of Inflation GDP Deflator (IFGDPUSA) provided by the World Bank as an annual percentage. We then used a time series of the exchange rate of the U.S Dollar versus the Canadian Dollar (USDCAD), also obtained from Bloomberg, to convert the original C1 Comdty time series into inflation adjusted Canadian Dollars between 2009 and 2019. We computed for each city $j \in \llbracket 1,10 \rrbracket$ the starting (planting) and ending (harvest) dates of the historical growing seasons between 2009 and 2019. Those dates are obtained by using the same method as described before for the future years in the simulations, except of course that there is only one climate path, which is the realized historical data from NOAA. For each city $j$ and for each year $i$ between 2009 and 2019, we compute a local price $p_{i}^{j}$ as the average of the inflation adjusted C1 Comdty expressed in Canadian Dollar over the two weeks located around the middle of the growing season. This is the time when corn farmers will sell their crop on the futures market and plan for storage. Since we thought that it was unrealistic and unnecessarily complicated to keep local prices for each city, we therefore defined the price of corn future $p_{i}$ in Ontario at year $i$ as the mean of the  values of $p_{i}^{j}$ for $j \in \llbracket 1,10 \rrbracket$. Finally $P$ from Equation 8 is computed as the mean of the values of  $p_{i}$ for $i \in \llbracket 1,10 \rrbracket$, between 2009 and 2010. We found $P=\$186.12$ CAN. We chose to work with a constant corn price in our study in order to focus exclusively on the impact of several climate change scenarios on the income of corn farms in Ontario. Indeed, it would have been difficult to evaluate the impact of corn price variability on farm income and differentiate its effect for the effects of climate change on the corn yield.
\end{itemize}

\noindent By computing $\mathscr{I}_{i}^{j}$ for each value of $W$ and taking the average over the 1500 paths that constitute a realization of the model, we obtain first Figure 25 to Figure 29. The x-axis represents the years of the simulation in the future, from 2020 to 2068 and the y-axis the farm income in Canadian Dollar. The influence of climate change on the income of the corn farms for the ten cities is subtle but very measurable. The income of many cities, but not all as we will see later, suffers under the scenarios $W=0$°C and $W=1$°C because they respectively represent a disappearance and a slowing down of the historical warning trend since 1970. Corn needs heat to grow and the CHU is an increasing function of heat so this result is not surprising. The scenario $W=2$°C represents a continuation of the historical climate trend so the income of the farms in most cities is stable. Since we have eliminated the technology trend in our computation of the yield paths for the future years, this result is not surprising. In the absence of a technology trend, the only way for the CHU, and thus the yield to increase is to get more heat and no extreme rainfall events that would interfere with the length of the growing season.

\begin{figure}
\includegraphics[width=150mm]{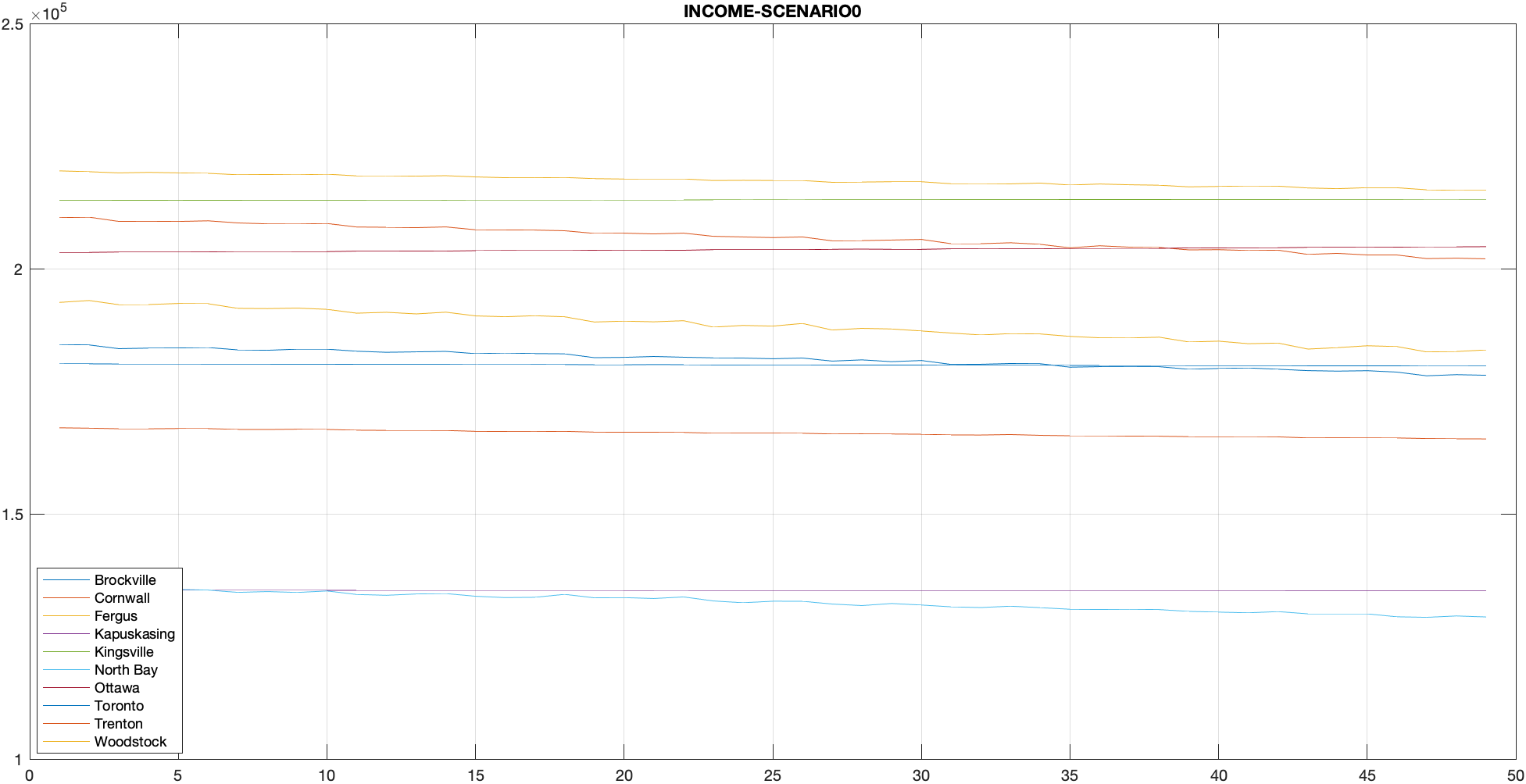}
\caption{Income for  $W=0$ °C}
\end{figure}
\begin{figure}
	\includegraphics[width=150mm]{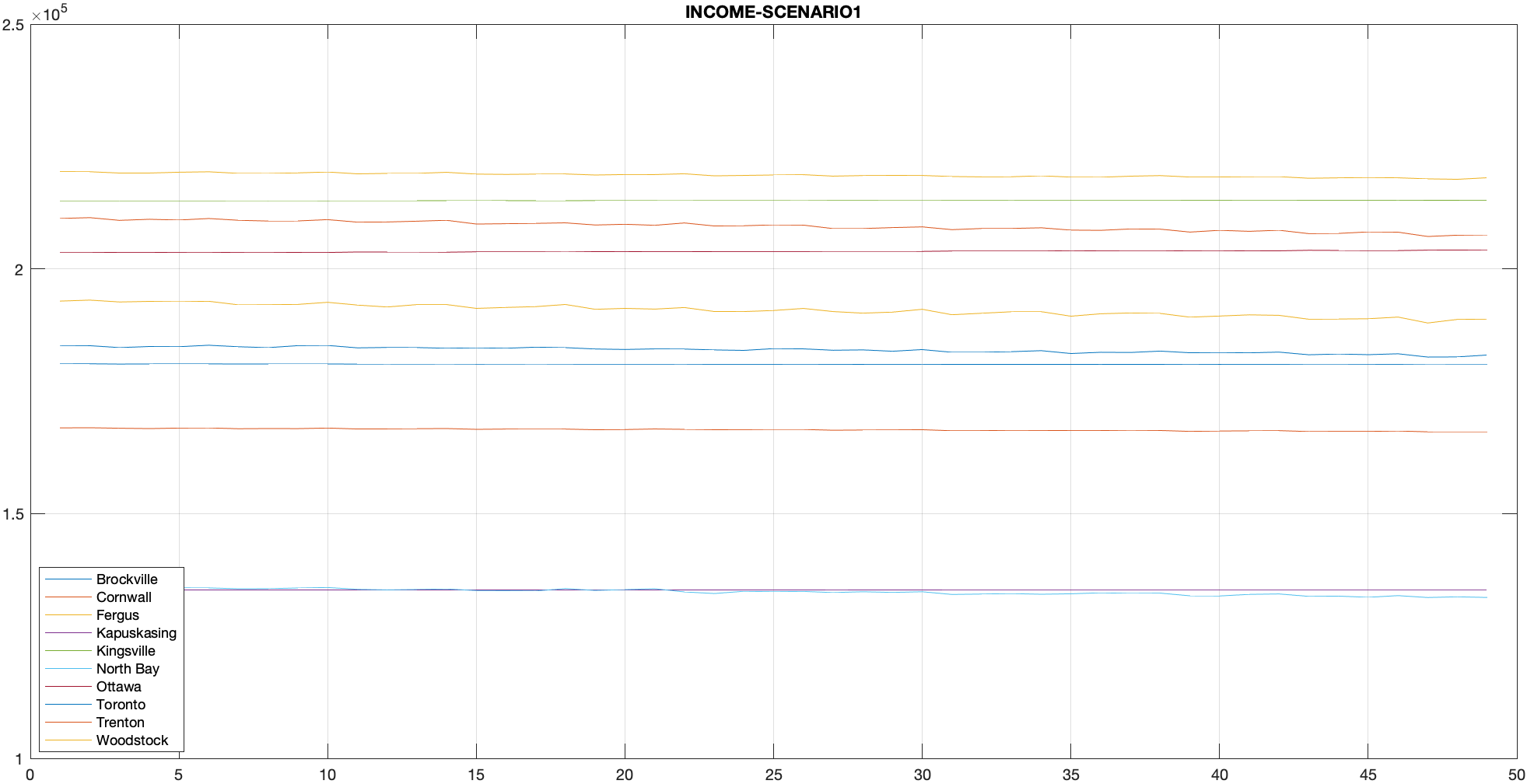}
	\caption{Income for  $W=0$ °C}
\end{figure}
\begin{figure}
	\includegraphics[width=150mm]{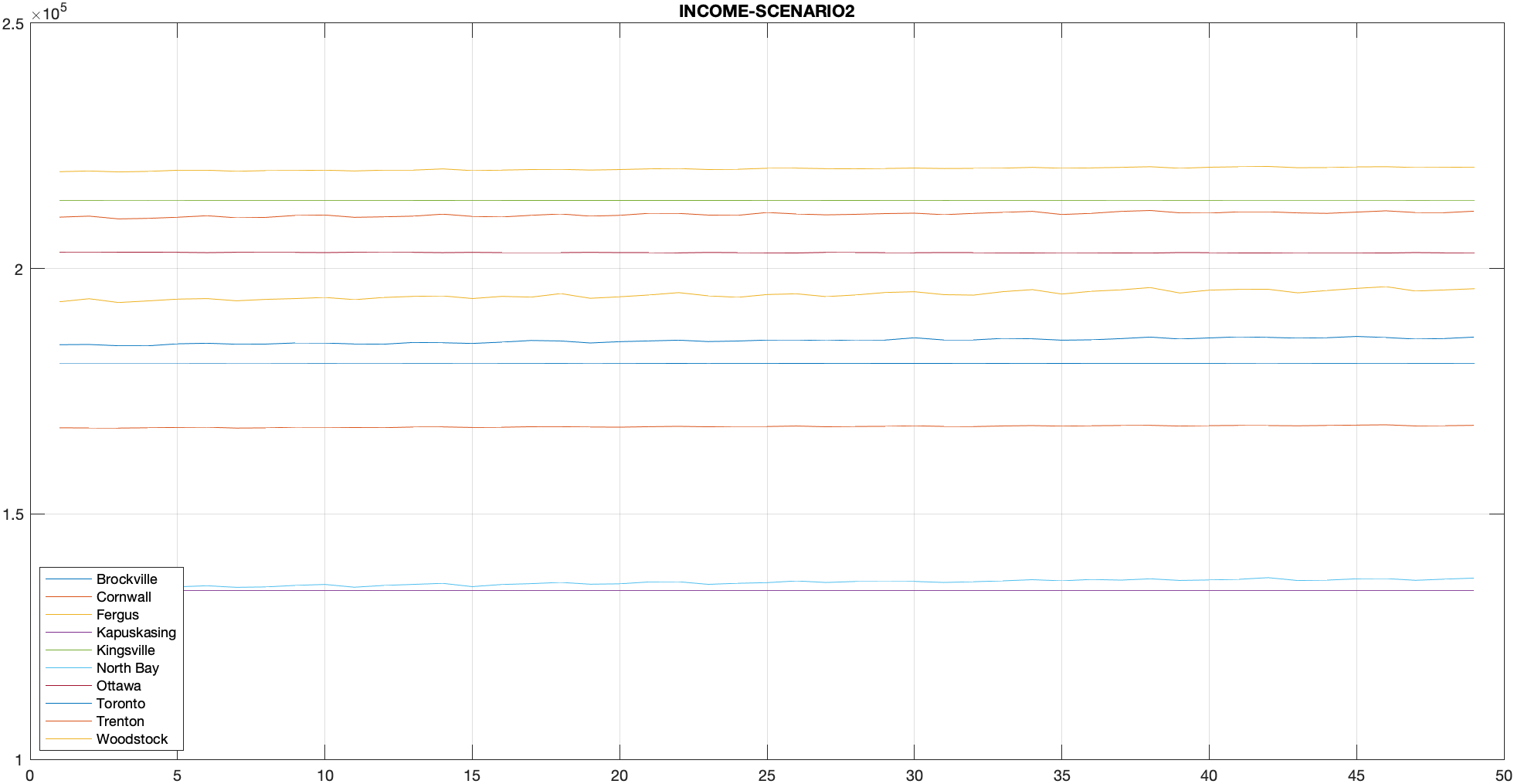}
	\caption{Income for  $W=0$ °C}
\end{figure}
\begin{figure}
	\includegraphics[width=150mm]{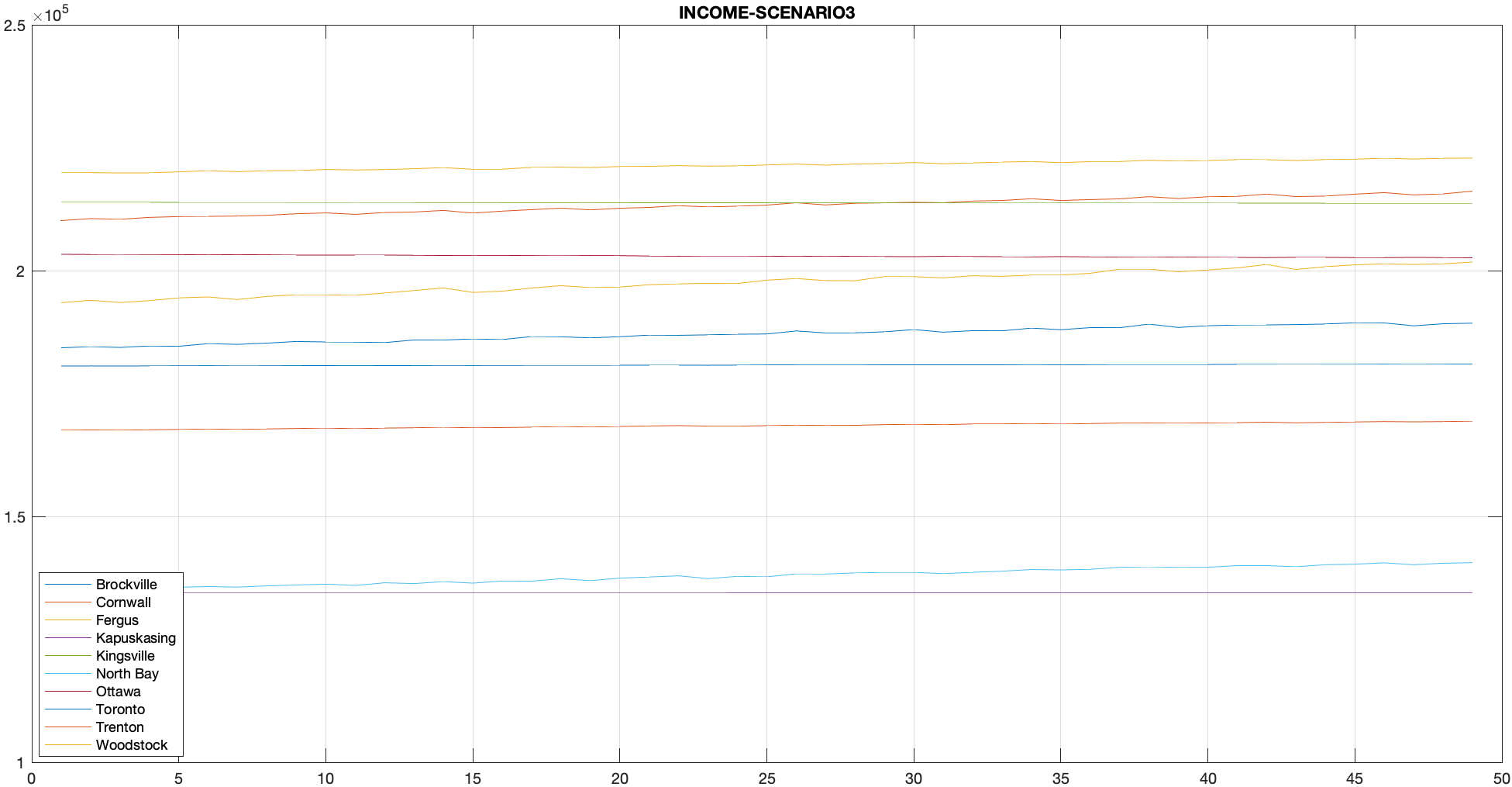}
	\caption{Income for  $W=0$ °C}
\end{figure}
\begin{figure}
	\includegraphics[width=150mm]{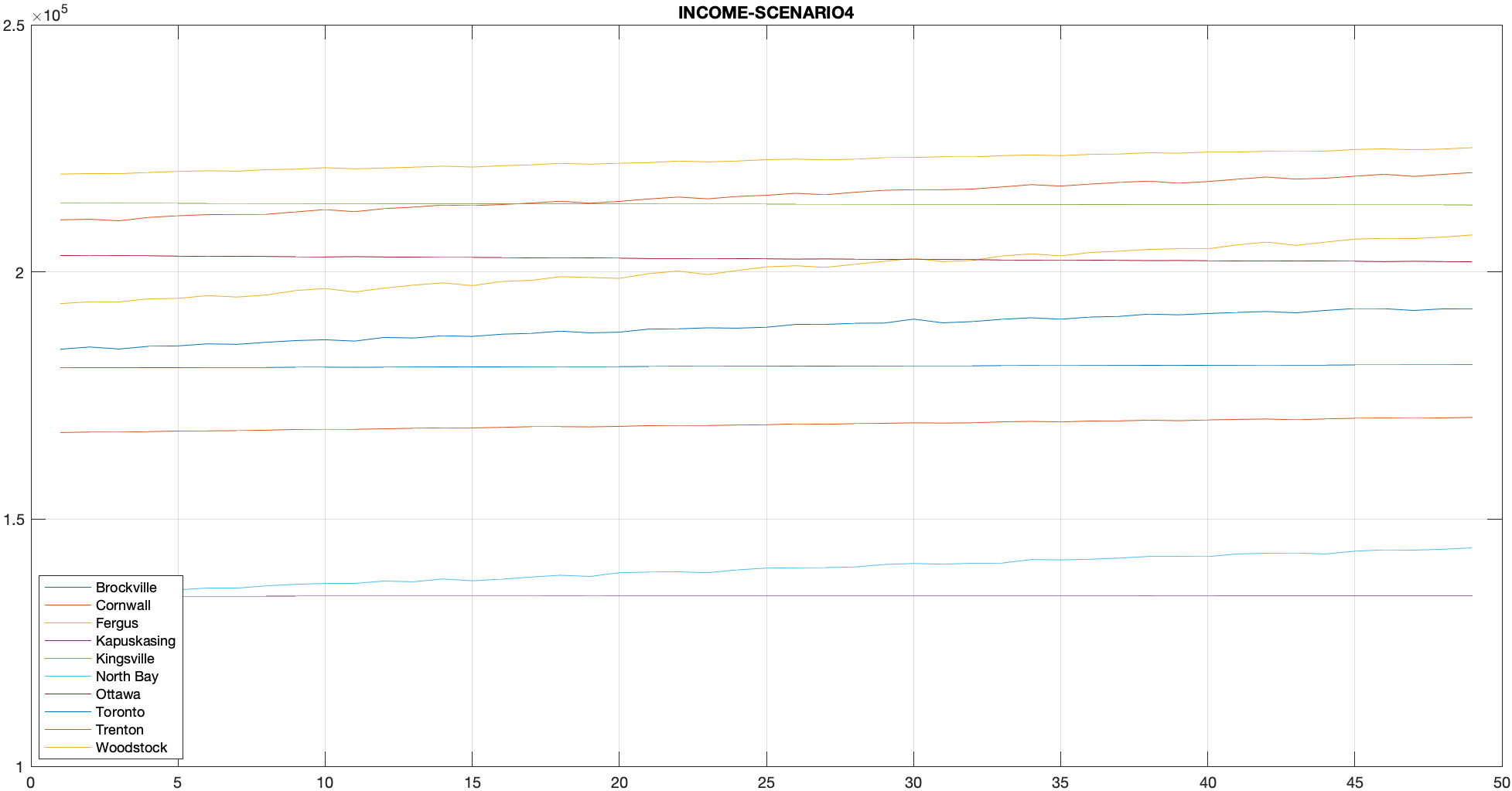}
	\caption{Income for  $W=0$ °C}
\end{figure}

\noindent For the scenarios  $W=3$°C and $W=4$°C, representing an acceleration of climate change, the income of the farms in most cities benefits from the extra heat that boosts the CHU and thus the yield. The fact that some cities do suffer and see a degradation of the income of the corn farms in their vicinity underlines the reality that the impact of climate change is more complex than merely increased average minimum and maximum temperatures. It also includes the possibility of extreme rainfall events that may shorten the corn growing season and interfere with the planting and harvest dates.\\

\noindent In order to better understand the impact of climate change under our five scenarios on every city, we have computed in Table 5 the difference between the farms averaged income over 1500 paths at the first year of the simulation (2020) and at the last year of the simulation (2068). As we have already said, reality is not as simple as the naive expectation that more warming is always a good thing for corn farmers. The influence of rainfall on the growing season makes it so some cities see their farms suffer a loss of expected income under the more extreme climate change scenarios. There are obvious gains under a scenario that includes more warming for North Bay in the north. Kapuskasing on the other hand does not seem to benefit much, but its corn industry is very small anyway and there were gaps in its historical time series for the yield, rendering the results of our simulations less precise for this particular city. Overall, cities in the North who would clearly benefit from a warming climate. Brockville and Cornwall to the East benefit as well in a spectacular fashion under the scenarios corresponding to the larger values of $W$, and so do Woodstock and Fergus to the West. Toronto and Trenton in the center of the province see increased income for their corn farms under more extreme climate change scenarios but Toronto seems to benefit less, maybe it is because the area is more urbanized generates its own micro-climate that is less susceptible to climate change. As a general observation, corn farming in Ontario seems to benefit from a warmer climate but there are notable exceptions. This underlines the importance of the local climate and the unpredictability of the impact of climate change on agriculture. Kingsville to the West sees a clear fall in the revenue of its corn farms as we consider more extreme climate scenarios and so does Ottawa to the East. The loss for Kingsville is modest as the climate gets warmer but Ottawa seems to follow the opposite trend as the rest of the province, netting big gains in the event of a complete stopping of climate change and large losses as the climate gets warmer under the more extreme scenarios. Those geographical disparities in the way that local ecosystems in Ontario react to climate change have also been demonstrated in Alberta in the work of  Dan and Williams (1985). They underline the financial risks associated with a warming climate. Indeed, while corn farming does generally benefit from higher average daily minimum and maximum temperatures, other factors like rainfall that controls the starting and ending dates of the growing season, as well as local characteristics of the climate can sometimes overpower the expected increase in daily CHU due to the higher temperatures and deliver a lower corn yield overall, thus damaging farm income.\\

\noindent We have also plotted in Figure 30 to Figure 40, each figure corresponding to one city, the evolution of income variation in Canadian Dollar on the y-axis as a function of the climate change scenario defined by the warming factor $W$ on the y-axis. Brockville, Cornwall, Fergus and North Bay see a massive increase of the income of their corn farms as we progress from an unlikely stop of climate change to a doubling of its warming trend. Kapuskasing does not seem to benefit much from a warmer climate, which is surprising. Toronto, Trenton and Woodstock benefit more modestly from the warmer climate. Ottawa and Kingsville defy expectations and see their farm income go down as we consider climate scenarios with a higher warming factor $W$.

\begin{table}[]
	\begin{center}
	
\begin{tabular}{|l|l|l|l|l|l|}
	\hline
	& $W=0$°C                         & $W=1$°C                         & $W=2$°C                        & $W=3$°C                        & $W=4$°C                         \\ \hline
	Brockville  & {\color[HTML]{000000} -6277.18} & {\color[HTML]{000000} -1952.15} & {\color[HTML]{000000} 1553.14} & {\color[HTML]{000000} 5017.80} & {\color[HTML]{000000} 8130.28}  \\ \hline
	Cornwall    & {\color[HTML]{000000} -8398.41} & {\color[HTML]{000000} -3359.17} & {\color[HTML]{000000} 1220.74} & {\color[HTML]{000000} 5979.81} & {\color[HTML]{000000} 9556.66}  \\ \hline
	Fergus      & {\color[HTML]{000000} -9687.55} & {\color[HTML]{000000} -3678.77} & {\color[HTML]{000000} 2657.68} & {\color[HTML]{000000} 8264.18} & {\color[HTML]{000000} 13898.55} \\ \hline
	Kapuskasing & {\color[HTML]{000000} -47.18}   & {\color[HTML]{000000} -22.94}   & {\color[HTML]{000000} 0.95}    & {\color[HTML]{000000} 22.40}   & {\color[HTML]{000000} 43.34}    \\ \hline
	Kingsville  & {\color[HTML]{000000} 248.43}   & {\color[HTML]{000000} 84.37}    & {\color[HTML]{000000} -75.36}  & {\color[HTML]{000000} -205.79} & {\color[HTML]{000000} -369.43}  \\ \hline
	North Bay   & {\color[HTML]{000000} -5893.04} & {\color[HTML]{000000} -2041.14} & {\color[HTML]{000000} 2160.78} & {\color[HTML]{000000} 5535.98} & {\color[HTML]{000000} 9095.75}  \\ \hline
	Ottawa      & {\color[HTML]{000000} 1177.21}  & {\color[HTML]{000000} 486.73}   & {\color[HTML]{000000} -95.18}  & {\color[HTML]{000000} -718.34} & {\color[HTML]{000000} -1270.15} \\ \hline
	Toronto     & {\color[HTML]{000000} -447.98}  & {\color[HTML]{000000} -159.29}  & {\color[HTML]{000000} 120.37}  & {\color[HTML]{000000} 378.84}  & {\color[HTML]{000000} 635.45}   \\ \hline
	Trenton     & {\color[HTML]{000000} -2266.69} & {\color[HTML]{000000} -788.77}  & {\color[HTML]{000000} 507.70}  & {\color[HTML]{000000} 1847.96} & {\color[HTML]{000000} 3072.64}  \\ \hline
	Woodstock   & {\color[HTML]{000000} -3963.35} & {\color[HTML]{000000} -1262.66} & {\color[HTML]{000000} 938.82}  & {\color[HTML]{000000} 2992.10} & {\color[HTML]{000000} 5345.16}  \\ \hline
\end{tabular}
	\end{center}
\caption{Yearly income variation forecasts at the 2068 horizon (Canadian Dollar).}
\end{table}

\begin{figure}[H]
	\begin{minipage}{0.5\textwidth}
		\centering
		\includegraphics[width=73mm]{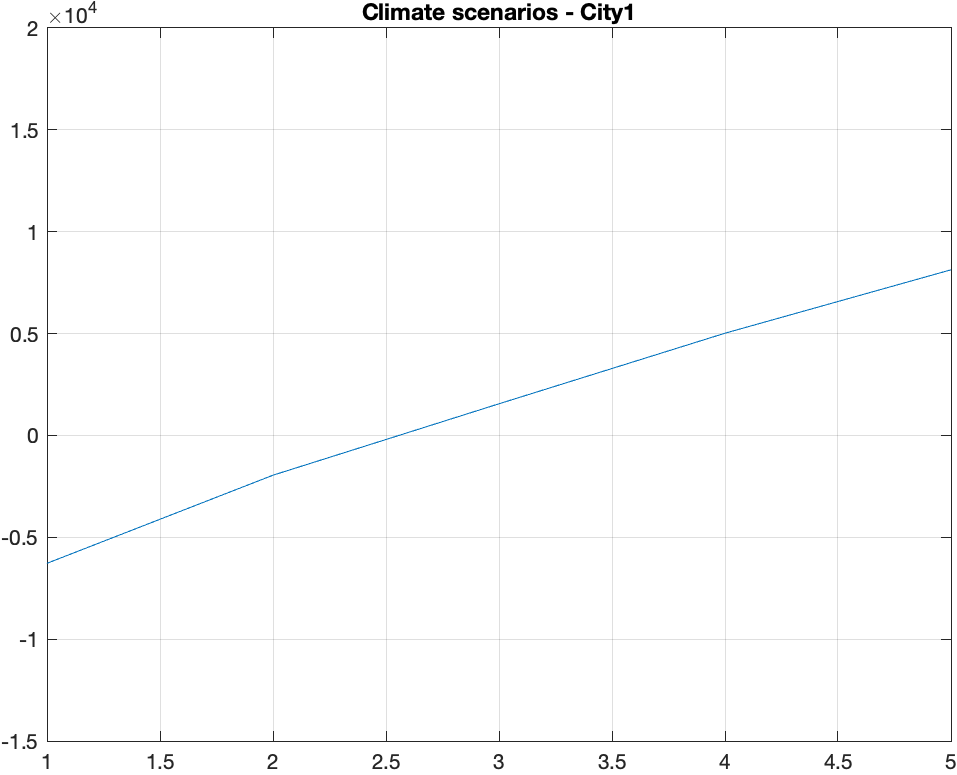}
		\caption{Brockville}
	\end{minipage}\hfill
	\begin{minipage}{0.5\textwidth}
		\centering
		\includegraphics[width=73mm]{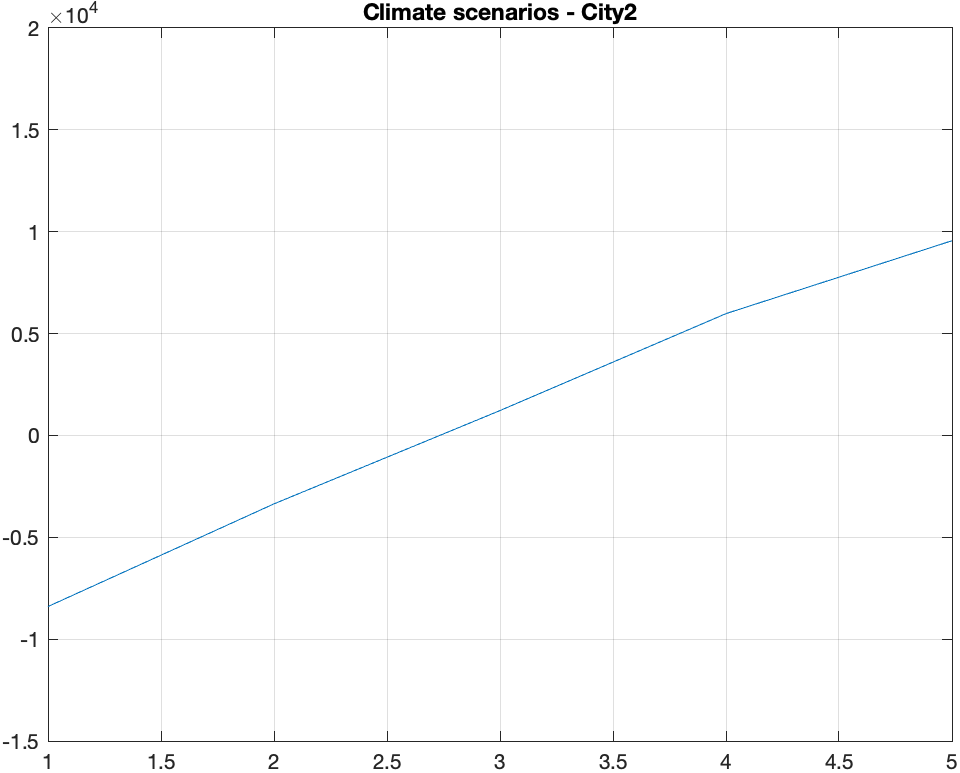}
		\caption{Cornwall}
	\end{minipage}\hfill
	\begin{minipage}{0.5\textwidth}
		\centering
		\includegraphics[width=73mm]{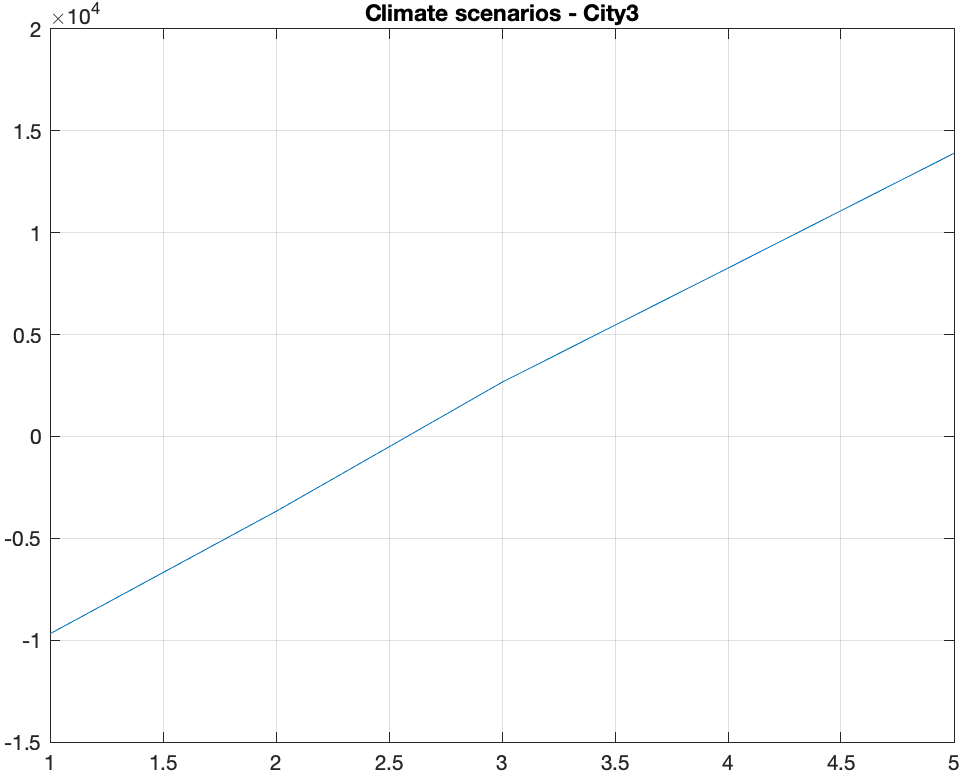}
		\caption{Fergus}
	\end{minipage}\hfill
	\begin{minipage}{0.5\textwidth}
		\centering
		\includegraphics[width=73mm]{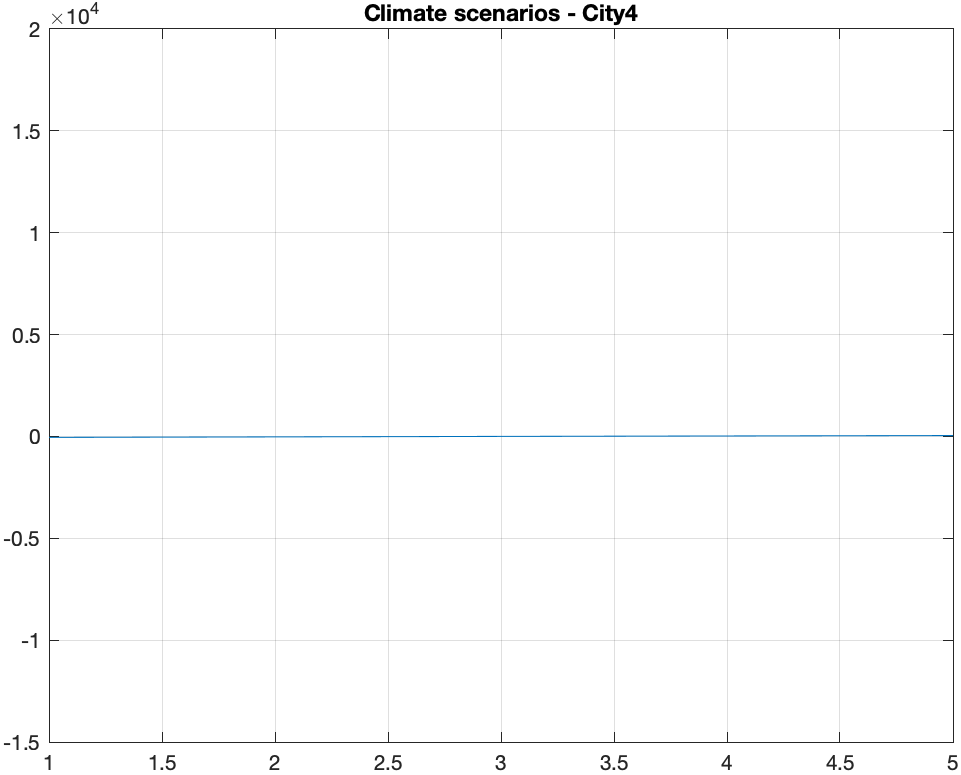}
		\caption{Kapuskasing}
	\end{minipage}\hfill
	\begin{minipage}{1\textwidth}
		\centering
		\includegraphics[width=73mm]{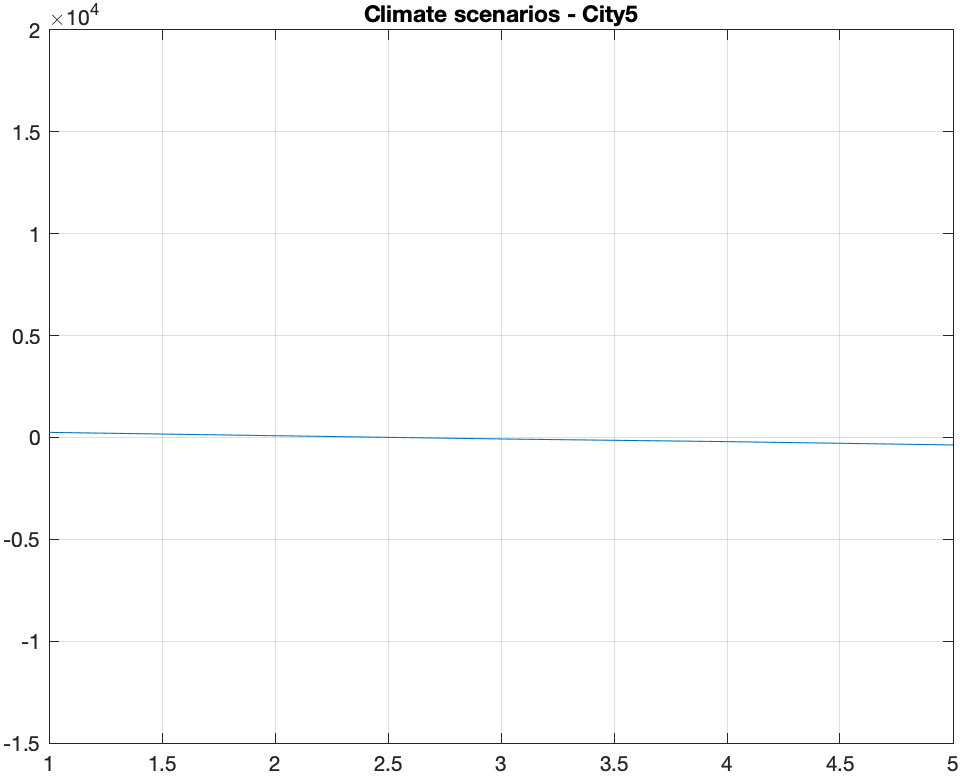}
		\caption{Kingsville}
	\end{minipage}\hfill
\end{figure}

\begin{figure}[H]
	\begin{minipage}{0.5\textwidth}
	\centering
	\includegraphics[width=73mm]{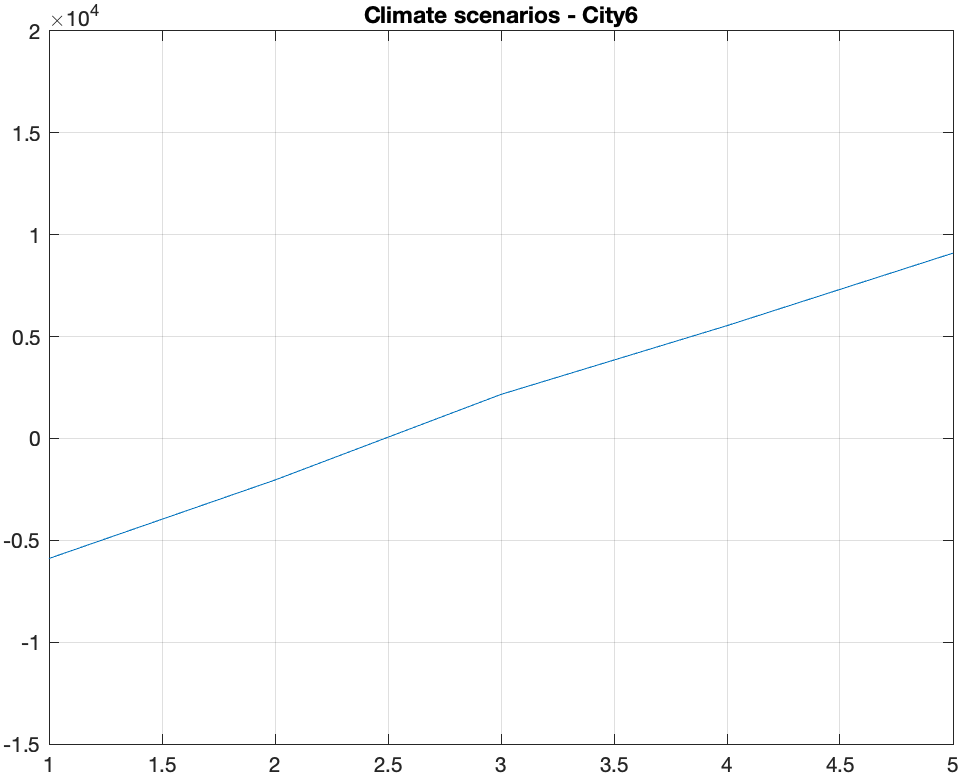}
	\caption{North Bay}
\end{minipage}\hfill
	\begin{minipage}{0.5\textwidth}
	\centering
	\includegraphics[width=73mm]{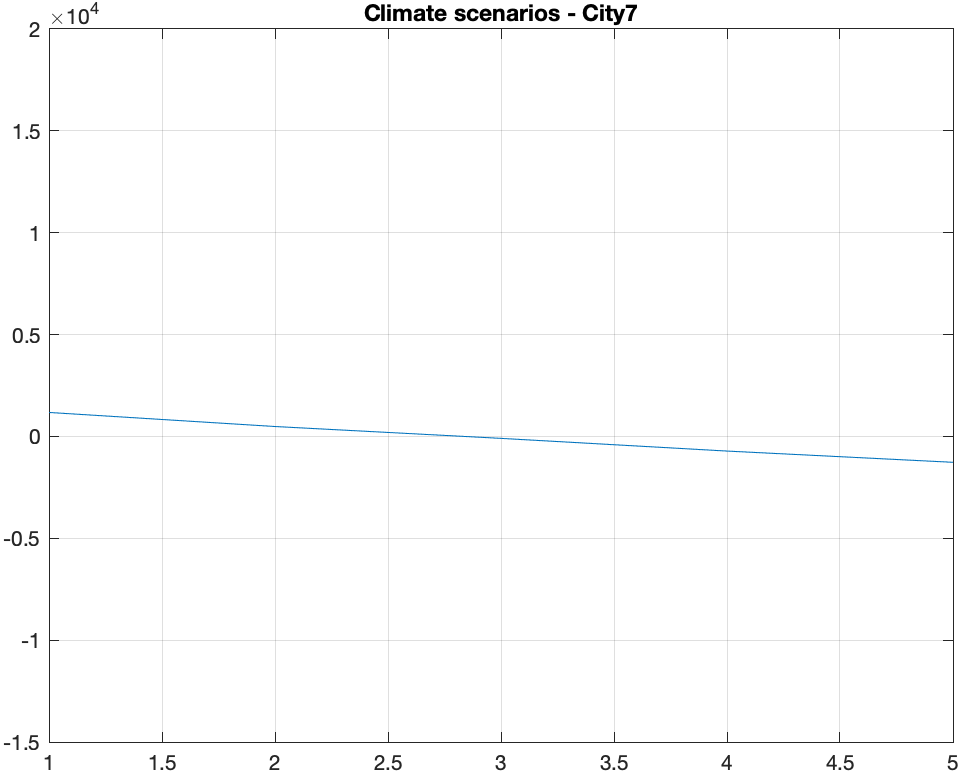}
	\caption{Ottawa}
\end{minipage}\hfill
	\begin{minipage}{0.5\textwidth}
	\centering
	\includegraphics[width=73mm]{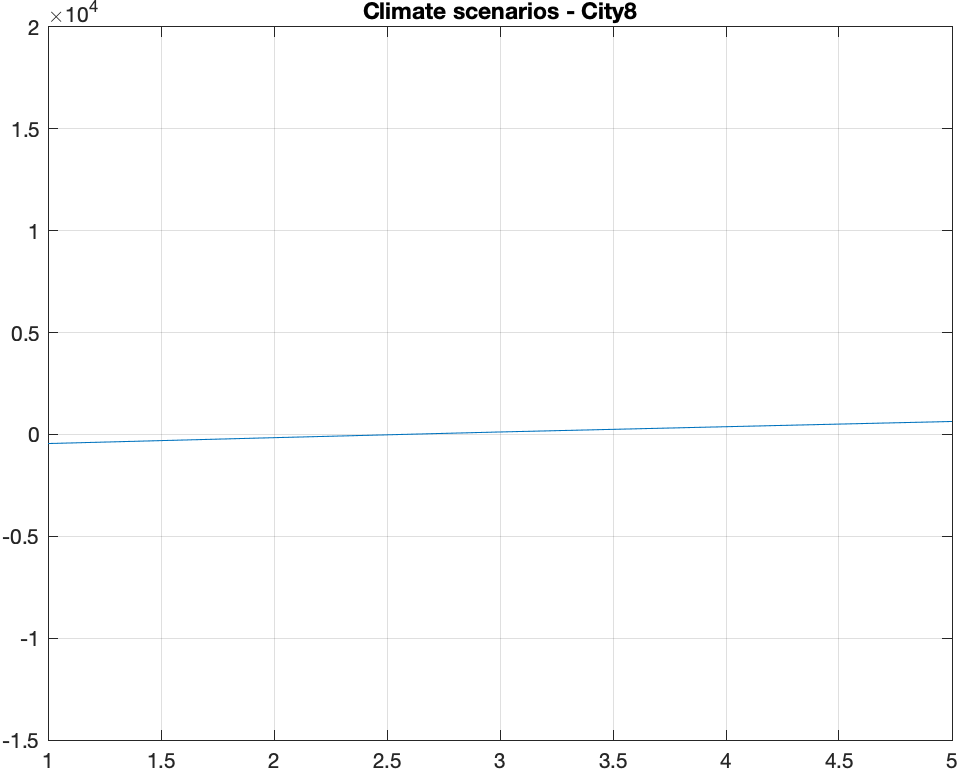}
	\caption{Toronto}
\end{minipage}\hfill
	\begin{minipage}{0.5\textwidth}
	\centering
	\includegraphics[width=73mm]{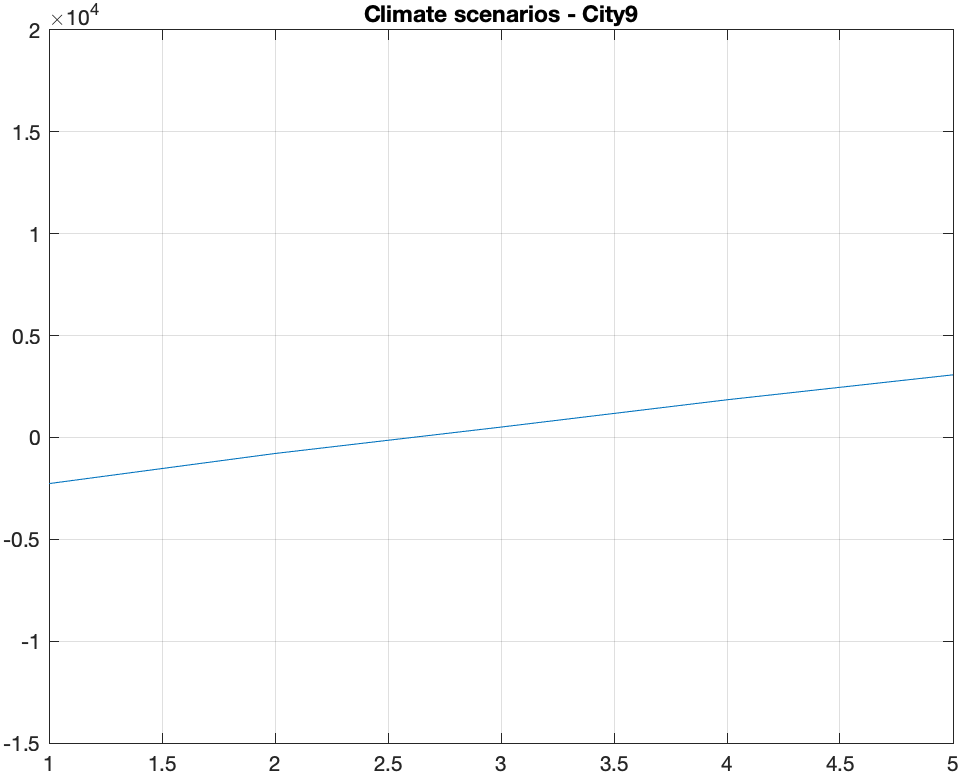}
	\caption{Trenton}
\end{minipage}\hfill
	\begin{minipage}{1\textwidth}
	\centering
	\includegraphics[width=73mm]{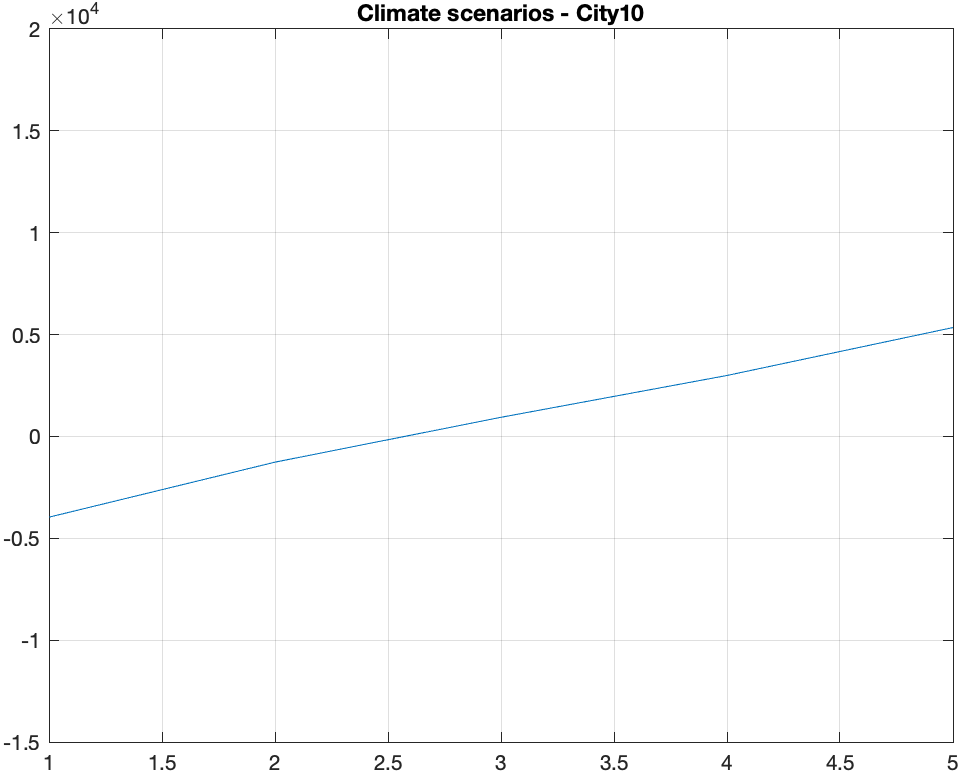}
	\caption{Woodstock}
\end{minipage}\hfill
\end{figure}

\section{Conclusions}

As a conclusion, we see that even a simple model of corn farm income like the one we have presented for the province of Ontario can produce very interesting results that underline the financial risks associated with climate change. Our model is not necessarily meant to be a very realistic depiction of the financial challenges encountered by corn farms in Ontario due to climate change but it still shows that climate change means uncertainty of income. It shows that the naive expectations (more heat equals more CHU and thus a better yield) are not always true because other factors, like rainfall, are at play. In Ontario, while more heat under a climate scenario that assumes an acceleration of the historical warming trends, both in minimum and maximum daily temperatures, tends to benefit corn farming and results in increased income for the corn farms in most areas of Ontario, there are notable exceptions. Those exceptions, like the region around Ottawa to West and the region around Kingsville to the East, have no obvious geographical explanation and seem to find their roots in the characteristics of the local climate, temperature and rainfall. This shows that climate change brings uncertainty in corn farm income and uncertainty means risk, which is expensive to handle from a financial point of view. Our simple model for modeling and forecasting corn farm income in Ontario could be used as a first step toward developing a more extensive credit risk framework, encompassing the use of options to counter financial risk created by the unpredictable climate change influence. Such an extended framework could include a modeling of corn future prices as well as interest rates and would open the possibility of computing the probability of default of a farm recipient of a loan under a given climate scenario. Our future research will build upon the simple model presented in this study and attempt to refine our understanding of the financial implications of climate change on the agricultural sector in order to better understand the threat that this phenomenon represents for the world economy and the stability of the global financial system as a whole.

\section*{Appendix A}
List of the ten cities in Ontario, representative of the corn growing counties, that we have used in our study and for which future climate paths were created from historical data. They are listed with the identification and location (GHCND-ID; latitude; longitude) of the nearest suitable weather station in the NOAA database that created the data that we have used:\\

\begin{table}[H]
	\centering
	\begin{tabular}{|l|l|l|l|}
		\hline
		\textit{City}        & \textit{GHCND-ID}    & \textit{latitude} &\textit{ longitude} \\ \hline
		Brockville  & CA006100971 & 44.6000  & -75.6667  \\ \hline
		Cornwall    & CA006101874 & 45.0167  & -74.7500  \\ \hline
		Fergus      & CA006142400 & 43.7333  & -80.3333  \\ \hline
		Kapuskasing & CA006073975 & 49.4167  & -82.4667  \\ \hline
		Kingsville  & CA006134190 & 42.0500  & -82.6667  \\ \hline
		North Bay   & CA006085700 & 46.3667  & -79.4167  \\ \hline
		Ottawa      & CA006105976 & 45.3833  & -75.7167  \\ \hline
		Toronto     & CA006158733 & 43.6833  & -79.6333  \\ \hline
		Trenton     & CA006158875 & 44.1167  & -77.5333  \\ \hline
		Woodstock   & CA006149625 & 43.1333  & -80.7667  \\ \hline
	\end{tabular}
\end{table}

\section*{Appendix B}
Historical corn yield data at the county level expressed in bushel per acre. When gaps in the data were presents, which happened in particular for the older dates and the northern cities of North Bay and Kapsukasing, we carried over the last valid entry in the time-series. We accounted for county mergers over time and the evolution of the legal structures of the counties or municipalities considered. Brockville is in the Greenville \& Leeds county (when they were separate, we took the average of both  entries), Cornwall is in Dundas, Glengarry and Stormont county (we took the average of the three entries when they were separate). Fergus is in Wellington county. For Kapuskasing and North bay we took the average value for Northern Ontario, which was the only suitable entry available in OMAFRA database. Kingsville is in Essex county. Toronto is in York county, Trenton is in Hastings county and Woodstock is in Oxford county.
\begin{table}[H]
\tiny
\begin{tabular}{|l|l|l|l|l|l|l|l|}
\hline
\rowcolor[HTML]{C0C0C0} 
1970             & 1     & 1971             & 2     & 1972             & 3     & 1973             & 4     \\ \hline
\rowcolor[HTML]{EFEFEF} 
Essex            & 82.0  & Essex            & 79.3  & Essex            & 92.2  & Essex            & 77.7  \\ \hline
Oxford           & 88.0  & Oxford           & 80.4  & Oxford           & 80.6  & Oxford           & 78.3  \\ \hline
\rowcolor[HTML]{EFEFEF} 
Wellington       & 80.7  & Wellington       & 84.4  & Wellington       & 74.1  & Wellington       & 75.9  \\ \hline
Hastings         & 73.0  & Hastings         & 65.3  & Hastings         & 72.1  & Hastings         & 81.6  \\ \hline
\rowcolor[HTML]{EFEFEF} 
York             & 80.6  & York             & 76.4  & York             & 78.1  & York             & 79.5  \\ \hline
Dundas           & 77.3  & Dundas           & 76.3  & Dundas           & 31.3  & Dundas           & 66.6  \\ \hline
Glengarry        & 87.4  & Glengarry        & 83.6  & Glengarry        & 40.6  & Glengarry        & 80.7  \\ \hline
Stormont         & 81.3  & Stormont         & 80.0  & Stormont         & 13.4  & Stormont         & 67.2  \\ \hline
\rowcolor[HTML]{EFEFEF} 
Grenville        & 80.0  & Grenville        & 82.6  & Grenville        & 35.8  & Grenville        & 73.3  \\ \hline
\rowcolor[HTML]{EFEFEF} 
Leeds            & 72.5  & Leeds            & 80.6  & Leeds            & 60.7  & Leeds            & 70.0  \\ \hline
Ottawa-Carelton  & 70.0  & Ottawa-Carelton  & 70.0  & Ottawa-Carelton  & 70.0  & Ottawa-Carelton  & 70.0  \\ \hline
\rowcolor[HTML]{EFEFEF} 
Northern Ontario & 65.0  & Northern Ontario & 65.0  & Northern Ontario & 65.0  & Northern Ontario & 65.0  \\ \hline
\rowcolor[HTML]{C0C0C0} 
1974             & 5     & 1975             & 6     & 1976             & 7     & 1977             & 8     \\ \hline
\rowcolor[HTML]{EFEFEF} 
Essex            & 68.9  & Essex            & 91.0  & Essex            & 80.0  & Essex            & 113.0 \\ \hline
Oxford           & 71.3  & Oxford           & 96.0  & Oxford           & 97.0  & Oxford           & 106.0 \\ \hline
\rowcolor[HTML]{EFEFEF} 
Wellington       & 40.2  & Wellington       & 84.0  & Wellington       & 74.0  & Wellington       & 88.0  \\ \hline
Hastings         & 72.5  & Hastings         & 55.0  & Hastings         & 71.0  & Hastings         & 71.0  \\ \hline
\rowcolor[HTML]{EFEFEF} 
York             & 65.5  & York             & 78.0  & York             & 77.0  & York             & 82.0  \\ \hline
Dundas           & 52.7  & Dundas           & 70.0  & Dundas           & 70.0  & Dundas           & 73.0  \\ \hline
\rowcolor[HTML]{EFEFEF} 
Glengarry        & 58.2  & Glengarry        & 85.0  & Glengarry        & 75.0  & Glengarry        & 67.0  \\ \hline
Stormont         & 44.5  & Stormont         & 69.0  & Stormont         & 66.0  & Stormont         & 63.0  \\ \hline
\rowcolor[HTML]{EFEFEF} 
Grenville        & 54.8  & Grenville        & 68.0  & Grenville        & 58.0  & Grenville        & 67.0  \\ \hline
\rowcolor[HTML]{EFEFEF} 
Leeds            & 42.5  & Leeds            & 45.0  & Leeds            & 68.0  & Leeds            & 50.0  \\ \hline
Ottawa-Carelton  & 70.0  & Ottawa-Carelton  & 70.0  & Ottawa-Carelton  & 73.0  & Ottawa-Carelton  & 52.0  \\ \hline
\rowcolor[HTML]{EFEFEF} 
Northern Ontario & 65.0  & Northern Ontario & 65.0  & Northern Ontario & 65.0  & Northern Ontario & 45.0  \\ \hline
\rowcolor[HTML]{C0C0C0} 
1978             & 9     & 1979             & 10    & 1980             & 11    & 1981             & 12    \\ \hline
\rowcolor[HTML]{EFEFEF} 
Essex            & 12.0  & Essex            & 103.0 & Essex            & 105.0 & Essex            & 105.0 \\ \hline
Oxford           & 88.0  & Oxford           & 96.0  & Oxford           & 109.0 & Oxford           & 109.0 \\ \hline
\rowcolor[HTML]{EFEFEF} 
Wellington       & 73.0  & Wellington       & 78.0  & Wellington       & 84.0  & Wellington       & 84.0  \\ \hline
Hastings         & 60.0  & Hastings         & 72.0  & Hastings         & 79.0  & Hastings         & 79.0  \\ \hline
\rowcolor[HTML]{EFEFEF} 
York             & 67.0  & York             & 82.0  & York             & 77.0  & York             & 77.0  \\ \hline
Dundas           & 74.0  & Dundas           & 84.0  & Dundas           & 84.0  & Dundas           & 84.0  \\ \hline
Glengarry        & 85.0  & Glengarry        & 87.0  & Glengarry        & 79.0  & Glengarry        & 79.0  \\ \hline
Stormont         & 70.0  & Stormont         & 78.0  & Stormont         & 79.0  & Stormont         & 79.0  \\ \hline
\rowcolor[HTML]{EFEFEF} 
Grenville        & 78.0  & Grenville        & 76.0  & Grenville        & 73.0  & Grenville        & 73.0  \\ \hline
\rowcolor[HTML]{EFEFEF} 
Leeds            & 72.0  & Leeds            & 72.0  & Leeds            & 69.0  & Leeds            & 69.0  \\ \hline
Ottawa-Carelton  & 88.0  & Ottawa-Carelton  & 79.0  & Ottawa-Carelton  & 82.0  & Ottawa-Carelton  & 82.0  \\ \hline
\rowcolor[HTML]{EFEFEF} 
Northern Ontario & 45.0  & Northern Ontario & 45.0  & Northern Ontario & 45.0  & Northern Ontario & 45.0  \\ \hline
\rowcolor[HTML]{C0C0C0} 
1982             & 13    & 1983             & 14    & 1984             & 15    & 1985             & 16    \\ \hline
\rowcolor[HTML]{EFEFEF} 
Essex            & 7.0   & Essex            & 91.0  & Essex            & 90.0  & Essex            & 125.0 \\ \hline
Oxford           & 104.0 & Oxford           & 105.0 & Oxford           & 110.0 & Oxford           & 112.0 \\ \hline
\rowcolor[HTML]{EFEFEF} 
Wellington       & 75.0  & Wellington       & 81.0  & Wellington       & 84.0  & Wellington       & 80.0  \\ \hline
Hastings         & 68.0  & Hastings         & 60.0  & Hastings         & 81.0  & Hastings         & 67.0  \\ \hline
\rowcolor[HTML]{EFEFEF} 
York             & 90.0  & York             & 73.0  & York             & 83.0  & York             & 86.0  \\ \hline
Dundas           & 99.0  & Dundas           & 81.0  & Dundas           & 89.0  & Dundas           & 96.0  \\ \hline
Glengarry        & 93.0  & Glengarry        & 78.0  & Glengarry        & 85.0  & Glengarry        & 90.0  \\ \hline
Stormont         & 90.0  & Stormont         & 80.0  & Stormont         & 73.0  & Stormont         & 90.0  \\ \hline
\rowcolor[HTML]{EFEFEF} 
Grenville        & 87.0  & Grenville        & 65.0  & Grenville        & 82.0  & Grenville        & 87.0  \\ \hline
\rowcolor[HTML]{EFEFEF} 
Leeds            & 67.0  & Leeds            & 64.0  & Leeds            & 63.0  & Leeds            & 70.0  \\ \hline
Ottawa-Carelton  & 88.0  & Ottawa-Carelton  & 75.0  & Ottawa-Carelton  & 84.0  & Ottawa-Carelton  & 89.0  \\ \hline
\rowcolor[HTML]{EFEFEF} 
Northern Ontario & 45.0  & Northern Ontario & 45.0  & Northern Ontario & 45.0  & Northern Ontario & 45.0  \\ \hline
\rowcolor[HTML]{C0C0C0} 
1986             & 17    & 1987             & 18    & 1988             & 19    & 1989             & 20    \\ \hline
\rowcolor[HTML]{EFEFEF} 
Essex            & 115.0 & Essex            & 130.0 & Essex            & 69.0  & Essex            & 123.0 \\ \hline
Oxford           & 108.0 & Oxford           & 128.0 & Oxford           & 100.0 & Oxford           & 114.0 \\ \hline
\rowcolor[HTML]{EFEFEF} 
Wellington       & 81.0  & Wellington       & 118.0 & Wellington       & 72.0  & Wellington       & 81.0  \\ \hline
Hastings         & 72.0  & Hastings         & 87.0  & Hastings         & 63.0  & Hastings         & 73.0  \\ \hline
\rowcolor[HTML]{EFEFEF} 
York             & 93.0  & York             & 110.0 & York             & 83.0  & York             & 83.0  \\ \hline
Dundas           & 86.0  & Dundas           & 115.0 & Dundas           & 94.0  & Dundas           & 111.0 \\ \hline
Glengarry        & 81.0  & Glengarry        & 115.0 & Glengarry        & 99.0  & Glengarry        & 104.0 \\ \hline
Stormont         & 85.0  & Stormont         & 110.0 & Stormont         & 97.0  & Stormont         & 101.0 \\ \hline
\rowcolor[HTML]{EFEFEF} 
Grenville        & 73.0  & Grenville        & 98.0  & Grenville        & 87.0  & Grenville        & 93.0  \\ \hline
\rowcolor[HTML]{EFEFEF} 
Leeds            & 76.0  & Leeds            & 84.0  & Leeds            & 84.0  & Leeds            & 78.0  \\ \hline
Ottawa-Carelton  & 71.0  & Ottawa-Carelton  & 110.0 & Ottawa-Carelton  & 95.0  & Ottawa-Carelton  & 109.0 \\ \hline
\rowcolor[HTML]{EFEFEF} 
Northern Ontario & 45.0  & Northern Ontario & 45.0  & Northern Ontario & 80.0  & Northern Ontario & 80.0  \\ \hline
\end{tabular}
\end{table}

\begin{table}[H]
\tiny
\begin{tabular}{|l|l|l|l|l|l|l|l|}
\hline
\rowcolor[HTML]{C0C0C0} 
1990                                                                               & 21    & 1991                                                                               & 22    & 1992                                                                               & 23    & 1993                                                                               & 24    \\ \hline
\rowcolor[HTML]{EFEFEF} 
Essex                                                                              & 118.0 & Essex                                                                              & 60.0  & Essex                                                                              & 130.0 & Essex                                                                              & 106.5 \\ \hline
Oxford                                                                             & 119.0 & Oxford                                                                             & 138.0 & Oxford                                                                             & 100.0 & Oxford                                                                             & 117.0 \\ \hline
\rowcolor[HTML]{EFEFEF} 
Wellington                                                                         & 93.0  & Wellington                                                                         & 122.0 & Wellington                                                                         & 69.0  & Wellington                                                                         & 97.0  \\ \hline
Hastings                                                                           & 86.0  & Hastings                                                                           & 64.0  & Hastings                                                                           & 84.0  & Hastings                                                                           & 81.0  \\ \hline
\rowcolor[HTML]{EFEFEF} 
York                                                                               & 90.0  & York                                                                               & 94.0  & York                                                                               & 81.0  & York                                                                               & 79.0  \\ \hline
Dundas                                                                             & 104.0 & Dundas                                                                             & 108.0 & Dundas                                                                             & 98.0  & Dundas                                                                             & 107.0 \\ \hline
\rowcolor[HTML]{EFEFEF} 
Glengarry                                                                          & 103.0 & Glengarry                                                                          & 110.0 & Glengarry                                                                          & 75.0  & Glengarry                                                                          & 91.0  \\ \hline
Stormont                                                                           & 102.0 & Stormont                                                                           & 102.0 & Stormont                                                                           & 89.0  & Stormont                                                                           & 97.2  \\ \hline
\rowcolor[HTML]{EFEFEF} 
Grenville                                                                          & 88.0  & Grenville                                                                          & 104.0 & Grenville                                                                          & 70.0  & Grenville                                                                          & 86.0  \\ \hline
\rowcolor[HTML]{EFEFEF} 
Leeds                                                                              & 88.0  & Leeds                                                                              & 104.0 & Leeds                                                                              & 99.0  & Leeds                                                                              & 80.0  \\ \hline
Ottawa-Carelton                                                                    & 107.0 & Ottawa-Carelton                                                                    & 110.0 & Ottawa-Carelton                                                                    & 103.0 & Ottawa-Carelton                                                                    & 103.0 \\ \hline
\rowcolor[HTML]{EFEFEF} 
\begin{tabular}[c]{@{}l@{}}Northern   Ontario\end{tabular}                       & 80.0  & Northern Ontario                                                                   & 80.0  & Northern Ontario                                                                   & 45.0  & Northern Ontario                                                                   & 69.5  \\ \hline
\rowcolor[HTML]{C0C0C0} 
1994                                                                               & 25    & 1995                                                                               & 26    & 1996                                                                               & 27    & 1997                                                                               & 28    \\ \hline
\rowcolor[HTML]{EFEFEF} 
Essex                                                                              & 134.0 & Essex                                                                              & 114.0 & Essex                                                                              & 126.0 & Essex                                                                              & 124.0 \\ \hline
Oxford                                                                             & 126.0 & Oxford                                                                             & 123.0 & Oxford                                                                             & 110.0 & Oxford                                                                             & 128.0 \\ \hline
\rowcolor[HTML]{EFEFEF} 
Wellington                                                                         & 116.0 & Wellington                                                                         & 118.0 & Wellington                                                                         & 101.0 & Wellington                                                                         & 98.0  \\ \hline
Hastings                                                                           & 108.0 & Hastings                                                                           & 87.0  & Hastings                                                                           & 98.0  & Hastings                                                                           & 89.0  \\ \hline
\rowcolor[HTML]{EFEFEF} 
York                                                                               & 103.0 & York                                                                               & 108.0 & York                                                                               & 98.0  & York                                                                               & 80.0  \\ \hline
Dundas                                                                             & 120.0 & Dundas                                                                             & 121.0 & \begin{tabular}[c]{@{}l@{}}Dundas, Glengarry\\   \& Stormont\end{tabular}          & 130.0 & \begin{tabular}[c]{@{}l@{}}Dundas, Glengarry\\   \& Stormont\end{tabular}          & 90.0  \\ \hline
Glengarry                                                                          & 117.5 & Glengarry                                                                          & 118.0 & \begin{tabular}[c]{@{}l@{}}Dundas, Glengarry\\   \& Stormont\end{tabular}          & 130.0 & \begin{tabular}[c]{@{}l@{}}Dundas, Glengarry\\   \& Stormont\end{tabular}          & 90.0  \\ \hline
Stormont                                                                           & 114.5 & Stormont                                                                           & 114.0 & \begin{tabular}[c]{@{}l@{}}Dundas, Glengarry\\   \& Stormont\end{tabular}          & 130.0 & \begin{tabular}[c]{@{}l@{}}Dundas, Glengarry\\   \& Stormont\end{tabular}          & 90.0  \\ \hline
\rowcolor[HTML]{EFEFEF} 
Grenville                                                                          & 119.0 & Grenville                                                                          & 106.0 & Grenville and Leeds                                                                & 109.0 & Grenville and Leeds                                                                & 94.0  \\ \hline
\rowcolor[HTML]{EFEFEF} 
Leeds                                                                              & 116.0 & Leeds                                                                              & 96.0  & Grenville and Leeds                                                                & 109.0 & Grenville and Leeds                                                                & 94.0  \\ \hline
Ottawa-Carelton                                                                    & 125.0 & Ottawa-Carelton                                                                    & 126.0 & Ottawa-Carleton                                                                    & 125.0 & Ottawa-Carleton                                                                    & 90.0  \\ \hline
\rowcolor[HTML]{EFEFEF} 
\begin{tabular}[c]{@{}l@{}}Northern   Ontario\end{tabular}                       & 84.0  & Northern Ontario                                                                   & 88.7  & Northern Ontario                                                                   & 111.7 & Northern Ontario                                                                   & 68.0  \\ \hline
\rowcolor[HTML]{C0C0C0} 
1998                                                                               & 29    & 1999                                                                               & 30    & 2000                                                                               & 31    & 2001                                                                               & 32    \\ \hline
\rowcolor[HTML]{EFEFEF} 
Essex                                                                              & 139.0 & Essex                                                                              & 117.0 & Essex                                                                              & 135.0 & Essex                                                                              & 98.0  \\ \hline
Oxford                                                                             & 125.0 & Oxford                                                                             & 137.0 & Oxford                                                                             & 105.0 & Oxford                                                                             & 114.0 \\ \hline
\rowcolor[HTML]{EFEFEF} 
Wellington                                                                         & 113.0 & Wellington                                                                         & 127.0 & Wellington                                                                         & 94.0  & Wellington                                                                         & 98.0  \\ \hline
Hastings                                                                           & 115.0 & Hastings                                                                           & 111.0 & Hastings                                                                           & 98.0  & Hastings                                                                           & 67.0  \\ \hline
\rowcolor[HTML]{EFEFEF} 
York                                                                               & 121.0 & York                                                                               & 119.0 & York                                                                               & 84.0  & York                                                                               & 98.0  \\ \hline
\begin{tabular}[c]{@{}l@{}}Dundas,   Glengarry\\ \& Stormont\end{tabular}          & 137.0 & \begin{tabular}[c]{@{}l@{}}Dundas, Glengarry\\   \& Stormont\end{tabular}          & 136.0 & \begin{tabular}[c]{@{}l@{}}Dundas, Glengarry\\   \& Stormont\end{tabular}          & 100.0 & \begin{tabular}[c]{@{}l@{}}Dundas, Glengarry\\   \& Stormont\end{tabular}          & 119.0 \\ \hline
\rowcolor[HTML]{EFEFEF} 
\begin{tabular}[c]{@{}l@{}}Grenville and    Leeds\end{tabular}                    & 125.0 & Grenville and Leeds                                                                & 116.0 & Grenville and Leeds                                                                & 81.0  & Grenville and Leeds                                                                & 75.0  \\ \hline
Ottawa-Carleton                                                                    & 131.0 & Ottawa-Carleton                                                                    & 134.0 & Ottawa-Carleton                                                                    & 77.0  & Ottawa-Carleton                                                                    & 106.0 \\ \hline
\rowcolor[HTML]{EFEFEF} 
\begin{tabular}[c]{@{}l@{}}Northern   Ontario\end{tabular}                       & 88.2  & Northern Ontario                                                                   & 111.0 & Northern Ontario                                                                   & 59.0  & Northern Ontario                                                                   & 65.7  \\ \hline
\rowcolor[HTML]{C0C0C0} 
2002                                                                               & 33    & 2003                                                                               & 34    & 2004                                                                               & 35    & 2005                                                                               & 36    \\ \hline
\rowcolor[HTML]{EFEFEF} 
Essex                                                                              & 80.0  & Essex                                                                              & 130.0 & Essex County                                                                       & 138   & Essex County                                                                       & 133   \\ \hline
Oxford                                                                             & 136.0 & Oxford                                                                             & 129.0 & Oxford County                                                                      & 130   & Oxford County                                                                      & 164   \\ \hline
\rowcolor[HTML]{EFEFEF} 
Wellington                                                                         & 123.0 & Wellington                                                                         & 117.0 & Wellington County                                                                  & 115   & Wellington County                                                                  & 138   \\ \hline
Hastings                                                                           & 116.0 & Hastings                                                                           & 115.0 & Hastings County                                                                    & 116   & Hastings County                                                                    & 94    \\ \hline
\rowcolor[HTML]{EFEFEF} 
York                                                                               & 109.0 & York                                                                               & 106.0 & \begin{tabular}[c]{@{}l@{}}York Regional\\   Municipality\end{tabular}             & 112   & \begin{tabular}[c]{@{}l@{}}York Regional\\   Municipality\end{tabular}             & 125   \\ \hline
\begin{tabular}[c]{@{}l@{}}Dundas,   Glengarry\\ \& Stormont\end{tabular}          & 118.0 & \begin{tabular}[c]{@{}l@{}}Dundas, Glengarry\\   \& Stormont\end{tabular}          & 122.0 & \begin{tabular}[c]{@{}l@{}}Stormont, Dundas and\\   Glenmont Counties\end{tabular} & 145   & \begin{tabular}[c]{@{}l@{}}Stormont, Dundas and\\   Glenmont Counties\end{tabular} & 153   \\ \hline
\rowcolor[HTML]{EFEFEF} 
\begin{tabular}[c]{@{}l@{}}Grenville and   Leeds\end{tabular}                    & 94.0  & Grenville and Leeds                                                                & 126.0 & \begin{tabular}[c]{@{}l@{}}Leeds and Grenville\\   United Counties\end{tabular}    & 123   & \begin{tabular}[c]{@{}l@{}}Leeds and Grenville\\   United Counties\end{tabular}    & 121   \\ \hline
Ottawa-Carleton                                                                    & 118.0 & Ottawa-Carleton                                                                    & 129.0 & Ottawa Division                                                                    & 138   & Ottawa Division                                                                    & 144   \\ \hline
\rowcolor[HTML]{EFEFEF} 
\begin{tabular}[c]{@{}l@{}}Northern   Ontario\end{tabular}                       & 102.0 & Northern Ontario                                                                   & 113.9 & \begin{tabular}[c]{@{}l@{}}Northern Ontario \end{tabular}                & 89.4  & \begin{tabular}[c]{@{}l@{}}Northern Ontario\end{tabular}                & 100.5 \\ \hline
\rowcolor[HTML]{C0C0C0} 
2006                                                                               & 37    & 2007                                                                               & 38    & 2008                                                                               & 39    & 2009                                                                               & 40    \\ \hline
\rowcolor[HTML]{EFEFEF} 
Essex County                                                                       & 153   & Essex County                                                                       & 142   & Essex County                                                                       & 138   & Essex County                                                                       & 161   \\ \hline
Oxford County                                                                      & 162   & Oxford County                                                                      & 151   & Oxford County                                                                      & 171   & Oxford County                                                                      & 153   \\ \hline
\rowcolor[HTML]{EFEFEF} 
\begin{tabular}[c]{@{}l@{}}Wellington    County\end{tabular}                      & 140   & Wellington County                                                                  & 117   & Wellington County                                                                  & 137   & Wellington County                                                                  & 120   \\ \hline
\begin{tabular}[c]{@{}l@{}}Hastings   County\end{tabular}                        & 104   & Hastings County                                                                    & 106   & Hastings County                                                                    & 141   & Hastings County                                                                    & 121   \\ \hline
\rowcolor[HTML]{EFEFEF} 
\begin{tabular}[c]{@{}l@{}}York Regional\\   Municipality\end{tabular}             & 119   & \begin{tabular}[c]{@{}l@{}}York Regional\\   Municipality\end{tabular}             & 103   & \begin{tabular}[c]{@{}l@{}}York Regional\\   Municipality\end{tabular}             & 138   & \begin{tabular}[c]{@{}l@{}}York Regional\\   Municipality\end{tabular}             & 123   \\ \hline
\begin{tabular}[c]{@{}l@{}}Stormont,   Dundas and \\ Glenmont Counties\end{tabular} & 148   & \begin{tabular}[c]{@{}l@{}}Stormont, Dundas and\\   Glenmont Counties\end{tabular} & 159   & \begin{tabular}[c]{@{}l@{}}Stormont, Dundas and\\   Glenmont Counties\end{tabular} & 146   & \begin{tabular}[c]{@{}l@{}}Stormont, Dundas and\\   Glenmont Counties\end{tabular} & 148   \\ \hline
\rowcolor[HTML]{EFEFEF} 
\begin{tabular}[c]{@{}l@{}}Leeds and   Grenville \\ United Counties\end{tabular}    & 133   & \begin{tabular}[c]{@{}l@{}}Leeds and Grenville\\   United Counties\end{tabular}    & 137   & \begin{tabular}[c]{@{}l@{}}Leeds and Grenville\\   United Counties\end{tabular}    & 123   & \begin{tabular}[c]{@{}l@{}}Leeds and Grenville\\   United Counties\end{tabular}    & 131   \\ \hline
\begin{tabular}[c]{@{}l@{}}Ottawa    Division\end{tabular}                        & 151   & Ottawa Division                                                                    & 157   & Ottawa Division                                                                    & 145   & Ottawa Division                                                                    & 148   \\ \hline
\rowcolor[HTML]{EFEFEF} 
\begin{tabular}[c]{@{}l@{}}Northern   Ontario \end{tabular}                & 88.6  & \begin{tabular}[c]{@{}l@{}}Northern Ontario \end{tabular}                & 86.3  & \begin{tabular}[c]{@{}l@{}}Northern Ontario \end{tabular}                & 77.8  & \begin{tabular}[c]{@{}l@{}}Northern Ontario \end{tabular}                & 83    \\ \hline
\end{tabular}
\end{table}

\begin{table}[H]
\tiny
\begin{tabular}{|l|l|llllll}
\hline
\rowcolor[HTML]{C0C0C0} 
2010                                                                                                    & 41                            & \multicolumn{1}{l|}{\cellcolor[HTML]{C0C0C0}2011}                                                                            & \multicolumn{1}{l|}{\cellcolor[HTML]{C0C0C0}42}    & \multicolumn{1}{l|}{\cellcolor[HTML]{C0C0C0}2012}                                                                            & \multicolumn{1}{l|}{\cellcolor[HTML]{C0C0C0}43}    & \multicolumn{1}{l|}{\cellcolor[HTML]{C0C0C0}2013}                                                                            & \multicolumn{1}{l|}{\cellcolor[HTML]{C0C0C0}44}    \\ \hline
\rowcolor[HTML]{EFEFEF} 
Essex County                                                                                            & 158                           & \multicolumn{1}{l|}{\cellcolor[HTML]{EFEFEF}Essex County}                                                                    & \multicolumn{1}{l|}{\cellcolor[HTML]{EFEFEF}163.3} & \multicolumn{1}{l|}{\cellcolor[HTML]{EFEFEF}Essex County}                                                                    & \multicolumn{1}{l|}{\cellcolor[HTML]{EFEFEF}165.1} & \multicolumn{1}{l|}{\cellcolor[HTML]{EFEFEF}Essex County}                                                                    & \multicolumn{1}{l|}{\cellcolor[HTML]{EFEFEF}162.5} \\ \hline
Oxford County                                                                                           & 176                           & \multicolumn{1}{l|}{Oxford County}                                                                                           & \multicolumn{1}{l|}{170.0}                         & \multicolumn{1}{l|}{Oxford County}                                                                                           & \multicolumn{1}{l|}{159.9}                         & \multicolumn{1}{l|}{Oxford County}                                                                                           & \multicolumn{1}{l|}{176.2}                         \\ \hline
\rowcolor[HTML]{EFEFEF} 
\begin{tabular}[c]{@{}l@{}}Wellington   County\end{tabular}                                           & 155                           & \multicolumn{1}{l|}{\cellcolor[HTML]{EFEFEF}Wellington County}                                                               & \multicolumn{1}{l|}{\cellcolor[HTML]{EFEFEF}140.1} & \multicolumn{1}{l|}{\cellcolor[HTML]{EFEFEF}Wellington County}                                                               & \multicolumn{1}{l|}{\cellcolor[HTML]{EFEFEF}143.4} & \multicolumn{1}{l|}{\cellcolor[HTML]{EFEFEF}Wellington County}                                                               & \multicolumn{1}{l|}{\cellcolor[HTML]{EFEFEF}153.1} \\ \hline
\begin{tabular}[c]{@{}l@{}}Hastings   County\end{tabular}                                             & 159                           & \multicolumn{1}{l|}{Hastings County}                                                                                         & \multicolumn{1}{l|}{158.0}                         & \multicolumn{1}{l|}{Hastings County}                                                                                         & \multicolumn{1}{l|}{116.8}                         & \multicolumn{1}{l|}{Hastings County}                                                                                         & \multicolumn{1}{l|}{139.1}                         \\ \hline
\rowcolor[HTML]{EFEFEF} 
\begin{tabular}[c]{@{}l@{}}York Regional\\   Municipality\end{tabular}                                  & 148                           & \multicolumn{1}{l|}{\cellcolor[HTML]{EFEFEF}\begin{tabular}[c]{@{}l@{}}York Regional\\   Municipality\end{tabular}}          & \multicolumn{1}{l|}{\cellcolor[HTML]{EFEFEF}170.0} & \multicolumn{1}{l|}{\cellcolor[HTML]{EFEFEF}\begin{tabular}[c]{@{}l@{}}York Regional\\   Municipality\end{tabular}}          & \multicolumn{1}{l|}{\cellcolor[HTML]{EFEFEF}151.4} & \multicolumn{1}{l|}{\cellcolor[HTML]{EFEFEF}\begin{tabular}[c]{@{}l@{}}York Regional\\   Municipality\end{tabular}}          & \multicolumn{1}{l|}{\cellcolor[HTML]{EFEFEF}138.3} \\ \hline
\begin{tabular}[c]{@{}l@{}}Stormont,   Dundas and\\ Glenmont Counties\end{tabular}                      & 165                           & \multicolumn{1}{l|}{\begin{tabular}[c]{@{}l@{}}Stormont, Dundas and\\   Glenmont Counties\end{tabular}}                      & \multicolumn{1}{l|}{149.6}                         & \multicolumn{1}{l|}{\begin{tabular}[c]{@{}l@{}}Stormont, Dundas and\\   Glenmont Counties\end{tabular}}                      & \multicolumn{1}{l|}{152.1}                         & \multicolumn{1}{l|}{\begin{tabular}[c]{@{}l@{}}Stormont, Dundas and\\   Glenmont Counties\end{tabular}}                      & \multicolumn{1}{l|}{153.7}                         \\ \hline
\rowcolor[HTML]{EFEFEF} 
\begin{tabular}[c]{@{}l@{}}Leeds and   Grenville\\ United Counties\end{tabular}                         & 148                           & \multicolumn{1}{l|}{\cellcolor[HTML]{EFEFEF}\begin{tabular}[c]{@{}l@{}}Leeds and Grenville\\   United Counties\end{tabular}} & \multicolumn{1}{l|}{\cellcolor[HTML]{EFEFEF}145.3} & \multicolumn{1}{l|}{\cellcolor[HTML]{EFEFEF}\begin{tabular}[c]{@{}l@{}}Leeds and Grenville\\   United Counties\end{tabular}} & \multicolumn{1}{l|}{\cellcolor[HTML]{EFEFEF}138.3} & \multicolumn{1}{l|}{\cellcolor[HTML]{EFEFEF}\begin{tabular}[c]{@{}l@{}}Leeds and Grenville\\   United Counties\end{tabular}} & \multicolumn{1}{l|}{\cellcolor[HTML]{EFEFEF}138.3} \\ \hline
\begin{tabular}[c]{@{}l@{}}Ottawa   Division\end{tabular}                                             & 170                           & \multicolumn{1}{l|}{Ottawa Division}                                                                                         & \multicolumn{1}{l|}{135.7}                         & \multicolumn{1}{l|}{Ottawa Division}                                                                                         & \multicolumn{1}{l|}{121.2}                         & \multicolumn{1}{l|}{Ottawa Division}                                                                                         & \multicolumn{1}{l|}{161.1}                         \\ \hline
\rowcolor[HTML]{EFEFEF} 
\begin{tabular}[c]{@{}l@{}}Northern   Ontario\end{tabular}                                     & 108.4                         & \multicolumn{1}{l|}{\cellcolor[HTML]{EFEFEF}\begin{tabular}[c]{@{}l@{}}Northern Ontario\end{tabular}}             & \multicolumn{1}{l|}{\cellcolor[HTML]{EFEFEF}91.7}  & \multicolumn{1}{l|}{\cellcolor[HTML]{EFEFEF}\begin{tabular}[c]{@{}l@{}}Northern Ontario\end{tabular}}             & \multicolumn{1}{l|}{\cellcolor[HTML]{EFEFEF}126.0} & \multicolumn{1}{l|}{\cellcolor[HTML]{EFEFEF}\begin{tabular}[c]{@{}l@{}}Northern Ontario\end{tabular}}             & \multicolumn{1}{l|}{\cellcolor[HTML]{EFEFEF}129.5} \\ \hline
\rowcolor[HTML]{C0C0C0} 
2014                                                                                                    & 45                            & \multicolumn{1}{l|}{\cellcolor[HTML]{C0C0C0}2015}                                                                            & \multicolumn{1}{l|}{\cellcolor[HTML]{C0C0C0}46}    & \multicolumn{1}{l|}{\cellcolor[HTML]{C0C0C0}2016}                                                                            & \multicolumn{1}{l|}{\cellcolor[HTML]{C0C0C0}47}    & \multicolumn{1}{l|}{\cellcolor[HTML]{C0C0C0}2017}                                                                            & \multicolumn{1}{l|}{\cellcolor[HTML]{C0C0C0}48}    \\ \hline
\rowcolor[HTML]{EFEFEF} 
Essex County                                                                                            & 185.0                         & \multicolumn{1}{l|}{\cellcolor[HTML]{EFEFEF}Essex County}                                                                    & \multicolumn{1}{l|}{\cellcolor[HTML]{EFEFEF}188.0} & \multicolumn{1}{l|}{\cellcolor[HTML]{EFEFEF}Essex County}                                                                    & \multicolumn{1}{l|}{\cellcolor[HTML]{EFEFEF}167.0} & \multicolumn{1}{l|}{\cellcolor[HTML]{EFEFEF}Essex County}                                                                    & \multicolumn{1}{l|}{\cellcolor[HTML]{EFEFEF}185.5} \\ \hline
Oxford County                                                                                           & 166.0                         & \multicolumn{1}{l|}{Oxford County}                                                                                           & \multicolumn{1}{l|}{178.0}                         & \multicolumn{1}{l|}{Oxford County}                                                                                           & \multicolumn{1}{l|}{175.5}                         & \multicolumn{1}{l|}{Oxford County}                                                                                           & \multicolumn{1}{l|}{185.0}                         \\ \hline
\rowcolor[HTML]{EFEFEF} 
\begin{tabular}[c]{@{}l@{}}Wellington   County\end{tabular}                                           & 145.0                         & \multicolumn{1}{l|}{\cellcolor[HTML]{EFEFEF}Wellington County}                                                               & \multicolumn{1}{l|}{\cellcolor[HTML]{EFEFEF}159.5} & \multicolumn{1}{l|}{\cellcolor[HTML]{EFEFEF}Wellington County}                                                               & \multicolumn{1}{l|}{\cellcolor[HTML]{EFEFEF}156.1} & \multicolumn{1}{l|}{\cellcolor[HTML]{EFEFEF}Wellington County}                                                               & \multicolumn{1}{l|}{\cellcolor[HTML]{EFEFEF}170.4} \\ \hline
\begin{tabular}[c]{@{}l@{}}Hastings   County\end{tabular}                                             & 140.5                         & \multicolumn{1}{l|}{Hastings County}                                                                                         & \multicolumn{1}{l|}{153.4}                         & \multicolumn{1}{l|}{Hastings County}                                                                                         & \multicolumn{1}{l|}{89.5}                          & \multicolumn{1}{l|}{Hastings County}                                                                                         & \multicolumn{1}{l|}{126.1}                         \\ \hline
\rowcolor[HTML]{EFEFEF} 
\begin{tabular}[c]{@{}l@{}}York Regional\\   Municipality\end{tabular}                                  & 149.4                         & \multicolumn{1}{l|}{\cellcolor[HTML]{EFEFEF}\begin{tabular}[c]{@{}l@{}}York Regional\\   Municipality\end{tabular}}          & \multicolumn{1}{l|}{\cellcolor[HTML]{EFEFEF}152.0} & \multicolumn{1}{l|}{\cellcolor[HTML]{EFEFEF}\begin{tabular}[c]{@{}l@{}}York Regional\\   Municipality\end{tabular}}          & \multicolumn{1}{l|}{\cellcolor[HTML]{EFEFEF}132.0} & \multicolumn{1}{l|}{\cellcolor[HTML]{EFEFEF}\begin{tabular}[c]{@{}l@{}}York Regional\\   Municipality\end{tabular}}          & \multicolumn{1}{l|}{\cellcolor[HTML]{EFEFEF}158.7} \\ \hline
\begin{tabular}[c]{@{}l@{}}Stormont,   Dundas and\\ Glenmont Counties\end{tabular}                      & 152.2                         & \multicolumn{1}{l|}{\begin{tabular}[c]{@{}l@{}}Stormont, Dundas and\\   Glenmont Counties\end{tabular}}                      & \multicolumn{1}{l|}{180.9}                         & \multicolumn{1}{l|}{\begin{tabular}[c]{@{}l@{}}Stormont, Dundas and\\   Glenmont Counties\end{tabular}}                      & \multicolumn{1}{l|}{153.9}                         & \multicolumn{1}{l|}{\begin{tabular}[c]{@{}l@{}}Stormont, Dundas and\\   Glenmont Counties\end{tabular}}                      & \multicolumn{1}{l|}{160.4}                         \\ \hline
\rowcolor[HTML]{EFEFEF} 
\begin{tabular}[c]{@{}l@{}}Leeds and   Grenville\\ United Counties\end{tabular}                         & 136.8                         & \multicolumn{1}{l|}{\cellcolor[HTML]{EFEFEF}\begin{tabular}[c]{@{}l@{}}Leeds and Grenville\\   United Counties\end{tabular}} & \multicolumn{1}{l|}{\cellcolor[HTML]{EFEFEF}155.0} & \multicolumn{1}{l|}{\cellcolor[HTML]{EFEFEF}\begin{tabular}[c]{@{}l@{}}Leeds and Grenville\\   United Counties\end{tabular}} & \multicolumn{1}{l|}{\cellcolor[HTML]{EFEFEF}133.9} & \multicolumn{1}{l|}{\cellcolor[HTML]{EFEFEF}\begin{tabular}[c]{@{}l@{}}Leeds and Grenville\\   United Counties\end{tabular}} & \multicolumn{1}{l|}{\cellcolor[HTML]{EFEFEF}131.4} \\ \hline
\begin{tabular}[c]{@{}l@{}}Ottawa   Division\end{tabular}                                             & 152.0                         & \multicolumn{1}{l|}{Ottawa Division}                                                                                         & \multicolumn{1}{l|}{184.7}                         & \multicolumn{1}{l|}{Ottawa Division}                                                                                         & \multicolumn{1}{l|}{152.3}                         & \multicolumn{1}{l|}{Ottawa Division}                                                                                         & \multicolumn{1}{l|}{143.4}                         \\ \hline
\rowcolor[HTML]{EFEFEF} 
\begin{tabular}[c]{@{}l@{}}Northern  Ontario \end{tabular}                                     & 37.4                          & \multicolumn{1}{l|}{\cellcolor[HTML]{EFEFEF}\begin{tabular}[c]{@{}l@{}}Northern Ontario\end{tabular}}             & \multicolumn{1}{l|}{\cellcolor[HTML]{EFEFEF}167.3} & \multicolumn{1}{l|}{\cellcolor[HTML]{EFEFEF}\begin{tabular}[c]{@{}l@{}}Northern Ontario\end{tabular}}             & \multicolumn{1}{l|}{\cellcolor[HTML]{EFEFEF}74.3}  & \multicolumn{1}{l|}{\cellcolor[HTML]{EFEFEF}\begin{tabular}[c]{@{}l@{}}Northern Ontario\end{tabular}}             & \multicolumn{1}{l|}{\cellcolor[HTML]{EFEFEF}91.9}  \\ \hline
\cellcolor[HTML]{C0C0C0}2018                                                                            & \cellcolor[HTML]{C0C0C0}49    &                                                                                                                              &                                                    &                                                                                                                              &                                                    &                                                                                                                              &                                                    \\ \cline{1-2}
\cellcolor[HTML]{EFEFEF}Essex County                                                                    & \cellcolor[HTML]{EFEFEF}167.0 &                                                                                                                              &                                                    &                                                                                                                              &                                                    &                                                                                                                              &                                                    \\ \cline{1-2}
Oxford County                                                                                           & 182.2                         &                                                                                                                              &                                                    &                                                                                                                              &                                                    &                                                                                                                              &                                                    \\ \cline{1-2}
\cellcolor[HTML]{EFEFEF}\begin{tabular}[c]{@{}l@{}}Wellington   County\end{tabular}                   & \cellcolor[HTML]{EFEFEF}168.0 &                                                                                                                              &                                                    &                                                                                                                              &                                                    &                                                                                                                              &                                                    \\ \cline{1-2}
\begin{tabular}[c]{@{}l@{}}Hastings   County\end{tabular}                                             & 134.0                         &                                                                                                                              &                                                    &                                                                                                                              &                                                    &                                                                                                                              &                                                    \\ \cline{1-2}
\cellcolor[HTML]{EFEFEF}\begin{tabular}[c]{@{}l@{}}York Regional\\   Municipality\end{tabular}          & \cellcolor[HTML]{EFEFEF}148.6 &                                                                                                                              &                                                    &                                                                                                                              &                                                    &                                                                                                                              &                                                    \\ \cline{1-2}
\begin{tabular}[c]{@{}l@{}}Stormont,   Dundas and\\ Glenmont Counties\end{tabular}                      & 158.8                         &                                                                                                                              &                                                    &                                                                                                                              &                                                    &                                                                                                                              &                                                    \\ \cline{1-2}
\cellcolor[HTML]{EFEFEF}\begin{tabular}[c]{@{}l@{}}Leeds and   Grenville\\ United Counties\end{tabular} & \cellcolor[HTML]{EFEFEF}131.7 &                                                                                                                              &                                                    &                                                                                                                              &                                                    &                                                                                                                              &                                                    \\ \cline{1-2}
\begin{tabular}[c]{@{}l@{}}Ottawa   Division\end{tabular}                                             & 164.3                         &                                                                                                                              &                                                    &                                                                                                                              &                                                    &                                                                                                                              &                                                    \\ \cline{1-2}
\cellcolor[HTML]{EFEFEF}\begin{tabular}[c]{@{}l@{}}Northern   Ontario \end{tabular}             & \cellcolor[HTML]{EFEFEF}125.0 &                                                                                                                              &                                                    &                                                                                                                              &                                                    &                                                                                                                              &                                                    \\ \cline{1-2}
\end{tabular}
\end{table}

\begin{small}
 
\end{small}

\begin{thebibliography} {1}
	
\bibitem{} Almaraz J.J., Mabood F., Zhou X., Gregorich E.G. and Smith D.L. (2008) "Climate Change, Weather Variability and Corn Yield at a Higher Latitude Locale: Southwestern Quebec", \textit{Climatic Change}, \textbf{88}, pp. 187-197, (\url{https://doi.org/10.1007/s10584-008-9408-y}).
	
\bibitem{} Andersson M., Bolton P. and Samama F. (2016) "Hedging Climate Risk", \textit{Financial Analysts Journal}, 72(\textbf{3}), pp. 13-32, (\url{https://doi.org/10.2469/faj.v72.n3.4}).
		
\bibitem{} Björnberg K.E., Karlsson M., Gilek M. and Hansson S.O. (2017). Climate and environmental science denial: A review of the scientific literature published in 1990–2015.\textit{ Journal of Cleaner Production} \textbf{167}, pp. 229-241 (\url{https://doi.org/10.1016/j.jclepro.2017.08.066}).

\bibitem{} Bootsma A., Gameda S., and McKenney D.W. (2005). Potential Impacts of Climate Change on Corn, Soybeans and Barley Yields in Atlantic Canada. \textit{Canadian Journal of Soil Science}, 85(\textbf{2}), pp. 345-357, (\url{https://doi.org/10.4141/S04-025}).

\bibitem{} Brinkmann W.A.R. (1979) "Growing Season Length as an Indicator of Climatic Variations?", \textit{Climatic Change}, \textbf{2}, pp. 127-138, (\url{https://doi.org/10.1007/BF00133219}).

\bibitem{} Cabas J., Weersink A. and Olale E. (2010) "Crop Yield Response to Economic, Site and Climatic Variables", \textit{Climatic Change}, \textbf{101}, pp. 599-616, (\url{https://doi.org/10.1007/s10584-009-9754-4}).

\bibitem{} Brinkman J., McKinnon P., Pitblado R. and Nichols I. (2008). A Study Of Starting Dates For CHU Accumulation in Ontario. \textit{Weather INnovations Incorporated}, pp. 1-15 (\url{http://www.weatherinnovations.com/docs/CORN-CHU-WIN%20Report-2008.pdf}).

\bibitem{} Dafermos Y., Nikolaidi M. and Galanis G. (2018). Climate Change, Financial Stability and Monetary Policy. \textit{Ecological Economics}, \textbf{152}, pp. 219-234 (\url{https://doi.org/10.1016/j.ecolecon.2018.05.011}).

\bibitem{} Dan G. and Williams V. (1985) "Estimated Bioressource Sensitivity to Climatic Change in Alberta, Canada", \textit{Climatic Change}, \textbf{7}, pp. 55-69, (\url{https://doi.org/10.1007/BF00139441}).

\bibitem{} Deryng D., Sacks W.J., Barford C.C.  and Ramankutty N. (2011). Simulating the Effects of Climate and Agricultural Management Practices on Global Crop Yield. \textit{Global Biogeochemical Cycles}, 25(\textbf{1}), pp. 1-18, (\url{https://doi.org/10.1029/2009GB003765}).

\bibitem{} Eastwood R., Lipton M. and Newell A. (2010).  Chapter 65 Farm Size. In \textit{Handbook of Agricultural Economics, Volume 4}, pp. 3323-3397, (\url{https://doi.org/10.1016/S1574-0072(09)04065-1}.

\bibitem{} Kaiser H.M., Riha S.J., Wilks D.S., Rossiter D.G. and Sampath R. (1993) "A Farm-Level Analysis of Economic and Agronomic Impacts of Gradual Climate Warming", "American Journal of Agricultural Economics", 75(\textbf{2}), pp. 387-398,  (\url{https://doi.org/10.2307/1242923}).

\bibitem{} Katz R.W. (1977) "Assessing the Impact of Climatic Change on Food Production", \textit{Climatic Change}, \textbf{1}, pp. 85-96, (\url{https://doi.org/10.1007/BF00162779}).

\bibitem{}Kolk A. and Pinkse J. (2004). Market Strategies for Climate Change.  \textit{European Management Journal}, 22(\textbf{3}), pp. 304-314, (\url{https://doi.org/10.1016/j.emj.2004.04.011}).

\bibitem{}  Kucharik C.J. (2006) "A Multidecadal Trend of Earlier Corn Planting in the Central USA", \textit{Agronomy Journal}, 98(\textbf{6}), pp. 1544-1550, (\url{https://doi.org/10.2134/agronj2006.0156}).

\bibitem{}  Kwabiah A.B., MacPherson M. and McKenzie D.B. (2003). Corn Heat Unit Variability and Potential of Corn Production in a Cool Climate Ecosystem, \textit{Canadian Journal of Plant Science}, 83(\textbf{4}), pp. 689-698, (\url{https://doi.org/10.4141/P02-127}).

\bibitem{} Liang B.C., MacKemie A.F., Kirby P.C. and  Remillard M. (1991). Corn Production in Relation to Water Inputs and Heat Units, \textit{Agronomy Journal}, 83(\textbf{5}), pp. 794-799, (\url{https://doi.org/10.2134/agronj1991.00021962008300050004x}).

\bibitem{} Lobell  D.B. and Field C.B. (2007) Global Scale Climate vs. Crop Yield Relationships and the Impacts of Recent Warming.  \textit{Environmental Research Letters}, 2(\textbf{1}), pp. 1-7, (\url{http://doi.org/10.1088/1748-9326/2/1/014002}).

\bibitem{} McDermid J., Fera S. and Hogg A. (2015). Climate change projections for Ontario: An updated synthesis for policymakers and planners. Climate Change Research Report Nº44 (CCRR44) \textit{Ontario Ministry of Natural Resources and Forestry, Science and Research Branch}, 1-40 (\url{http://www.climateontario.ca/scripts/MNR_Pub/publication_summary.php?pubId=59}).

\bibitem{} Sacks W.J., Deryng D., Foley J.A. and Ramankutty N. (2010) "Crop Planting Dates, An Analysis of Global Patterns", \textit{Global Ecology and Biogeography}, 19(\textbf{5}), pp. 607-620, (\url{https://doi.org/10.1111/j.1466-8238.2010.00551.x}).

\bibitem{} Smit B., Brklacich M., Stewart R.B., McBride R., Brown M. and Bond D. (1989) "Sensitivity of crop yields and land resource potential to climatic change in Ontario, Canada", \textit{Climatic Change}, \textbf{14}, pp. 153-174, (\url{https://doi.org/10.1007/BF00142725}).

\bibitem{} Tol R.S.J. (2009) "The Economic Effects of Climate Change", \textit{Journal of Economic Perspectives}, 23(\textbf{2}), pp. 29-51, (\url{https://doi.org/10.1257/jep.23.2.29}).

\bibitem{} Wang J., Mendelsohn R., Dinar A. and Huang J (2010). How Chinese Farmers Change Crop Choice To Adapt To Climate Change. \textit{Climate Change Economics}, 1\textbf{3}, pp. 167-185  (\url{https://doi.org/10.1142/S2010007810000145}).
\end{thebibliography}
\end{document}